# Anyone Can Code:

# Algorithmic Thinking

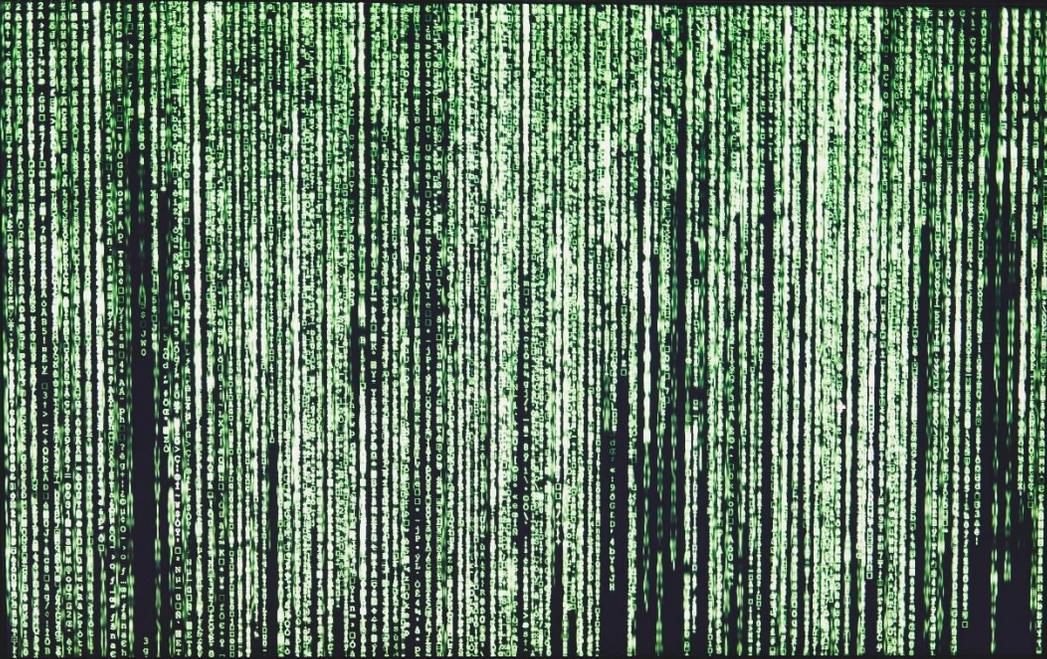

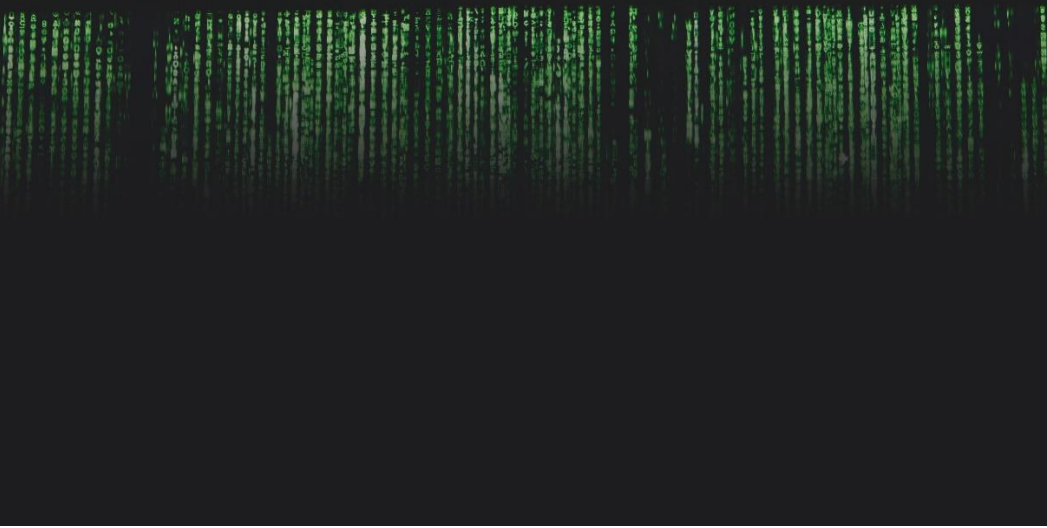

Ali Arya

# Anyone Can Code:

# Algorithmic Thinking

Ali Arya

First Edition © 2023

Cover Image by Marcu Spiske, free to use from Pexels.com

# Table of Content

# Preface

I will never forget the blank look on some of my students' faces when they were given a programming task and simply stared at the screen not knowing how and where to start. It took me a while to realize the problem was not that the development tool they were using was not friendly or the programming language was too complex. They had a solution in their mind, and they could remember (or look up) the language syntax. But there was a link missing between what was in their mind and a working program. This book is about that missing link, Algorithmic Thinking; how to organize our thoughts into good algorithms that can be easily translated into good programs. In other terms, in order to write a program, we need to first come up with the right algorithm (step-by-step process that computer has to follow), and to design good algorithms, we need to think algorithmically.

When I wrote *Anyone Can Code: The Art and Science of Logical Creativity*, I had a few things in mind as the key parts of helping readers learn programming. One of them was the notion of modularization, which became the foundation of the book and how it is organized. Another key feature of the book was to use multiple languages so the reader can see how code in different languages shares the same basic concepts while having syntax (and occasional functionality) differences. Algorithmic Thinking was also one of those basic concepts I wanted to discuss in that book. Writing good programs is about thinking logically and algorithmically, finding out the steps we need to take to achieve our goal and solve a problem.

While I introduced Algorithmic Thinking in the first *Anyone Can Code* book, I had to spend the main bulk of the book discussing the craft of programming, how to use programming languages, and organize a program. The examples I discussed often had simple algorithms, although some were quite complex due to the many modules they had. The goal was to show how complex programs can be designed using a collection of simple modules. I found it too ambitious of a goal to also delve into details of algorithmic thinking and how to design good (and complex) algorithms. That goal became the inspiration to write the second book in the *Anyone Can Code* series: Algorithmic Thinking.

We all come up with and use algorithms every day to achieve objectives and solve problems. My approach in this book is primarily based on using common problem-solving approaches and applying them to algorithmic thinking to design programs. I also continue using modularization as a key element. Any design involves combining simpler modules to make more complicated ones. Algorithm design is no exception. I introduce algorithmic thinking through a series of examples grouped based on the complexity of modules of code and data they are using. Similar to the previous *Anyone Can Code* book, I use multiple languages to demonstrate the algorithms. This seems like the natural choice to me, as the main point is to show that the algorithm is the key, not the language.

With the recent advances in artificial intelligence, it may seem to some of us that computer programming is an obsolete skill that now can be automated. But my belief is that new tools are always invested to make things easier, but they won't replace creativity. Algorithmic thinking is about creative problem solving and that is what we will continue to need. Intelligent tools may implement our solutions, but it will be up to us to come up with the right solutions and make sure the implementation is also right. Those old enough to remember the days of command-line compilers know better what I'm talking about when I say tools change. But we, humans, are still around doing what we do best. Solve problems.

# Notes

## Code Examples

Most code examples in this book are either in Python or C/C++, as they are very common and typical languages programmers use these days. Python and C/C++ examples are tested on Python version 3.7 and Microsoft Visual Studio 2019, respectively.

C and C++ are two different languages, but C++ is considered an object-oriented extension to C. While almost any C program can be compiled with a C++ compiler, C++ offers new ways of doing things such as input/output and memory management, and there are some behind-the-scenes or syntax differences in the way C and C++ compilers work. For example, `struct` keyword is used in both languages to define a structure. In C, your structure cannot include functions but in C++ it can. C++ structures are basically classes, but their members are public by default. Also, in C, `struct` is not a regular type, so if you have defined one, to define a variable you still need to use the keyword `struct`:

```
struct Student {

};

//… later in the code

struct Student s1;
```

You can use `typedef` keyword to make that a type:

```
typedef struct Student {

};

…

Student s1;
```

In C++, using `typedef` is not necessary and `struct` itself defines a new type:

```
struct Student {

}

…

Student s1;
```

Most modern C/C++ compilers deal with both languages. If your source file has extension .c, compilers follow C rules. If you use .cpp extension, the C++ rules will be followed. There may also be small differences based on each specific compiler and the version of C/C++ standard they use. For example, some compilers allow `main()` function to be `void` and some don't. All examples in this book work as .cpp files with Microsoft Visual Studio. Even if the examples do not use C++ features, I recommend using a .cpp extension to avoid compatibility issues. The variations from C to C++, or from Visual Studio to other development programs such as Apple Xcode are beyond the scope of this book.

If you have access to a digital copy of this book, you can copy/paste the code examples, but the pasted code may not be easily usable due to word-processing effects such as numbering and curly quotation marks. Also, note that the code within the book may have some details removed so you can focus on the parts relevant to the topic. Again, this may mean that copied code into your programming environment will not work. Instead, please find all examples with all their details, ready to compile and run, at https://ali-arya.com/anyonecancode. **Download ACC2-Examples.zip and look for README.txt in the main folder or in each of three folders: C for standard C/C++, SDL for graphics C/C++, and Python.**

## Signage and Numbering

Throughout the book, I use the following signs to show special text:

Practice and exploratory tasks

Notes, key points, and other extra information

Reflective and review questions

Code examples that form a full program are numbered, such as Example 2.1, which is the first example in Chapter 2. They use the suffixes a, c, and p to identify the language (algorithm, C/C++, and Python, respectively), such as Example 3.2p.

Figures, charts, and tables are called Exhibits and have similar chapter-based numbering.

# Chapter 1: Introduction

## 1.1. At the Restaurant

If you have recently started programming, you are probably familiar with the terrible feeling when we look at the computer screen, not knowing where to start and what to do. You are given a programming task but have no clue how searching for a word, sorting a list, or calculating a Physics formula are related to strangely named functions and keywords in the programming language you are supposed to use. You have seen the examples on the video tutorial or in your lecture notes or textbook, you know there are ways to create variables and write a function but can't make the connection to the given problem. You even know how to solve the given problem yourself, just not how to tell the computer to do it.

There is no need to be disappointed if something like that happens. There is a missing link and once you learn it, you will find the transition from your solution to a working program a lot easier. The key word there, though, is "once you learn it." It involves a certain way of thinking and presenting thoughts that you need to learn. Here is an example.

It's your birthday and you have invited all your friends to dinner at your favourite restaurant. They have brought you presents and, as the host, the dinner is on you. The bill arrives and wow: that is more than you expected. You have no intention of embarrassing yourself in front of your friends but want to have a quick check to make sure it is correct. Glancing at the bill while smiling at your friend's joke, you do a quick calculation. Well, too bad for you, it adds up. You pay and hope you are getting good presents.

Adding numbers is not a particularly complicated task and if we are fast, then we can do it even when pretending to listen to someone. But can you describe how you add a series of numbers to an imaginary person who doesn't have a clue and only knows how to add two clearly identified numbers?

Give it a try and describe the process clearly so it can be followed by that person. Obviously, if we say, "add them all," it's not going to work. It should involve only adding "two" numbers. So, let's look at the following description:

```
1.   Take the first number,
2.   Keep adding the others.
```

This description seems better as it involves only adding two numbers: the others (coming one by one) and the sum that we are building. This sum value is implied though, it is not mentioned specifically so that imaginary person may complain by saying "add the other number to what?" Also, keep adding until when?

So, let's try another version:

1.  In addition to the numbers you have, imagine a new number that is the "sum,"
2.  Assume sum is equal to zero,
3.  Read a new number from the list,
4.  Add it to the sum,
5.  If there is any number left on the list, go to step 3.

The above description is a set of instructions or steps that you have to follow to calculate the sum of numbers. Each step is simple and clear enough to be done by the person following the instructions. It may seem non-intuitive and too complicated to you at first. You may wonder why we can't just say "add them all." The reason is the lack of clarity and using implied information. It is not clear where the numbers are coming from, and it implies that we add them and store the result somewhere. Real-world problems that you want to solve (or you want to instruct others to solve) are not always simple. Complicated tasks have various steps, and you should be able to describe those steps; an ability that takes practice, starting with simple ones.

In the special case of programming, computers don't have the insight and assumptions you have. For example, doing the task of adding two numbers, you simply "remember" the result. But a computer needs to be told to keep that data in a memory location with a name or known address. So, computer programs have to say exactly what the computer has to do. Even the most advanced Artificial Intelligence (AI) programs are made of clear instructions, although their output may not be completely known in advance.

*Add tip and tax to the bill calculation process.*

As another example of how computers need specific instructions and don't have our assumptions, imagine that you have two data items called `first` and `second`. As a result of sorting or other operations, you need to swap the values of these two, i.e., if the values were `first=X` and `second=Y`, now you want them to be `first=Y` and `second=X`. A solution that may come to our mind quickly may be to simply state the new values:

```
first  = Y [1]

second = X
```

But this solution assumes that the values are known at the time of giving these instructions (that is what we call **hard-coded** value[2]), as opposed to swap "any value they have at the time". Another solution is to set the value of each variable to the other one's value:

```
first  = second

second = first
```

This solution works for `first`, but by the time we get to the next line, `first` has already lost its initial value and is now equal to `second`, so they both will have the same value (`second` won't change). The trick is to remember that before we set `first` to a new value, we need to "remember" its old value. When doing this in our mind, we implicitly remember the old value and use it when changing `second`. But computers need to be told specifically.

```
old_first = first

first  = second

second = old_first
```

The new variable, `old_first`, is a temporary one and the above lines are an example of what we commonly use to swap values of data items.

A description such as the above examples is called an **algorithm,** a set of steps we follow, each with an instruction to perform, in order to solve a problem. Algorithms are the foundation of computer programming and the missing link for many beginner and intermediate programmers. Jumping from what you know to a program can be too difficult and confusing. An algorithm is a midway step that allows you to gather your thoughts, know what you need to do, and then easily translate that into the language your computer understands. This book is about writing algorithms.

---

[1] Remember that in programming, wen we say `first=Y`, it doesn't mean `first` and `Y` are equal. It means "`set the value of first to Y`" which is what we call the **assignment operation**.

[2] Such data is called "**constant**" as opposed to "**variable**" that is a piece of data can change.

---

---

## 1.2. Algorithms

My cat, Aztec, loves to go for a walk around the neighbourhood, especially in the morning. He also likes to have his breakfast early and to have us around, paying attention to him. Over time, Aztec has developed a morning ritual, a fairly predictable process that gets him what he wants. It starts, to our complete annoyance, with him waking me and my wife up by scratching the bed and if necessary, some meowing. Once I'm up and call him, he jumps on the bed and rests on my belly a little. These are usually enough to get me out of bed and to the kitchen to give him his dry food. Aztec prefers his wet food, though, so he nibbles on his dry one and then goes to the kitchen patio door and scratches that. I open the door, but he won't go out, maybe just for a couple of seconds to get some air. He looks up at me and meows, letting me know that he is not going anywhere yet. If necessary, he repeats this a couple of times until I get the can of wet food, open it for him, and he eats half of it. Then he goes to the patio door again and scratches. I let him out, and usually, after 10-15 minutes, he comes back and finishes his food. I call this his scratch-and-meow ritual.

Every morning, Aztec solves a problem: how to get company, food, and some fresh air. In order to solve this problem, he has developed a solution in the form of a series of well-defined steps that follow and depend on each other, and are known to achieve the expected outcome. Cats are not the only intelligent beings who develop such solutions. We humans frequently do that too. Not including Aztec, who doesn't usually wear any clothes, laundry day in our home involves getting clean clothes for us parents, and our son. We have developed a system to make sure there is a central control (with kids you need that) and also some individual responsibility (with kids you need that too). Our system is called fold-and-pile. The person or persons who get the clothes out of the drier, fold and stack them into three piles (with or without help). Each person is then responsible for getting their pile and putting the clothes in their drawers.

A more systematic and universal multi-step solution is what we all learn (or should learn) in school: addition using digits. When we add two numbers, we start by adding the right-most digits (the first column) and writing the result below them. If the result is 10 or more, we keep the right-most digit and "carry" the second digit (always a one) over to the next column. We then continue these steps all the way to the last column of digits on the left. I challenge you to remember long division and define the steps for that.

Scratch-and-meow, fold-and-pile, and carry-over are examples of what we, more formally, call an **algorithm**. What makes all these solutions similar is that they are (1) based on a sequence of clear steps, and (2) they are known to achieve a certain goal. Each step in an algorithm is an instruction that is clear to and doable by the person executing it.

To understand algorithms better, it helps if we look at an example that is not an algorithm. In math, we are told that the shortest distance between two points (in Euclidian geometry, at least) is a straight line. So, if we want to go from point A to point B as soon as possible, our first choice is to move straight. The go-straight solution can be defined as these steps: (1) identify A and B, (2) find the straight line, (3) move along the straight line. But there is no guaranty that this solution will work. In fact, in most cases, it won't because there are obstacles on our way. Still, go-straight is a good starting point that can be the base for a solution which we form by adding details and variations. For example, if we see an obstacle, we go around it and then continue. Go-straight is not an algorithm because it lacks determinism and reliability. It is what we call a **heuristic** or **rule-of-thumb**, another common problem-solving approach.

---

*Algorithms are sets of clear instruction that can solve a problem or perform a task.*

---

Algorithms are a fundamental part of our lives, even when we don't notice them. They allow us to break a complicated task into smaller steps, see how those steps are related as a process, and make sure that we achieve our goals as long as external and unexpected factors don't affect our process. For an algorithm to be successful (correct), the steps need to be properly related to each other, so each can prepare us for the next. We sometimes call this the **logical flow**. Remember that logic is about the rules of arranging statements in a way that they can build on each other and support an argument. So, thinking in terms of algorithms can help us be more logical, and logical people are more skilled at defining algorithms.

Algorithms are not limited to human (or cat) use, though. One of their advantages is that they can help build machines that perform tasks on their own, especially computers.

## 1.3. Programs

As shown brilliantly in the opening scene of the movie *2001: A Space Odyssey*, making tools and machines is one of the characteristic abilities of humans that allowed us to actively control our environment. Many machines, from a simple lever to a complex vehicle, work manually, i.e., there is a human operator using them. But there are also machines that perform certain actions automatically. The process that these automated machines have to go through can be defined through various means such as mechanical, hydraulic, or electrical elements. An input generally initiates the process, and each element performs its action and activates the next element. A good example of such an automated process is a chain reaction, as shown in Exhibit 1.1.

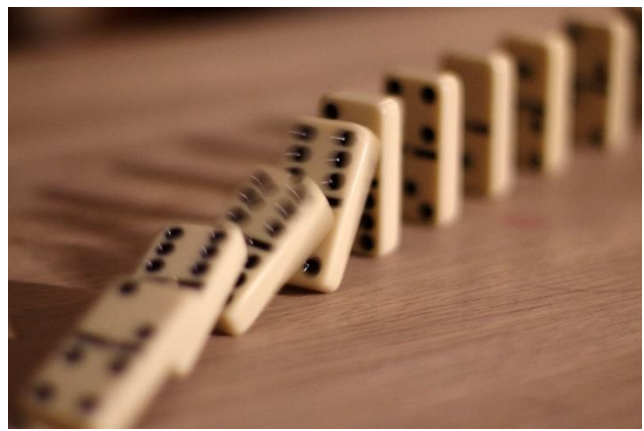

**Exhibit 1.1. The Domino Effect Chain Reaction**

https://en.wikipedia.org/wiki/Domino_effect#/media/File:Dominoes_falling.jpg

The Domino Effect is a simple chain reaction where each step is a simple action: falling down. Other chain reactions can be more complicated and include combinations of mechanical, electrical, chemical, optical, and other elements with time-consuming actions. Such sequences of actions are good demonstrations of the algorithms. A series of clear steps, taken in a certain order, to achieve a defined result. Another interesting and related example is an automaton (Exhibit 1.2). Generally speaking, automata are machines that operate on their own. While many of our modern-day machines are self-operating, this concept was not common in pre-industrial societies. They are mentioned in many ancient historical records, for example, from Greece, Egypt, and China. Jewish stories talk about King Solomon creating a throne with mechanical animals who interacted with him. Using wind and water, going through tubes and other mechanisms, ancient inventors managed to design devices that would make sound and move. Such devices were the original Cuckoo Clocks[1]. Science and engineering advances in medieval and renaissance eras increased attention and interest in automata and resulted in a wide range of products from clocks to figurines capable of writing and dancing. Later on, automata lead the way to robots who are now used commonly in many industries and even households.

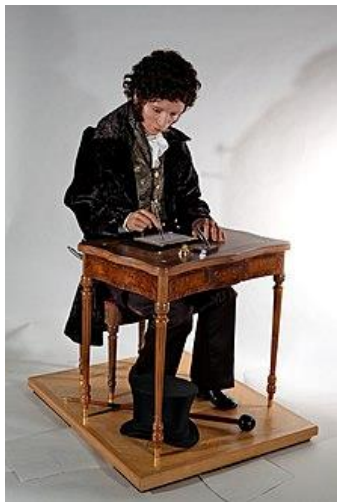

**Exhibit 1.2. Automaton**

---

[1] *The Rise and Fall of Alexandria: Birthplace of the Modern World* by Justin Pollard and Howard Reid

Automata and robots differ in the fact that the former is designed to look like it is working on its own, and the latter is simply a tool for performing a desired task instead of human operators. But both implement the concept of algorithms, as a sequence of actions leads to the final result. For example, the slow opening of a spring eventually reaches a point when it triggers a switch which then starts the flow of air, followed by the movement of gears, and through the complicated combination of these gears, the movement of arms of a mechanical person. In other terms, automata and robots follow **programs**, i.e., implementation of an algorithm through specific elements made to perform each step of the algorithm. If the algorithm is asking for "waiting for 5 seconds," a spring that slowly opens and triggers a switch in 5 seconds is an implementation of that algorithm. Another implementation of that algorithm could use a container that fills up with water and overflows in 5 seconds.

Machines that run a program (implement an algorithm) have the sequence of actions defined in some form. Sometimes, this sequence is embedded within the structure of the machine and cannot be changed without structural modifications. Many automata were in this category. Some music boxes, on the other hand, can play different songs using metal or paper tapes with homes in them. These holes specify the note to play, and mechanical elements in the box turn this into sound. These boxes later lead to record and tape players and demonstrate a limited case of changing the program. Here, the "functionality" of the device doesn't change, but its "content" changes. An abacus or early electronic calculators are also machines that perform a functionality (calculation) but they have no built-in sequence of automated actions nor any stored content. In other terms, they don't run a program, although they may help us manually implement an algorithm.

A **computer** is also a machine that helps with calculation, but it runs programs, in this case, **a sequence of instructions implementing an algorithm**. For example, the following instructions can tell a computer to calculate the average of two numbers:

Algorithm

```
1.   Get the first number
2.   Get the second number
3.   Add two numbers
4.   Divide the results by two
```

Program (implementation of the above algorithm in Python language)

```
1.   num1 = 4
```

```
2.    num2 = 3
3.    sum = num1+num2
4.    avg = sum / 2
```

*A program is an implementation of an algorithm in a specific language. A computer is a device that runs programs.*

The ability to run a program is what makes a computer different from a calculator[1]. The ability to change its program relatively easily is what makes it different from an automaton. Note that having a keyboard and screen is not required for computers, neither is being digital or even electronic. There are analog electronic[2] and even mechanical computers, although they are not used often and not considered in this book. Computers traditionally, but not necessarily, follow the Von Neumann model (Exhibit 1.3) that defines a computer as a **Central Processing Unit (CPU)** that reads instructions in the program and executes them, plus some form of input/output devices and a memory unit. Programs and other information reside in the memory and are accessed by the CPU.

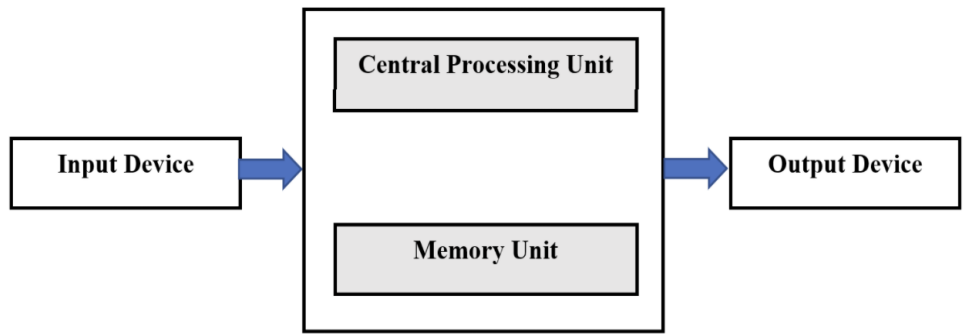

**Exhibit 1.3. Von Neumann Model of a Computer**

---

[1] I am only talking about simple old calculators here. Recent calculators are in fact small computers and can be programmed.

[2] Digital electronics uses a binary system meaning that each point in the electrical circuit can only have two discrete acceptable values, while in analog electronics values are continuous.

Programs can be written in different **programming languages** due to reasons such as programmer's preference or system requirements that suit or enforce a certain language. Some programing languages such as Python and C/C++ are human readable. To be executed by a computer, they need to be translated into a machine-readable language, that is designed for different types of CPU. The discussion of the languages and the reasons to choose them is beyond the scope of this book. The choice of language affects the final properties of the program, such as size, speed of execution, ease of development, and supported features. But at the end of the day, all programs are implementation of algorithms. While programming does require the knowledge of specific languages, it is, before anything, the art and science of finding the right algorithm.

## 1.4. Programming and Algorithmic Thinking

Imagine the common programming task of finding the maximum value in a series of numbers, let's say, 10, 1, 13, and 8. If we are doing this ourselves, we simply look at the numbers and recognize the largest one, 13. But remember two problems that I mentioned at the start of this chapter: (1) not all tasks are as simple, and (2) computers need to be told what to do. To deal with these problems, we need to come up with a description of what to do that has the following characteristics:

- It is clear and not open to interpretations and assumptions. For example, if we add two numbers and intend to use the results in future, the description says where (or under what name) to keep that data. Or if we are supposed to perform tasks that are similar, it specifies what to do and under what conditions or how many times.
- It works for any other set of numbers of any length, at least within a certain range. This means we don't assume we have 4 numbers, or they are 10, 1, 13, and 8.

Here is a possible set of steps to find the maximum value:

```
1.  In addition to the numbers you have, imagine a new number that is
    called "max,"
2.  Assume max is equal to the first number,
3.  Read a new number from the list,
4.  If max is smaller than the number, then set max to the number,
5.  If there is any number left on the list, go to step 3.
```

This algorithm is not limited to any set of numbers and seems generally clear. But does it rely on any assumptions? Look at line 3. It assumed we know how to "read a new number." It could be that we have a user who types the numbers, and then the instruction at line 3 could mean show a prompt and ask the user to enter a new value. On the other hand, we may have an array of data and in that case, we need to point to a new member of that array. In other terms, working with an array, we need to remember which member of the array we are working with. Just like max, this information about "which member" has to be stored somewhere and processed later. Let's call that "index" and modify the algorithm to use it:

```
1.  In addition to the numbers you have, consider a new number that is
    called "max,"
2.  Assume max is equal to the first number,
3.  Consider a new number that is called index,
4.  Assume index is zero,
5.  Read a new number from the list at index,
6.  If max is smaller than the number, then set max to the number,
7.  Increment index
8.  If there is any number on the list at index, go to step 5.
```

Do you see any clarity issues in line 5? Can the number be used in line 6 if we don't store it somewhere with a name? The reference in line 6 to "the number" may seem very clear to us but for a computer, it may not be as clear. Subsequently, when you start writing the actual program in a specific language, this vague reference may cause you problems if it is not about a clear piece of data. Recall that in programming terminology, we refer to data that can change as **variable**. We usually identify variables with names and values, although they also have types and memory addresses. Here is a clearer version of our algorithm and its Python and C/C++ programs[1]:

### *Example 1.1a*

```
1.  Consider a new data called Max and set it to the first list value,
2.  Consider a new data called Index and set it to zero,
3.  As long as Index is less than the array length, repeat the next lines
```

---

[1] The suffixes a, p, and c at the end of an Example number mean algorithm, Python, and C/C++, respectively.

```
4.      Set a new data item called Number to the list value at Index,
5.      If Max is smaller than Number, then set Max to Number,
6.      Increment Index
```

### *Example 1.1p*

```
1.  Data = [10, 1, 13 , 8]  # array of 4 data
2.  Max = Data[0]
3.  for Index in range(4):
4.      Number = Data[Index]
5.      if Max < Number:
6.          Max = Number
```

### *Example 1.1c*

```
1.  int Data[ ] = {10, 1, 13 , 8};
2.  int Max = Data[0];
3.  for (int Index=0; Index < 4; Index++) {
4.      Int Number = Data[Index];
5.      if (Max < Number)
6.          Max = Number; }
```

Lines 3 to 6 in the algorithm implement an iteration or loop, i.e., a part of the program that repeats. Lines 1 and 2 set the basic element of the loop: a variable that defines the condition.

You see that once the algorithm is clearly defined, writing the program in any language is fairly easy and simply needs the knowledge of syntax in that language:

- C/C++ programs need to define the type of variables, but Python programs don't.
- C/C++ programs end each instruction with a ; but Python programs don't need that either.
- The syntax for creating an array in Python and C/C++ is somewhat different but they are mostly similar and follow the same logic.
- The syntax for creating a loop in Python and C/C++ is also different but they follow the same logic: Index is the key part.
- In C/C++, we initialize it at the start of the whole loop, check the condition at the start of each iteration, and increment Index at the end of each iteration. That is what `for` keyword does. The increment and compare are explicit and can in fact be other operations.

- In Python, `for` keyword does something similar by defining Index and a range between zero and 4. The increment and compare parts are implicit and built into the operation. This is easier but less flexible.

The process of replacing vague and ad-hoc solutions with clear algorithms is called **Algorithmic Thinking**. You may also hear terms such as Logical Thinking and Logical Creativity. They are not exactly the same but overlap significantly. Logical thinking emphasizes a type of thinking that is based on logic and reason, rather than emotions or chance. Logical creativity adds the component of creative problem-solving and novel ideas and designs. When talking about algorithmic thinking, we are focusing on the well-defined sequence of steps. These steps are logically related and commonly result in creative ways of solving a problem or designing a system.

Algorithmic thinking is not limited to computer programming, but in this book, I consider it as the process of algorithm design for computer programs. In other terms, there are two aspects to algorithmic thinking in software design:

- Thinking of computer programs as solutions that are based on an algorithm,
- A well-defined way of thinking for designing those algorithms and programs. My goal is to give you tools and methods that help with that design process.

As demonstrated in the above example, algorithms define the logic of the program while language syntax controls how to implement that logic. Separating these two is particularly helpful for many beginner and intermediate programmers.  They get confused by language syntax and struggle with it, unable to get their thoughts into a working program, sometimes not even knowing where to start. Algorithmic thinking allows us to understand and visualize our program before even starting with any confusing syntax. Once we know what the program does in full clarity, it will be a lot easier to translate that into any specific language.

---

*Do we need the variable Number in Examples 1.1? It may add value in some cases where we have to call a function to get the number from a database, but if we have Data[Index], then maybe not.*

---

Algorithmic thinking is a skill to learn if you intend to write computer programs. For those who want to jump into coding and write a program quickly, this may sound like an extra thing to do. We can't run algorithms and see results. That is why many people skip them. The result, unfortunately, is a weak foundation both for the program we are writing and for our software design abilities. Without a strong foundation, sooner or later, you find yourself looking at that blank screen not knowing where to start. The good news is that algorithmic thinking has its roots in our everyday activities and many existing problem-solving techniques can help us strengthen our ability to think algorithmically. I will review these in the next three chapters but there is one more thing that I need to introduce before wrapping up the Introduction.

*Algorithmic thinking is a well-defined way of designing programs through establishing algorithms and using a series of tools and methods discussed in this book.*

Just like we break down a task into a series of clear steps within an algorithm, we break down large and complex systems into simpler parts, or modules. It is difficult, or impossible, to write a single algorithm to describe how a word processor or a game works. But once you define them as a set of modules, you will find it easier to have a group of algorithms each describing one part of your program. **Modularization** is another key concept in programming and software development, closely related to and almost inseparable from algorithmic thinking. Together, they are the foundations of this book and my proposed approach to software design and development. Chapters 5 to 8 show how a modular approach can help us develop more complex algorithms and programs. But first, let's have a closer look at what modularization is.

## 1.5. Modularization

I studied engineering in my undergrad years. I feel very proud and happy about the foundation it gave me. Designing a complex object by assembling simpler parts is a common engineering process. When you plan to design a machine, you hardly design everything from scratch. The art of design is to see how to reuse existing components (or modules) and design new ones only if necessary. This saves you time and also allows relying on proven expertise that is gone into the design of those existing components. Nothing demonstrates this better than LEGO ©: We have simple basic components that build parts, and then parts can be used in different projects.

Such a component-based (modular) design is not unique to engineering. Architects, industrial designers, and artists do that too. Arches, doors, and windows are modules of architectural design, just like legs are reused modules for a chair[1].

Software Engineering is an area of computer science and information technology that applies engineering processes and insight to software development. Reuse-based software process is one of those but designing software programs as a combination of simpler modules is not limited to any specific process. Almost all programs revolve around modules of **Data** and **Code**. While these two are not commonly separable, Data refers to the information in a program while the word Code refers to the operations. A line like "`int x;`" in C is indeed a line of code but is mostly creating a piece of data and doesn't show what we do with it.

Consider a task such as double-clicking on a filename (say an image) and opening the file to show the contact. This uses a module of data corresponding to an image, and also another module of data for a file of any type. It also uses a code module for opening a file, another for parsing image data, and finally a last one for showing the image on screen. It is possible to write an algorithm for the whole process but is more effective and reusable to divide it into modules and design the algorithm in a modular way.

### *Example 1.2*

Algorithm for Double-Click on Image filename

```
1.   Open the file
2.   Read the image data
3.   Show the image
```

Algorithm for Opening a file

```
1.   Read the filename
2.   Pass the filename to operating system module for File Operations
3.   Open a stream of data
```

---

[1] I use the terms component and module interchangeably here. But they are not the same. In software development, module is a more general term that can be simple or very complex, data or code or both. A software component on the other hand, is usually a set of related data and functions.

```
4.   Read bytes until the end of file
5.   Close the stream of data
```

As shown in the above example, algorithms can be defined for modules of code that together make more complicated modules. We can also define algorithms for modules of code that all process the same data. For example, an image can be represented using a module called ImageData with the following members:

- Width and
- Height
- Filename
- Bits per pixel
- Array of pixels data

Modules of code and their algorithms can then be defined for all operations that need to be performed on an image, such as opening, reading, and showing. Throughout the book, I use modularization as the means of progressing through algorithms from simple to more complicated. We start with algorithms that work on **simple (a.k.a. basic or atomic) data**. These are data types commonly defined in programming languages that cannot be divided into smaller parts, such as integer, float, and character. Then, I move to algorithms that work with **compound (a.k.a. aggregated) data**. These are made of combining atomic data into modules, for example arrays of simple types and user-defined types (such as Image above). I continue this to hierarchies of modular code (modules made of other modules) and their algorithms, and then finally discuss algorithms that use objects and classes. But before all that, I will take a quick look at algorithmic thinking in our daily lives. This allows us to use what we have already learned and apply it to what we are learning, computer programming. Similarly, later I will discuss creativity and creative problem-solving that introduces new techniques for algorithmic thinking.

*Modularization allows grouping related pieces of code and data to allow better management of the program. Algorithms can be defined based on what types of modules they use, just like they are defined based on the actions they perform.*

*What other examples of modular design can you think of, in programming or beyond that?*

## Highlights

- An algorithm is a set of clear instructions to solve a problem.
- Computer programs are instructions that implement an algorithm in a specific language to be executed by a computer.
- Algorithmic thinking is a well-defined way of designing programs through establishing algorithms and using a series of tools and methods discussed in this book.
- Modularization allows grouping related pieces of code and data to allow better management of the program. Algorithms can be defined based on what types of modules they use, just like they are defined based on the actions they perform.

# Chapter 2: Algorithmic Thinking in Daily Life

## Overview

In our everyday lives, we solve problems regularly. They may seem trivial, and we may not even notice, but problem-solving skills exist in all of us to some level. In many of these cases, we subconsciously develop algorithms (step-by-step solutions), although there are other forms of problem-solving such as intuition, that go beyond the scope of this book. The best way to learn about algorithm development in programming is to revisit how we develop algorithms in daily life and make a conscious effort to mindfully think about and in terms of algorithms when solving common problems. This allows us to see how algorithms are built and used. So, in this chapter, I will look at some examples of algorithmic thinking in our daily lives.

## 2.1. Sequential Execution

In the Introduction, I reviewed some simple algorithms for mathematical problems. Math was an obvious choice for algorithmic thinking, but we follow algorithms in other daily activities too. Cooking is another good example. Let's consider frying an egg described in the following example algorithm:

***Example 2.1***

```
1.   Warm up the frying pan.
2.   Add oil.
3.   Break the egg.
4.   Let it cook (to the level you want).
5.   Flip and cook the other side (if you want).
```

Notice that our simple algorithm is flexible to your special preferences. Some of us may be happy with this but some may want more details, for example, how many eggs? how much oil? cook for how long? The instruction set above tells us what the operations are, but it has missing information. Programs perform operations on and with information. Move, how far and in what direction. Add, what number to what number? Increase, by how much? A good recipe will give you such information in addition to the operations. Similarly, a program or an algorithm are made up of operations and information, code and data.

Let's take a look at a more complicated recipe[1]:

***Example 2.2***:

<u>Ingredients</u>

- `1 ½ cups all-purpose flour`
- `3 ½ teaspoons baking powder`
- `¼ teaspoon salt, or more to taste`
- `1 tablespoon white sugar`
- `1 ¼ cups milk`
- `1 egg`
- `3 tablespoons butter, melted`

<u>Directions</u>

1. `In a large bowl, sift together the flour, baking powder, salt and sugar.`
   `Make a well in the centre and pour in the milk, egg and melted butter; mix`
   `until smooth.`
2. `Heat a lightly oiled griddle or frying pan over medium-high heat. Pour or`
   `scoop the batter onto the griddle, using approximately 1/4 cup for each`
   `pancake. Brown on both sides and serve hot.`

Here, we start with our data (what ingredients and how much of each). Then, we give the code (what to do with them and in what order). Compare this to the algorithm and programs of Examples 1.1, where we did the same thing.

*Code and data (operation and information) are not really separate. Operations (as in "what to do") is, in fact, a form of information. Information, on the other hand, is itself created using operations ("int Max = Data[0];" in Example 1.1c).*

---

[1] https://www.allrecipes.com/recipe/21014/good-old-fashioned-pancakes

---

The above simple algorithms demonstrate a basic characteristic of computer programs that we refer to as **sequential execution**. Programs and algorithms are executed (run) step by step. That is the default behaviour and necessary for the causal relation between instructions in each step; each operation depends on what was done before and prepares for what is coming after, and the program flows linearly forward. However, this default behaviour doesn't always apply. For one thing, there are parallel processors and multi-threading that allow programs to run as parallel branches. But those are beyond the scope of this book, and each branch in them is still sequential. Sequential execution, as in the simple linear one-directional flow of operations, gets more complex in almost any program through three mechanisms: selection (decision-making), iteration (repeating), and function (a module of code). These mechanisms are based on three programming constructs, i.e., a certain arrangement of instructions. Let's explore these in some daily life examples.

---

*Linear sequence is the simplest programming construct. More complex constructs like selection, iteration, and function call add complexity to the program to provide new features.*

---

## 2.2. Selection

My wife and I are Eric Clapton fans, and his *Wonderful Tonight* was our wedding song. There is part in the songs that says:

*It's late in the evening.*

*She's wondering what clothes to wear.*

*She puts on her makeup.*

*And brushes her long blonde hair.*

I am sure all of us have experienced the "wondering what clothes to wear" part of the evening he is describing[1]. Our experience probably follows something like the following example:

***Example 2.3v1***

```
1.   Think of where I'm going
2.   If it is a date
3.       Then dress fancy
4.   If it is to work
5.       Then dress semi formal
6.   If it is the gym
7.       Then wear work-out clothes
8.   If none of those
9.       Then put on something casual
```

In this algorithm, the execution no longer flows sequentially and in one direction. It branches out into multiple directions based on certain conditions. In other terms, we make decisions and select which way to go and what operations to perform, based on the information we have. Each branch can actually get more complicated with further decisions to make. See version 2 of the previous example to see how our step 2 can have follow-up decisions:

***Example 2.3v2***

```
1.   Think of where I'm going
2.   If it is a date
3.       If it is a first date
4.           Then dress fancy in what I think I look the best
5.       If not the first date
6.           Then dress nice; what I look good and my date seems to like
7.   If it is to work
8.       Then dress semi-formal
9.   If it is the gym
```

---

[1] In case you like Taylor Swift better: "First date. Can't wait. … Wondering what to wear."

```
10.       Then wear work-out clothes
11.   If none of those
12.       Then put on something casual
```

Many of us deal with seemingly trivial problems like "what to wear" and "what to eat" all the time. We do solve them through a process that may not look as organized as these algorithms. And we may spend a long time and get confused and frustrated. The ability to structure our process with algorithmic thinking helps us avoid the waste of time and frustration. It helps us see the options more clearly, have a better understanding of the situation and our criteria, and take actions more efficiently and effectively.

---

*Find some other examples of selection-based algorithms.*

---

## 2.3. Iteration

In the previous example, I showed how the linear and one-directional flow of execution can change when we define "forks" using the selection mechanism. While the flow is still sequential, as in step-by-step, it no longer follows a single direction and can have multiple paths. Now let's look at another example that the flow gets even more complicated by going "back".

Imagine you are a schoolteacher and want to choose the tallest student in a class to wave a flag in your sporting event. All of us can look at a group and identify the tallest person without thinking much about algorithms, but in fact we are following one, maybe even subconsciously. We are probably scanning the group noticing a tall person (taller than those next to them), keep that person in mind and scanning the rest. If we see someone taller, we consider that the tallest, and keep scanning for someone even taller. This is how the Max algorithm in the previous chapter worked.

***Example 2.4v1***

```
1.   Consider the first person as the tallest
```

```
2.   Check the next person.
3.   If taller than the current tallest
4.      Then consider them the tallest
5.   If more people, go to Step 2
```

This algorithm works based on **iteration** (repeating), which is to define a loop in your algorithm/program, i.e., a set of steps that are executed and then repeated as many times as necessary. In this case, the loop is built by steps 2-4 that will be repeated until there is no person left.

If you are dealing with a small group, one quick scan will give you the tallest. But when working with large groups, you need a clear method otherwise you can get confused and make mistakes. Algorithmic thinking is the tool that gives you that clear method. One of the ways to make the algorithm clearer is to define iteration better. In the above case, the iteration happens by looping through steps 2-4, but when looking at the steps, we cannot see any loops until we get to the end. Another, clearer, version of defining an iteration is shown below:

***Example 2.4.v2***

```
1.   Consider the first person as the tallest
2.   While there is another person:
3.      Check the next person.
4.      If taller than the current tallest
5.         Then consider them the tallest
```

The difference between these two ways of defining the loop (go to vs. while) is that v2 shows the boundary of the loop in a clearer way and is forward-looking (we say from here we repeat) but in v1 the boundary of the loop is not easy to detect, and the go-to statement works backward. In general, the use of go-to statements is considered problematic in programming and the dominant programming paradigms, such as Structured Programing, recommend avoiding it.

*Find other examples of iteration-based algorithms.*

Another example of iteration is searching through a phone book to find someone's number. If it is an old-fashioned printed book, the algorithm you follow will be similar to Example 2.4 with these differences:

- No need to start with an initial guess
- Instead of person we have name entry
- Instead of comparing to the next person, we compare to the name we are looking for
- Instead of considering the tallest, we pick the related phone number

If you have a smartphone, the device is doing the search for you when you enter the name but if you enter partial name (for example, the first letter) then it shows you the matching names, you scan them, and then tap on the correct one. You notice that the operation is always following a certain pattern:

- Have an initial action if needed that is not going to be repeated
- Check a condition
- Perform an action

---

*How is selection used in iteration? Can we have iteration without selection?*

---

Iteration is a very useful construct in algorithms and programs, because in many cases we are dealing with a series of similar operations performed on different information. The power of computers, in fact, lies in their ability to perform precise operations very quickly. This power shows its real value when we are dealing with many operations.

Iterations can also be combined. Imagine you are lining up a group of students and decide to sort them by height, shorter ones closer to the front and taller ones in the back. How would you do that? There are probably multiple ways but one common one will be to look for the tallest and put them in the back, then look for the tallest of the remaining people and put that person in line, and continue until everyone is lined up. This method is simple because it relies on "finding the tallest" algorithm that we have already developed. The sort algorithm will look like Example 2.5v1.

*Example 2.5v1*

```
1.  While there are people in the unsorted group
2.      Find the tallest
3.      Add the tallest at the end of the sorted group
4.      Remove tallest from the unsorted group
```

Looking up the details of "find the tallest" algorithm and replacing line 2, we will have the full algorithm.

*Example 2.5v2*

```
1.  Get the sorted group
2.  Make an empty unsorted group
3.  While there are people in the unsorted group
4.      Consider the first person as tallest
5.      While there is another person:
6.          If next person taller than the current tallest
7.              then consider them the tallest
8.      Add the tallest at the end of the sorted group
9.      Remove tallest from the unsorted group
```

Example 2.5 adds two lines to Example 2.4: at the start it defines a new loop to repeat finding the tallest, and at the end of that loop, it puts the tallest in a new group. Note that the tallest person we find is removed from the unsorted group, so in the next iteration we won't consider them. Alternatively, you could put them at the back of the group and in the next iteration, start after them. Either way, the sorted people will not be considered in the search for the tallest in each iteration.

## 2.4. Combining Iteration and Selection

We all have learned to perform basic arithmetic operations (add, subtract, multiply, and divide). Mathematic operations are good examples of algorithmic problem-solving. Think of an addition to start. How do we add 5 and 3? Adding two one-digit numbers is the basic operation in addition. We do it in different ways, maybe counting up from the first number, and counting as many times as the second number: 5 plus 1 is 6, 6 plus 1 is 7, and 7 plus 1 is 8. We could also use our fingers. Either way, for most of us, this operation is now so basic that we have memorized all the results. But what about numbers that have more than one digit, say 27 and 15? That is where we use carryovers. Here is an algorithm. See if it matches how you add:

***Example 2.6***

<u>Addition (results shown for 27 and 15)</u>

```
1.   Start from the right-most digits of both numbers. 6 and 5
2.   Add the two digits and get the result. 11
3.   If the result is less than 10
4.       Then write it under.
5.   Else (If the result is 10 or more)
6.       Then write only its right digit and consider the left digit (has to
     be a 1) as carryover. 2 and a 1 as carryover
7.   Now move to left and get the next digits of the two numbers. 2 and 1
8.   Add the two digits and get the result. 3
9.   If there is a carryover
10.      Then add it to the result. 4
11.  Else (If the result is less than 10)
12.      Then write it under. If the result is 10 or more, write only its
     right digit and consider the left digit (has to be a 1) as a
     carryover. 42
13.  If there are more digits
14.      Then go to step 4
```

Subtraction is very similar to addition, except that instead of carryover to the left digit, we borrow from it if needed (for example if we are subtracting digit 5 from digit 3). So is multiplication; instead of adding digits, we multiply them and then use the carryover.

*Write Example 2.6 using a while loop.*

*Write an algorithm to find the digits of a given number. Hint: Divide the number by 10 and consider the remainder, then continue.*

*See if you can write the algorithm for subtraction and multiplication.*

## 2.5. Functions

Looking at Example 2.5, you may have noticed that v1 looked much simpler as we did not include the details of how to find the tallest person in a group. This version of the algorithm assumes we know how to find the tallest person and don't need to worry about describing that process. In other terms, we have a stand-alone module for that. **Modularization** is a key in any type of problem-solving (including computer programming) because it breaks down a complex problem into smaller and more manageable ones. Sometimes these modules themselves include yet smaller ones to form a hierarchy, where we have the main problem at the top and at each level, we have more but smaller problems (or tasks to solve the higher problem). The most common modules in computer programming are **functions**[1] (for code) and **structures** and **arrays** (for data). We also have **objects** that combine code and data. I will discuss algorithms using these modules in more detail in later chapters. For now, let's have a quick look at the first (and probably the most commonly used) one: function. We will see how a function can help make an algorithm easier to design and manage.

One of the things that make weekend mornings enjoyable is the freedom to sleep in, eat late breakfast, and do things as we feel like[2]. But in workdays, we usually have a morning routine. Such a routine is made of a series of tasks, each with its own way of doing. If you don't want to end up with the situation my wife calls "headless chickens running around," then you want to have your routine and tasks well-defined. What simplifies the routine is to define it in term of tasks and not every single step we take. Here is an example of a possible routine:

***Example 2.7***

<u>Morning Routine</u>

```
1.   Personal grooming
2.   Making coffee
3.   Packing lunch while coffee is brewing
4.   Breakfast
5.   Dressing up
6.   Getting out
```

---

[1] In older texts, functions were called sub-routines, as opposed to routine or which was the main program.

[2] That is, at least, in an ideal world. If you have kids, for example, things may not be as simple as that.

Each one of these tasks, in turn, has its own steps. For example, eating breakfast can be something like this:

<u>Breakfast</u>

1.  `Toast bread`
2.  `Get butter and jam from the fridge`
3.  `Pour coffee`
4.  `Eat`

When designing your morning routine, you don't want to get into these details though, as it complicates the routine and may change day by day. Having a module (function) called `Breakfast` allows us to define the routine without worrying about what we eat for breakfast and how. Those can be defined later or decided by our spouse or parent!

*Functions are the basic modules of code that allow ease of management and also re-use.*

Another example can be preparing for travel. I lived in Vancouver for many years before moving to Ottawa and starting a family here. I still have many favorite people and places there, so if I plan to go for a summer family vacation, there are so many things we want to do. Then there are flight and hotel to be booked, and pet-sitters to be arranged. The process is quite complicated, but modularization helps. The algorithm may look like Example 2.8.

***Example 2.8***

1.  `Select dates`
2.  `Book flight`
3.  `Book hotel`
4.  `Book pet sitter`
5.  `Plan activities`

The algorithm is designed based on the assumed priorities. We select the dates first because there is no point to go when my friends are not there, or my son is still going to school. Finding good flights is trickier than good hotels, so that is done next. Other things come after in the order of priority we assume. Each of these five steps is a module with its own algorithm. For example, selecting the dates requires coordinating among the family and also with people we are visiting, not to mention checking the calendar of events in Vancouver for possible festivals or other scheduled attractions. Of course, the result of each module may affect the previous steps. For example, if a much cheaper flight can be found if we go one day earlier, then step 1 has be revisited. Having well-defined modules makes this process easier because we know what to do or repeat.

*Break down a task you perform to a series of modules, and then write the algorithm for each. See how this modularization helps manage the task.*

*Can you re-use some of the modules in another task?*

# Highlights

- Code and data (operation and information) are not really separate. Operations (as in "what to do") is, in fact, a form of information. Information, on the other hand, is itself created using operations ("int Max = Data[0];" in Example 1.1c). But for conceptualization and program management, it helps to think of these two distinct aspects of the program.

- Linear sequence is the simplest programming construct. More complex constructs like selection, iteration, and function call add complexity to the program to provide new features.

- Functions are the basic modules of code that allow ease of management and also re-use.

- Exhibit 2.1 shows how three constructs we discussed (selection, iteration, and function) affect the flow of execution in an algorithm or program. While the process is still step-by-step, each of these constructs causes a certain deviation from the linear one-directional flow.

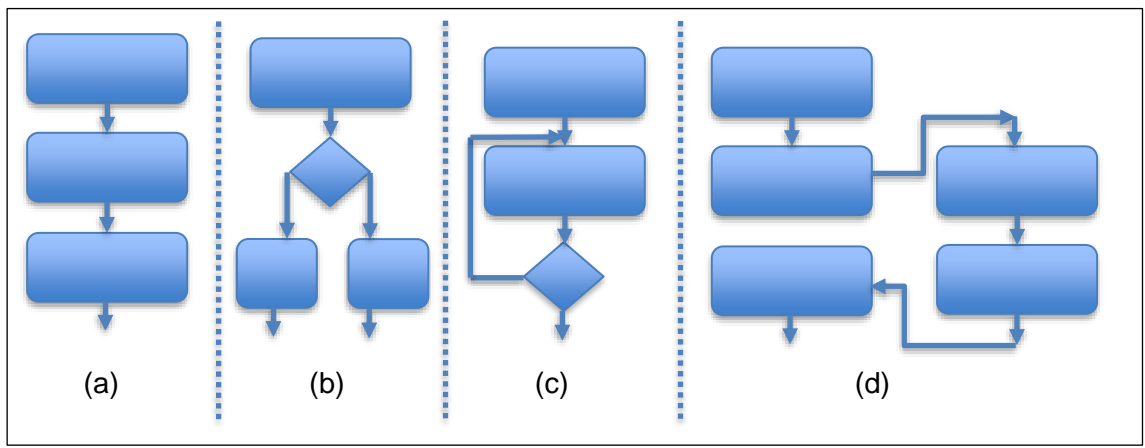

**Exhibit 2.1. Algorithm Flow (a) Linear, (b) Selection, (c) Iteration, (d) Function Call**

# Chapter 3: Algorithmic Thinking in Simple Programs

## Overview

In the previous chapter, I reviewed some examples of algorithmic thing in daily tasks. They were a good start and could help you understand the concept and see that it is not as strange as you may have thought. In this chapter, I will focus on simple programs and see how algorithmic thinking helps develop them. Due to the simplicity of the examples, one may find algorithmic thinking an unnecessary process. "Why bother? I know how to write the code for this," you may say. But the point is to learn the right process through simple examples so we can apply it in more complicated ones that will be hard to do without proper thinking. Hopefully, you can see the value of algorithmic thinking even in the simpler examples too, as they streamline the programming and help you make sense of what you code better. Sensemaking in software development is mainly about knowing why we code in a certain way and what happens if we make a change. It is essential in a good software design and also in finding defects and improving the program. For this chapter, our focus will be on search and sort as two very important groups of algorithms. But before we talk about them, it is time to look at how we represent an algorithm.

## 3.1. Representing Algorithms

Chapters 1 and 2 included many examples of different algorithms. To represent them, I used a textual description. Such a representation can be very effective as it doesn't require any previous knowledge, conventions, or tools. But it may not be the most effective way to show an algorithm. Unstructured textual description can be vague and open to personal interpretation. For example, when we say "get the number," where and how do we do that? If we have to work with multiple numbers, should we keep using words such as "the first number", "the second number", etc., or there are better, more clear, ways? In a program, we define variables and use named functions and clear syntax. That clarity makes a program executable by a machine. An algorithm is not meant to be used by a machine (at least, not yet) but it can be inspired by the notations used in programming languages for the purpose of increased clarity. Such a semi-structured easier-to-read algorithm representation is called **pseudocode**.

Let's revisit Example 1.1a:

```
1.   Consider a new data called Max and set it to the first list value,
```

```
2.   Consider a new data called Index and set it to zero,
3.   Consider a new data called Number and set it to the list value at
     Index,
4.   If Max is smaller than Number, then set Max to Number,
5.   Increment Index
6.   If Index is less than the array length, then go to step 3.
```

A more structured and formal version of this algorithm tries to avoid casual descriptions and be more "program-like":

**Example 3.1**

```
1.   Max = list[0]
2.   Index = 0
3.   Number = list [Index]
4.   If Max < Number
5.      Max = Number
6.   Index = Index + 1
7.   If Index < array_length
8.      Go to step 3.
```

Note that the new algorithm (1) doesn't follow the real syntax of any existing programming language although it looks like Python code[1] without some specifics like using : after `if`, and (2) doesn't include details that are needed in a real program. But it follows some common conventions such as using `[ ]` to show arrays/lists of data and = for assigning values. It doesn't show where `array_length` is defined though, as it is a less important detail. Example 3.1 is a well-written pseudocode, while Example 1.1a may also be considered a pseudocode but not a very good one. Unfortunately, there is no standard for writing pseudocode and different people use different styles which are lessor more similar to actual programs. In this book, I use a format inspired by Python, i.e., regions of code are indented and there is no standard symbol at the end of the lines. Depending on the case, I may have more or less textual descriptions, but I try to avoid them as much as possible.

---

[1] In fact, Python is designed to look like natural language.

Pseudocode is not the only way to represent an algorithm. Another common method is a graphical format called **flowchart**. Graphical representations can be more appealing and easier to understand, although they are less code-like and may require special tools and skills or take more time. In a flowchart, the algorithm will be shown using nodes (for what happens) and lines (showing the flow). The nodes generally follow standard shapes: Terminal, Process, Decision, and Input/Output are the most common ones, as illustrated in Exhibit 3.1.

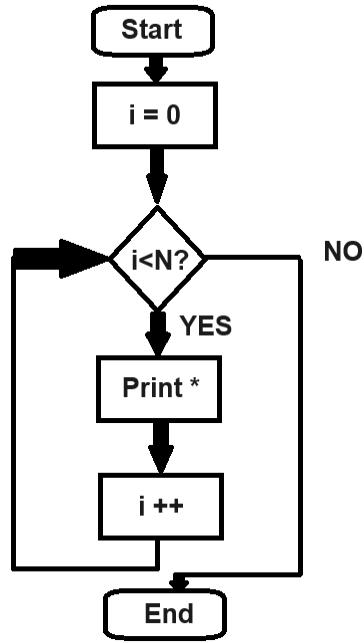

**Exhibit 3.1. Example of a Flowchart showing how to print N starts (*)**

---

*Pseudocode and flowchart are two common ways of showing an algorithm.*

---

In this book, I primarily use pseudocode for algorithms as I find them more helpful for beginner programmers. You will see flowcharts in some examples both to make sure you are familiar with their use and also to take advantage of their visual clarity in some cases.

## 3.2. Getting Started

Let's get started with implementing some of the algorithms we reviewed in Chapter 1 and 2. Before we move forward, I need to mention a couple of things about my programming examples in this book:

- I use a "p" at the end of an example number to mean it is in Python. A "c" means the example is in C or C++. These two languages are fairly compatible as far as this book is concerned, except for the use of classes that exist only in C++. Please see my note about C and C++ at the start of the book. I will not get into their specific differences and base my examples on C++.

- This book is not to teach the craft of programming, language syntaxes, and other details of that sort. Appendix A provides a quick review of C/C++ and Python.

- Note that in C/C++, your code needs to be part of a function, but in Python it doesn't. Some of my C/C++ examples (such as 3.2c) may not show the function. You are required to add that yourself if testing the code.

- Many programming languages require the use of external libraries for various functionality. These are identified in the program using `#include` in CC++ and `import` in Python. When dealing with common libraries, I may skip showing those. For example, to use `cin` and `cout` in C++, you have to use the following lines:

```
#include <iostream>

using namespace std;
```

### 3.2.1. Iteration with Definite Loops

I started the book with the restaurant bill example. Let's take a look at a possible implementation of this in two common languages:

***Example 3.2p***

```
1.  sum = 0
2.  for n in range(3):     #assume 3 items
3.      price = int(input("Enter a price: "))
4.      sum += price
5.  print(sum)
```

*Example 3.2c[1]*

```
1.  int sum = 0
2.  for(n=0; n<3; n++)      { //assume 3 items
3.      int price;
4.      cout << "Enter a price: ";
5.      cin >> price;
6.      sum += price; }
7.  cout << sum;
```

This simple algorithm may seem very trivial, but it is the foundation of one of the most commonly used programming patterns that we discussed in Chapter 2: iteration. Looking at the two examples, you see that the syntax may be different, but the structure remains the same and has the following elements:

- Initialization of variables used inside and outside the loop
- A condition or count to control the loop
- A set of actions that repeat
- Some actions that happen after the loop

The variables are also very important. Those like `price` are used in each iteration but their value is changing and doesn't need to be saved. Such variables are created inside the loop. On the other hand, the value of `sum` at the end of an iteration is used in the next iteration (we add the new number to the `sum` of previous ones). Creating a new `sum` variable in each iteration will result in losing the old value. So, we define it before the loop. If a variable needs a value the first time we use it, then we give it an initial value when we are defining it (line 1).

This basic loop structure can be used for any form of **accumulation**, i.e., a series of elements combining together one after another. Let's follow the basic loop structure to calculate factorial, i.e., the multiplication of all numbers from 1 to a given number. This is another form of accumulation but instead of adding, we are multiplying.

---

[1] To run this code, you need to copy it inside a function that is called in your program. You can create a new main() and add it there.

***Example 3.3p***

```
1.  factorial = 1
2.  n = int(input("Enter an integer number: "))
3.  for n in range(1,n):
4.      factorial *= n
5.  print(factorial)
```

As you can see, there are some differences, but we are following the same structure: `factorial` is the value that is needed throughout the iterations, so it is defined at the start. Note that the neutral/initial value for addition is 0, but for multiplication is 1. In the previous example, the new numbers were entered by the user at each iteration. Here the for loop has a variable (loop index) that can be used directly as the new number. However, we need a new initial variable `n`[1].

*The loop index is commonly used as a variable inside your loop code.*

## 3.2.2. Indefinite Loops

Another variation to the common loop pattern is when we don't know how many times we have to repeat the code. This is where we traditionally use `while` loops, which run as long as a certain condition is true. Example 3.4 shows this variation. Notice that `while` loop requires a condition. There are multiple ways to define that condition. My preferred one is to create a new variable. Here, I call it `run` as it tells us we need to continue running the loop. To start, it is set to `True` so we can run the loop. If anything happens that requires us to end the loop, we set run to `False`.

***Example 3.4p***

---

[1] In Python, variables such as `n` are identifiers (names or tags) associated with a value. Unlike C/C++, we don't declare a type for variables and their type is set when they get a value. But it doesn't mean that we don't need to be aware of their type. The function `input()` returns a string of text but we want the variable `n` to be numeric in line 3. So, we need to use the function `int()` to convert it. I personally prefer the C/C++ approach as it make types more obvious.

```
1.  sum = 0
2.  run = True
3.  while run==True:        #assume 3 items
4.      price = int(input("Enter a price (0 to end): "))
5.      if price == 0:
6.          run = False
7.      sum += price
8.  print(sum)
```

This way of defining the condition is based on what I call the **Golden Rule of Programming #1**. There are a couple of more "golden rules" that I will discuss later. The rule #1 says: "**If you have a piece of information, then make a variable for it**." In this case, running the loop is a piece of information, so it has its own variable. This rule helps immensely in simplifying your programming as it gives structure to your data, and once the data is properly defined, the code follows as we will see in the next chapter. Alternatively, you could use price as your condition and say `while price!=0`. Then you would need to define `price` before the loop and give it a non-zero initial value. This method is simpler, as we have one less variable and it saves you two lines of code (3.b and 3.b.i), but it is less structured and flexible. If later, you want to have multiple reasons to end the loop, you will have to change the condition. Using `run`, you can simply set it to `False` anywhere else. It is also easier to design because you don't need to think which existing variable can serve as condition. You follow the golden rule #1 and define a new variable for your new information.

*When using iterations, it is important to note which part of the code goes inside the loop (is repeated) and what stays outside the loop (executed only once).*

*Golden rule #1 helps you define the variables you need.*

*Try converting while and for loops.*

## 3.2. Search Algorithms

Searching is one of the most common programming algorithms. In many cases, we need to look for an item among a set of items that has unique attributes. The example that comes to mind first, for many of us, is probably web search. When looking for information, we frequently "Google it." Search engines go through a long list of indexed web pages to look for the words we have entered in the search (or address) bar of our browsers. There are various search algorithms, combined with indexing and archiving methods, all aimed at making the search process more efficient and showing the results faster. But these methods cannot be understood without understanding the most basic one first: **linear search**. After explaining this basic search method, I will discuss a variation called **binary search**.

### 3.2.1. Linear Search

The linear search is very similar in structure to the "fins the tallest" algorithm that we saw earlier. It goes through a list and compares. The difference is that to find the tallest (or max, or min, of a list), we compare to the tallest we have seen so far. If we find someone taller, we replace the tallest with the new person. In linear search, we always compare to the search target. Once we find a match, we return the index of the match (e.g., match found at item #3). Then, there can be other information associated with that item that we can use. For example, if we are searching through a phone book, the search word is the name. Once it is found (at, say, entry number 15), we use the phone number for that number.

Let's see how we can write the algorithm and Python code for a simple phone book. Imagine we have two lists: names and phone numbers. Think of these as two columns in a spreadsheet that share the same index. At row (index) 5, for example, the columns (lists) have the name and the phone number of the same person, respectively.

**Example 3.5a**

```
1.   Read names (from a database, or get from user or any other method)
2.   Read numbers
```

```
3.   Get search_name
4.   For all names on the list
5.       If name is equal to search_name
6.           Print corresponding number
```

**Example 3.5p**

```
1.   names = ["ali", "john", "kim", "“jane"]
2.   numbers = ["555-1000", "555-2000", "555-3000", "555-4000"]
3.   search_name = input("Enter a name: ")
4.   for index in range( len(names) ):
5.       if names[index] == search_name:
6.           print( names[index]+" "+numbers[index] )
```

Running this program, you will notice that if there are more than one match, all of them will be shown. You may choose to print only the first match by using a `break` statement after printing. This statement will end the current loop. A similar statement is `continue` which ends the current iteration of the current loop, skipping any remaining code (none in this example) and going to the next index. These statements exist in Python and C/C++, and many other programming languages, with similar functionality.

The linear search algorithm is fairly simple to understand and code, but it is not very efficient. If you are working on a long list, this algorithm can take a long time, even with a fast computer. In computer science terms, we say that the algorithm has a complexity at the order of N (number of items in the list), or O(N). That means the maximum time is linearly proportional to the list size. Note that the item we are looking for may be the first one on the list. In that case, we find it immediately. But it may be the last one, and then the search will take the maximum time.

## 3.2.2. Binary Search

To reduce the complexity and time for search operation, other algorithms have been suggested. One example is **binary search**, which works if the list is sorted. Making a sorted list can happen in two ways: (1) we add new items to the list at the right location. For example, if the list is [1,5], and we want to add 2, then we add it after 1, to have [1,2,5]. This method requires finding the right place every time we add a new item. (2) We sort the list once all data is entered. I will discuss the algorithms for both of these two options in the next section, but to understand how binary search works, let's just assume we have a sorted list, like the one in Exhibit 3.2, that shows the list of ID numbers for students registered in a course. The search involves finding out if a certain student is registered.

| 10045 | 10067 | 10089 | 10101 | 10156 | 10178 | 10199 | 10200 | 10205 |
|-------|-------|-------|-------|-------|-------|-------|-------|-------|

**Exhibit 3.2. An Example of Student IDs**

The binary search is based on the simple idea of constantly checking the middle of the range. If the middle item is what we are looking for, then we are done. Otherwise, we continue the search only in the upper or lower half because the list is sorted. This constantly saves search time because we narrow down the range. The complexity of binary search is said to be of the order of log(N), O(log(N)), where log is the logarithm function and log(N) is significantly lower than N, especially as N increases.

The algorithm and Python code are shown in Example 3.6.

***Example 3.6a***

```
1.  get data_list
2.  input x to search for
3.  low = 0 (lowest index in the range)
4.  high = length-1 (highest index in the range)
5.  start searching
6.  while searching
7.      mid = (low + high)/2
8.      if x == data_list[mid]
9.          print mid
10.         end searching
11.     else
12.         if low == high       (end of search)
13.             print "not found"
14.             end searching
15.         if x > data_list[mid] (x is on the right side)
16.             low = mid + 1
17.         else                  (x is on the left side)
18.             high = mid - 1
```

***Example 3.6p***

```
1.  data_list = [1,4,6,9,13,25,67,99]
```

```
2.    x = int( input("enter the number to search: ") )
3.    low = 0
4.    high = len(data_list) - 1
5.    searching = True
6.    while searching == True :
7.        mid = int((low + high)/2)
8.        if x == data_list[mid]:
9.            print("found at ",mid)
10.           searching = False
11.       else:
12.           if low == high :
13.               print ("not found")
14.               searching = False
15.           if x > data_list[mid]:
16.               low = mid + 1
17.           else:
18.               high = mid - 1
```

Note how this algorithm follows the golden rule #1. The idea is to always check the middle and then adjust the range. So, we have three pieces of information: start and end to define the range and its middle. Three variables are defined to hold the information. Then the loop structure is built: The action that has to be repeated (comparing) in put inside the loop and initial code (defining the initial range) is outside the loop. All that is left is to add to the loop what we need to do after a comparison to update the range.

*Find the algorithm for Hash table-based search.*

## 3.3. Sorting Data

In the previous section, I mentioned that binary search is faster but requires a sorted list. Sorting is a common task both in programming and in daily life. In section 2.3, I showed a simple sort algorithm. It was based on another which determined the tallest person. Note that the algorithm has two loops, one inside another. Before we get to the sort code, let's understand this structure better.

When I was a child, our elementary school had very strict hygiene and dress code. Once we week, the school principal would line up the kids and check if we are following the rules for having proper uniform and clipped fingernails. He would follow an algorithm: for each student, check the uniform, and for each finger, check to see if fingernail was clipped. This means he had two iterations/loops: one for each student and then as part of that another for each finger. So, for example, my left pinkie could be finger #1 of student #1 (Because of my name, I was usually the first). Each fingernail check could be identified by two numbers: which iteration of the first loop and which iteration of the second loop. The structure is called a **nested loop**, as one loop is nested inside (or is part of) another. Nested loops are used frequently in programming when we need to repeat an action over two variables. Another example is the multiplication table. Each multiplication has two operands. To create the table, we need to go from 1 to 9 for the first operand, and for each number go from 1 to 9 for the second, to the total of 9x9=81 results.

Now, let's look at Example 3.7 that show sorting an array of data from largest to smallest using a pair of nested loops.

### Example 3.7a

```
1.  get the sorted group
2.  make an empty unsorted group
3.  While there are people in the unsorted group
4.      Consider the first person as tallest
5.      While there is another person:
6.          If next person taller than the current tallest
7.              then consider them the tallest
8.      Add the tallest at the end of the sorted group
9.      Remove tallest from the unsorted group
```

### Example 3.7p

```
1.  unsorted = [6.5, 4.5, 6, 5, 7]
2.  sorted = []
3.  while len(unsorted)!=0:
4.      tallest = 0
5.      for i in range(len(unsorted)):  #Easier to use a for loop
6.          if unsorted[i] > unsorted[tallest]
7.              tallest = i
```

```
8.        sorted.append(unsorted[tallest])
9.        unsorted.remove(unsorted[tallest])
```

*Do the sorting without a separate sorted array (in-place sorting)*

*for (int i = 0; i < 7; i++)*

    *for (int j = i; j < 7; j++)*

    *{*

        *if (data[i] < data[j])*

        *{*

            *int temp = data[j];*

            *data[j] = data[i];*

            *data[i] = temp;*

        *}*

    *}*

*\*\*\**

*Find the algorithm for Bubble sort.*

## 3.4. Notes on Problem-Solving

In the previous sections, I introduced some of the most basic algorithms that we use commonly in many programs. Despite the differences, they shared the use of two important concepts:

- Iteration construct as a basic pattern in many programs. It is a key element of good algorithm design to notice what is repeated (the loop), what needs to be done prior to that (the initialization), and how many times we repeat the loop (the condition).
- Defining the right variables (data). I introduced the first golden rule of programming that guides you to define the main variables in your program. Having the right data elements, as we will see in the next chapter, is essential in designing clear and manageable algorithms and avoiding confusion about what to do; the code simply processes those variables.

Designing an algorithm (or a program) is a problem-solving task: we aim to develop a set of instructions that start from some input and result in a desired output. Problem-solving is not limited to computer programming though, so it makes sense to get ideas from many studies on creativity and problem-solving in general. There are many sources to study these topics and exploring the details is beyond the scope of this book, but three important skills (or strategies) for problem-solving are worth mentioning here. I did use them in the above examples, and you will most likely benefit from them, especially when dealing with more complicated algorithms. These skills are decomposition, abstraction, and pattern recognition[1]:

- **Decomposition**: Break down a task into smaller ones. For example, sorting a list is made of (1) finding the largest number, (2) moving it to another list, and (3) repeating these tasks until everyone is moved to the sorted list. Another common decomposition example is to divide a program into Input, Process, and Output parts. Having separate code modules such as functions for each of these parts will help organize the algorithm/program. Example 2.8 demonstrated such an organization with multiple functions when writing a calculator program. Applying decomposition method in software development results in what is usually called **modularization**, which I will discuss in detail in the next chapter.

---

[1] The combination of these three skills and algorithms is sometimes called Computational Thinking.

- **Abstraction**. Objects, environments, and problems include many details that may not matter for the specific purpose we have. For example, in the search example (3.5) we don't really care where the list is coming from. So, we hard-coded it in the code. The data could be read from a database or entered by user, but the details are irrelevant to the search algorithm. So are things such as reporting the results or even what type of data we are searching. The algorithm removes the need for such details and deals with the general search approach and the elements of the problem space that affect our search method. The concept of removing irrelevant details and coming up with representations of the subject that holds only the important elements is called abstraction or simplification. Abstract representations may not be fully practical and need details to be applied. For example, hard-coded data is almost always useless. But the algorithm serves the purpose of showing the search method. When applying this method, we add the details as needed[1].

- **Pattern recognition**. A pattern is a relationship between elements in space and/or time. The human brain has evolved to be good at pattern recognition due to its importance in effective and quick decision-making. For example, a fast-approaching object is a commonly dangerous pattern. Recognizing it quickly can save our lives, while spending time analyzing the details such as color and texture of the object can delay our decision. Or when entering a room in school, rows of chairs facing a screen is a pattern telling us we have entered a classroom. Noticing it will help us quickly choose the right behaviour (for example, quietly sitting in the back). Pattern recognition is related to abstraction as it also removes the irrelevant details and focuses on important aspects and finding the common solution for that set of important aspects. Our factorial example (3.3) has a very similar pattern to the sum example (3.2): starting with an initial number and adding/multiplying numbers to it. Noticing this common pattern, and not getting stuck on irrelevant details such as where the number is coming from, could help us easily use the same solution with minor modifications[2].

In the next chapter, I will start with the first golden rule and expand that into a generic system of algorithm design. I then demonstrate how the selection and iteration constructs, and also the three problem-solving methods above, can be used to design algorithms for solving various problems.

---

[1] Abstraction is related and sometimes equivalent to modeling. A model is an abstract representation of a system, focusing on some common aspects among similar instances and after removing details.

[2] Another concept, related to pattern recognition, is analogy, i.e., finding the similarity between two cases so we can use the same solution.

## Highlights

- Pseudocode and flowchart are two common ways of showing an algorithm.
- The loop index is commonly used as a variable inside your loop code.
- When using iterations, it is important to note which part of the code goes inside the loop (is repeated) and what stays outside the loop (executed only once).
- Golden rule #1 helps you define the variables you need.
- Search and sort are common algorithms that use selection and iteration.

# Chapter 4: Data-Centred and Modular Algorithm Design

## Overview

In the previous chapters, I discussed some examples of algorithmic thinking and algorithm design in daily life and simple programs. These examples helped identify a series of important methods and guidelines for designing algorithms. In particular, we defined three groups that are essential in almost any algorithmic work:

1. The first golden rule of programming is our starting point to design an algorithm as it tells us to identify the main data and give it a clear identity (by defining a variable for it). All your program/algorithm will follow as it performs operation on the data.
2. Selection, iteration, and function constructs are the building blocks of any program or algorithm. They help structure the operations (code).
3. Decomposition, pattern recognition, and abstraction are the main problem-solving skills that we use to define specific algorithms.

In this chapter, I first propose the data-centred approach to software or algorithm design as an extension to the first rule. It is a systematic way to start with your main data and come up with the main structure of the algorithm needed to process that data. Then, I will show how we use the constructs and skills in groups 2 and 3 above to implement that data processing. Throughout the chapter, we also see a series of common design patterns, which together with skills in group 3 can help design various algorithms that have common structures. For this chapter, I limit the examples to those with simple data. In the next two chapters, we will see how more complex algorithms require more complex data organized into data structures and objects. I will use games to demonstrate some of the concepts being discussed. So, I'm going to start the chapter with a brief review of simple computer games and how their programs are structured.

## 4.1. Simple Games

In Chapter 4, I talked about the iteration construct and how it is made of an initialization, a loop condition, and a series of actions to be repeated. Iterations are commonly used in many programs/algorithms but more than that, the whole program can be structured as a main loop.

Organizing the program around a main loop is one of the most common features of computer games, which I will discuss with examples of text-based and graphics games.

### 4.1.1. Simple Text-based Games

Turn-based games demonstrate the use of loops very clearly. Think about chess, for example. The game can be described simply as:

```
Set up the board

While there is a possible move

     Player 1 moves

     Player 2 moves
```

Let's say we are playing 20 Questions[1]. Again, the game is structured around a loop that this time has a max number of times it can repeat:

```
Answerer chooses something

For twenty rounds,

     Questioner asks something

     Answerer provides a response

     If questioner has guessed what the answerer had chosen,

          The game ends and questioner wins

     If this is question 20 and questioner has not guessed yet,

          The game ends and answerer wins
```

In general, almost all games even single player ones follow a similar pattern:

```
Set up the game.

     Define some main data items like scores and positions

     Initialize variables

While the game has not ended,

     Player takes an action.
```

---

[1] For simplicity, imagine there is only one set of questions. The algorithm can just repeat as many times as we want with players switching their roles.

```
Some result happens.

A win/lose condition is checked to see if the game ends.
```

We refer to this common structure as **Main Loop** or **Game Loop**. Establishing this loop is the first step in designing the algorithm for any game or other interactive program[1]. Example 4.1p is the Python code for a guessing game played against the computer.

### *Example 4.1p*

```
1.   import random
2.   quit = False
3.   number = random.randint(1, 10) #replace 1 and 10 with any other limits
4.   while quit == False:
5.       usernum = int ( input("enter a number: ") )
6.       if usernum < number :
7.           print("go higher")
8.       elif usernum > number:
9.           print("go lower")
10.      else:
11.          print("you win!")
12.          quit = True
```

The game has three main pieces of information:

- The number to be guessed,
- The number(s) that the user guesses,
- Whether or not the game is ending (or continuing).

We define three variables for these. We could have designed our game without using the `quit` variable, but the algorithm will be cleaner and easier to understand and change this way. Rule #1 is not aimed at designing an optimal program but an easier one to understand and write. It gives us a clear set of steps to follow. Once we are more comfortable with algorithmic thinking and software design, we can aim at optimizing our program.

---

[1] By "interactive" I mean any program that is designed based on repeated interactions: user performs an action, the program processes that action, the user receives some form of result.

We initialize `quit` and `number` before the loop starts as their values are needed. The variable `usernum`, on the other hand, doesn't need an initial value because the first time we use it in the loop is when it gets a value from user.

*The basic structure of almost all games includes an initialization followed by a main/game loop where the main action happens in a repeated way: user input and results.*

## 4.1.2. Simple Graphics Game

Now, let's consider a graphical game. Our game has a very simple design illustrated in Exhibit 4.1. We use arrow keys to move an object in 2D space, avoiding some moving objects (enemies) and picking up some stationary objects (prizes). The game is over when we pick up all prizes or are hit by an enemy.

While a simple design, this game demonstrates the basic structure of many games. Typical graphics-based computer games can be simplified into a combination of three types of objects: player (controlled and moved by the player), moving objects (controlled and moved by the game, such as enemies), and stationary objects (such as environmental items). Each object has a shape (a 2D image or a 3D model) and a location (2D or 3D coordinates)[1]. These objects are placed within an environment (2D background or 3D space). With this simplified notion, and following golden rule #1, our starting point will be to define variables for the pieces of information we have.

---

[1] In 3D, objects also have orientation, but I don't discuss 3D examples in this book.

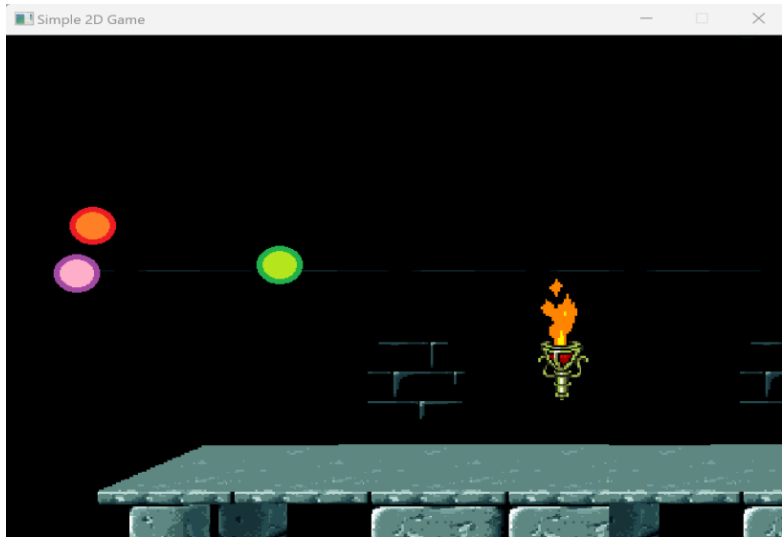

**Exhibit 4.1. Simple 2D Game. Move the player object (purple/left circle), avoid enemy (red/middle circle) and pick up the prize (green/right circle).**

---

*Graphics games include three types of objects: player, stationary, and moved by the game. The main loop involves updating the information related to these and then drawing them.*

*Can you think of a game and see how it is made of these three types of objects? How does the updating work?*

---

To write this game in Python or C/C++, we need an extra library as none of those languages comes with standard graphics support. In this book, I use SDL and Graphics.py for C/C++ and Python, respectively, as they are simpler. More professional and feature-rich options are OpenFrameworks and PyGame. For the purpose of algorithmic thinking and algorithm design, you won't notice any difference apart from syntax and function names. Please refer to Appendices A and B for a quick review.

To simplify, I assume one enemy and one prize. Example 4.2a shows the algorithm. Following rule #1, we start by defining the main variables, which in this case are the location of all objects and `quit` (as in Example 4.1). Then we have the main loop where we use and change our variables (remember Three How Questions). The positions of player and enemy are changed by the arrow keys and automatically, respectively. Finally, we use this information to determine if the game is over, i.e., the player has hit the enemy or the prize. The algorithm uses decomposition to separate three groups of actions: initialization (outside the loop), updating the variables (inside the loop), and drawing objects (inside the loop). Such a separation of updating and drawing is very useful both for designing the algorithms as they are different tasks and for later implementation as they can be done by different people or different methods. For example, when using Unity and similar game engines, drawing objects is generally done by the engine while the game developer is in charge of updating them. In fact, such engines commonly hide the main loop altogether and only provide update functions to be filled out by developers.

***Example 4.2a***

```
1.   playerX = 0
2.   playerY = 0
3.   enemyX = 100
4.   enemyY = 100
5.   prizeX = 200
6.   prizeY = 200
7.   quit = False
8.   win = False
9.   While quit == False
10.      Check collision:
11.      If playerX==enemyX and playerY==enemyY
12.          win = False
13.          quit = True
14.      If playerX==prizeX and playerY==prizeY
15.          Win = True
16.          quit = True
17.      Check keyboard for player movement:
18.      If Right key is pressed
19.          playerX ++
20.      If Left key is pressed
21.          playerX --
```

```
22.     If Up key is pressed
23.         playerY --
24.     If Down key is pressed
25.         playerY ++
26. Move enemy:
27.     enemyX += random
28.     enemyY += random
29.     Draw objects at new locations
30. If win == True
31.     Show win message
32. Else
33.     Show lose message
```

   Note how the algorithm above uses abstraction to skip details such as how we read the keyboard to detect arrow keys or how we show objects and results. Such matters are generally dependent on the language and library we use, and do not affect the overall algorithm. Examples 4.2p and 4.2c show the Python and C code using Graphics.py and SDL, respectively.

### Example 4.2p

```
1.  # initialization of the game
2.  window = GraphWin('Face', 600, 600) # give title and dimensions
3.
4.  # Loop until the user clicks the close button.
5.  quit = False
6.  win = False
7.
8.  #player
9.  playerX = 0
10. playerY = 0
11. playerR = 30    #radius
12. player = Circle(Point(playerX,playerY), playerR)
13. player.setFill("yellow")
14.
15. #enemy
16. enemyX = 100
17. enemyY = 100
18. enemy = Image(Point(enemyX, enemyY), "enemy.png")
19.
```

```
20.  #prize
21.  prizeX = 200
22.  prizeY = 200
23.  prize = Image(Point(prizeX, prizeY), "prize.png")
24.
25.  #draw in the order we want to see on screen
26.  prize.draw(window)
27.  enemy.draw(window)
28.  player.draw(window)
29.
30.  # -------- Main game Loop -----------
31.  while not quit:
32.      # UPDATE part of the main loop
33.
34.      #check if we hit the prize object
35.      distanceX = abs(playerX - prizeX)
36.      distanceY = abs(playerY - prizeY)
37.      if distanceX<10 and distanceY<10:
38.              prize.move(1000,1000)
39.              quit = True
40.              win = True
41.
42.      #check if we hit the enemy. End the game if so
43.      distanceX = abs(playerX - enemyX)
44.      distanceY = abs(playerY - enemyY)
45.      if distanceX<10 and distanceY<10:
46.              quit = True
47.
48.      #move the player
49.      playerMoveX = 0
50.      playerMoveY = 0  #remember that positive y is down
51.      needsMove = False
52.      key = window.checkKey()
53.
54.      if key=='Escape':
55.              quit = True
56.      if key=='Up':
57.              playerMoveY -= 1
```

```
58.     if key=='Down':
59.             playerMoveY += 1
60.     if key=='Right':
61.             playerMoveX += 1
62.     if key=='Left':
63.             playerMoveX -= 1
64.     playerX += playerMoveX
65.     playerY += playerMoveY
66.
67.     enemyMoveX = random.randint(-2,2)
68.     enemyMoveY = random.randint(-2,2)
69.
70.     # DRAW part of the main loop
71.     #in graphics.py the move() function actually moves and draws
72.     player.move(playerMoveX, playerMoveY)
73.
74.     #enemy needs drawing
75.     enemy.move(enemyMoveX,enemyMoveY)
76.
77.     time.sleep(0.05)
78. #after the loop ends
79. window.close()
80. if win == True:
81.     print("you win!")
82. else:
83.     print("you lose!")
```

## Example 4.2c

```
1.   int main(int argc, char* argv[])
2.   {
3.       ///////////////////////////////////////////////////////////
4.       // INIT part of the game
5.       // things that are done only once at the start (initialization)
6.       ///////////////////////////////////////////////////////////
7.
8.       //init SDL library and graphics window
```

```
9.        SDLX_Init("Simple 2D Game", 640, 480, true);
10.
11.       //game map (background)
12.       SDLX_Bitmap* map = SDLX_LoadBitmap("map.bmp");
13.
14.       SDLX_Bitmap* player = SDLX_LoadBitmap("Player.bmp");
15.       int playerX = 0;
16.       int playerY = 0;
17.
18.       SDLX_Bitmap* enemy = SDLX_LoadBitmap("Enemy.bmp");
19.       int enemyX = 100;
20.       int enemyY = 180;
21.
22.       SDLX_Bitmap* prize = SDLX_LoadBitmap("Prize.bmp");
23.       int prizeX = 200;
24.       int prizeY = 200;
25.
26.       bool quit = false;
27.       bool win = false;
28.
29.       ////////////////////////////////////////////////////////////
30.       //--------Main game Loop---------- -
31.       // things that are done repeatedly
32.       ////////////////////////////////////////////////////////////
33.       while (!quit)
34.       {
35.          ////////////////////////////////////////////
36.          // UPDATE
37.          ////////////////////////////////////////////
38.
39.          //check if we hit the prize
40.          int distanceX = abs(playerX - prizeX);
41.          int distanceY = abs(playerY - prizeY);
42.          if (distanceX < 10 && distanceY < 10)
43.          {
44.             prizeX = prizeY = 1000;
45.             quit = true;
46.             win = true;
```

```
47.        }
48.        //check if we hit the enemy
49.        distanceX = abs(playerX - enemyX);
50.        distanceY = abs(playerY - enemyY);
51.        if (distanceX < 10 && distanceY < 10)
52.        {
53.            quit = true;
54.        }
55.
56.        //check if there is any movement
57.        //player controlled by keyboard
58.        SDLX_Event e;
59.        //keyboard events
60.        if (SDLX_PollEvent(&e))
61.        {
62.            if (e.type == SDL_KEYDOWN)
63.            {
64.            //press escape to end the game
65.            if (e.keycode == SDLK_ESCAPE)
66.                quit = true;
67.            //update player
68.            if (e.keycode == SDLK_RIGHT)
69.                playerX++;
70.            if (e.keycode == SDLK_LEFT)
71.                playerX--;
72.            if (e.keycode == SDLK_UP)
73.                playerY--;
74.            if (e.keycode == SDLK_DOWN)
75.                playerY++;
76.            }
77.        }
78.
79.        //update enemy
80.        enemyX += rand() %5 - 2;
81.        enemyY += rand() %5 - 2;
82.
83.        ///////////////////////////////////////////
84.        // DRAW
```

```
85.          ////////////////////////////////////////
86.          //note the order (last one will be drawn on top)
87.          SDLX_DrawBitmap(map, 0, 0);
88.          SDLX_DrawBitmap(prize, prizeX, prizeY);
89.          SDLX_DrawBitmap(enemy, enemyX, enemyY);
90.          SDLX_DrawBitmap(player, playerX, playerY);
91.
92.          //Update the screen
93.          SDLX_Render();
94.
95.          //now wait for loop timing
96.          SDLX_Delay(1);
97.      }
98.
99.      if (win)
100.         printf("you win");
101.     else
102.         printf("you lose");
103.     return 0;
104. }
```

Two important points are noteworthy in the above examples:

- The main loop is divided into Update and Draw modules. Separating these two functions allows us to deal with updating variables and drawing game objects separately. For example, we can continue to update the same way, but replace the drawing module to do the visual aspects of the game in a different style. In some case, these two modules can be implemented by different people or even supported through different libraries[1].

---

[1] Many game engines provide the library modules for drawing, or even hide them (and the main loop) completely from programmer who only needs to write the update code.

Modify the code to use separate functions for Update and Draw. What parameters need to be passed to these functions? Can we use global variables instead?

- The main loop also controls the timing of any game object that changes its location or shape. Each iteration of the main loop creates a new frame of the game where objects move, animations advance, and other changes happen. I'll show an example of animating objects in Section 5.1.1.

*How can the timing of all events in the game be controlled by the frames? Hint: A game that has 60 frames per second uses 1/60th of second for each frame.*

## 4.2. Data-Centred Algorithm Design

Computer programs are almost always data processing instructions. You can imagine cases when the program simply involves actions with no data but in most of those cases, there is implied data. For example, when we say, "turn the light on when the user presses the button," the implied data includes which button, which light, and possibly what color. Identifying the data that we want processed and what process needs to be done on it will help us greatly when designing an algorithm. I refer to this approach to software/algorithm design as **Data-Centred**, i.e., a design process that starts and is organized based on important data items. To streamline this design process, I introduced what I called the first golden rule of programming: for any piece of information, define a variable. This rule helps you identify and highlight your main data items. Then you can see what you should do with them.

Imagine you are designing a driving game where the player car needs to be refueled after a certain amount of driving. The important information we need to manage this aspect of the game are (1) the current amount of fuel, and (2) the maximum capacity of our fuel tank. So, the program should start by defining two variables for these and giving them initial values. Then, as the program continues and the player drives, two things need to happen: (1) if necessary, these variables have to change, and (2) at some point in the program, they have to be used. The amount of fuel should decrease in the main loop as the car moves. It should be used by giving alarms to the player when it gets too low and making the car stop when the tank gets empty. The tank capacity doesn't change but its value is used to limit the current fuel amount when refueling.

*Data-centred algorithm design is an approach to algorithmic thinking where data is the key concept and everything is organized around that concept.*

Let's look at a couple of game examples to demonstrate the data-centred approach.

### 4.2.1. Attacking Enemy

Let's imagine an example where we are designing an algorithm for a game program in which a non-player character (NPC) is to attack the player, as opposed to random movement the enemy had in our simple game of Example 4.2. The player can move freely around the two-dimensional (2D) screen and the NPC has to move towards the player character and hit it. This action is an example of a general pattern of moving towards a target, which can be used in many cases. At first, we may wonder how we can move an object towards another. The first step is to think about information: what is the information in this task? The answer is "location." We need to know where the objects are. On a 2D screen, this information can be stored in a pair of coordinates, X and Y. For two objects, we will have (X1,Y1) and (X2,Y2). So, the first step is to define these variables. To change the location of an object, we can simply say: X1=X2 and Y1=Y2. These instructions will move object 1 to the location of object 2, as shown in Exhibit 4.2.

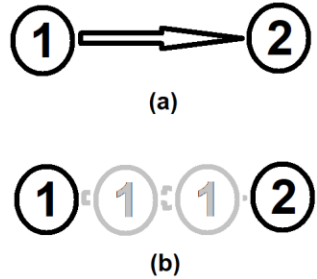

**(a)**

**(b)**

**Exhibit 4.2. Object Movement. (a) jumping from 1 to 2, (b) gradually moving from 1 to 2.**

But this "jumping to destination" is not quite the same as moving towards a target. What we need is to show the movement that happens over time. Movement over time means speed, so we have another piece of information, speed of movement, and another variable, V1. If we define V1 as the amount of movement for object 1at each step, then we will have X1=X1+V1 and Y1=Y1+V1. Note that in a 2D space, every movement consists of independent movements along X and Y axes. But how do the above instructions take into account that we need to move towards a target? They identify the amount of movement but not the direction. That is when the target coordinates come in. Consider each axis separately. On the X axis, we move to right if object 2 is on the right side of object 1 and we move to left if it is on the left. Similar movement can be done for Y axis, and we will have the first version of our algorithm:

***Example 4.3***

```
1.   X1 = 0
2.   Y1 = 0
3.   V1 = 1
4.   X2 = random number on screen
5.   Y2 = random number on screen
6.   while X1!=X2 and Y1!=Y2
7.       if X1<X2
8.           X1+=V1
9.       else
10.          X1-=V1
11.      if Y1<Y2
12.          Y1+=V1
13.      else
14.           Y1-=V1
15.      display objects
16.      delay 1 second
```

Line 15 allows us to show the movement on screen at a speed that is visible by the viewer. Displaying and also how to initialize the coordinates are details irrelevant to the main algorithm, so we don't need to deal with them at this time.

In the above example, we used all our algorithm design guidelines:

- We identified the main variables and structured the algorithm around them.
- We used selection and iteration.
- We recognized the "move towards" pattern and worked on an abstract version that doesn't deal with irrelevant details. We also decomposed the task into modules and considered random, display, and delay as parts of the program that can be done elsewhere.

---

*Modify the code in 4.2p and 4.2c using the algorithm in 4.3. See section 4.5 for an example.*

*Use OpenFrameworks and make a simple game*

*Use PyGame and make a simple game*

*Use separate speeds for X and Y*

---

Example 4.3 demonstrates the data-centred approach as it clearly performs three important types of operation on our data: initialize, use, change:

- We initialize all variables to a meaningful value at the start[1].
- We use X2 and Y2 to determine the direction and V1 to determine the amount of movement. We also use all coordinates for displaying.

---

[1] Technically, any variable needs to be initialized only if the first time we use it, it needs to have a value (for example, V1 in Example 4.3. If the first time we use it, the variable is getting a value, then an initial value is not needed.

- We change the value of X1 and Y1 in a loop to perform the movement until we reach the target. X2, Y2, and V1 don't change.

Once these are taken care of, we can deal with details such as random, display, and delay which usually depend on the programming language and various libraries we are using and also the rest of the game and how it is structured.

Based on the above example, the **data-centred design** can be formulated as the following steps:

- **Visualize your program**. If the program has no clear visual output, use any form of diagram or figure to show what are the parts of the program. This step is essential in letting you see the data and operations. See Exhibit 4.3 for some other examples of visualizing an algorithm/program.
- Identify the **main data items** and define variables for them (golden rule #1).
- For each data item, ask three questions: (1) how is it initialized? (2) how is it used? (3) how is it changed? I call these **Three How Questions (3HQ)**.
- Write the **main algorithm** based on the answers to the above questions.
- Add **details** for other parts of the program.

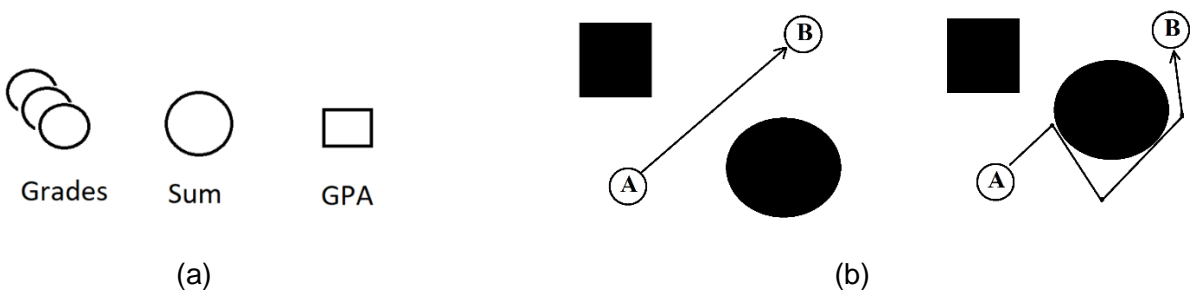

Exhibit 4.3. Program Visualization. (a) GPA calculation, (b) obstacle avoidance.

*Data-centred design starts by identifying and visualizing data elements and then asking three how questions (3HQ).*

## 4.2.2. Collision Detection

In games, the problem of detecting if two objects have reached (hit) each other is usually called **Collision Detection**. The sample code 4.1 ends the movement when it detects a collision between the moving object and its target, determined by the loop condition in line 6 (`X1!=X2 and Y1!=Y2`). This method is a good start and may work in some cases but has two major issues:

- If the speed is greater than 1, then the object may end up moving back and forth around the target. For example, if `X2=6` and `X1=5`, with a speed of 2, the object will never "reach" its target and keeps bouncing between 5 and 7. This problem can be solved by calculating the distance between two objects and ending the movement if the distance is less than a certain amount. As we can see in our program visualization, if we are "close enough" to the target, we can stop and this will avoid bouncing back and forth, and is more practical. Following our data-centred approach, after we identify a new piece of information, we define a new variable, initialize it, update it in the loop (as we move) and use it as the condition for the loop.

```
distance = square_root( (X2-X1)^2 + (Y2-Y1)^2 )  [1]
while distance < limit
     distance = square_root( (X2-X1)^2 + (Y2-Y1)^2 )
...
```

- Objects do not occupy only one point and their coordinates are generally based on the coordinates of one of their important points (usually centre or top-left in 2D space). Depending on their sizes and the direction of movement, objects may reach and hit each other (collide) at different coordinates, as illustrated in Exhibit 4.4. Calculating distance may not be a suitable solution here, as we need to know the distance from which point on the target. A better solution may be detecting if a point is "inside" the area representing the target. In 2D games, the rectangular image of the object is the simplest way of representing this area. In many cases, the object does not fill the whole rectangle (or box in 3D), so we commonly refer to this area as **Bounding Box**, a box surrounding the object and simplifying collision detection[1]. Assuming that XY coordinates of the object correspond to top-left, we will need two new pieces of information, width and height (W and H), to define the bounding box, and the corners will be (X,Y), (X+W,Y),(X,Y+H), and (X+W,Y+H). For a point to be inside the box, its x needs to be between X and X+W, and y between Y and Y+H. We will define new variables for width and height, initialize them, and use them in the loop condition (they don't usually change during the program, but they could).

---

[1] All programming languages provide library functions for math operations such as square root. `^2` denotes "to the power of two". If such operator or function is not available, we can simply have `(X2-X1)*(X2-X1)`.

```
while X1>X2 and X1<X2+W2 and Y1>Y2 and Y1<Y2+H2
```

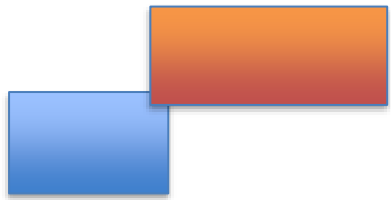

**Exhibit 4.4. Collision Detection. Top-left, or centre, points are far from each other but there is collision.**

---

*What are the main data elements for collision detection using bounding boxes?*

*Add proper collision detection to the simple game of Examples 4.2*

---

## 4.3. Nested Loops

In Sections 4.1 and 4.2, I introduced the Data-Centred Design and the Main Loop pattern as a common example, plus some other common algorithms such as following and collision detection. Using the pattern recognition skills and familiarity with a few common patterns will help us easily find solutions to our algorithm problems. Another common algorithm pattern involves nested loops, i.e., loops that contain another loop. This pattern happens when we have multiple variables changing together.

---

[1] When the object shape does not fill the whole rectangle, there are more complicated ways to detect collision. Bounding box is simple and usually good enough both in 2D or 3D. We also have Bounding Sphere in 3D, although Bounding Circle is not really used in 2D.

Let's consider a simple one-loop example: setting all members of a list to zero. As I discuss in more details in the next chapter, using lists (arrays of members) usually involves iteration as we create a loop to perform the same action on all members[1]. The set-to-zero algorithm will look like the following lines:

```
For i=0 to i=array_length

   Data[i] = 0
```

This iteration involves one data that is changing: the index to members, `i`, because each member is identified using one number (the index). If instead of a list we have a table, then each member is defined using two numbers, row and column. Let's show the item at row `i` and column `j` with `Data[i][j]`. To set all data to zero, we first need a loop that goes over all rows:

```
For i=0 to i=row_size
```

Then, within that loop and for each value of `i`, we will need another loop to try all values of `j`. Finally, within the second loop, we set the data to zero.

```
For i=0 to i=row_size

   For j=0 to j=column_size

         Data[i][j] = 0
```

Note that the operation `Data[i][j]=0` is repeated `row_size*column_size` times:

```
Data[0][0] = 0 (for i=0)

Data[0][1] = 0

Data[0][2] = 0

…

Data[0][colun_size-1] = 0
```

[1] Lists or arrays are groups of data that usually have similar role, such as grades for all students or location of all objects. So, we usually perform the same actions on them.

```
Data[1][0] = 0 (for i=1)

Data[1][1] = 0

…
```

Instead of setting to zero, we could do anything else with the data such as printing or other processes. But the algorithm to go over two variables stay the same. They didn't even need to be a table. We could have two sets of data that need to be processed together in all combinations, such as days of week and hours of day in a weekly event calendar. The structure above is called a **nested loop** and is a very common algorithmic pattern when dealing with two or more variables at the same time.

*Try printing all combination of a set of first names and a set of last names*

Another example is image manipulation. Imagine that we have an image, or an area of screen, and we want to set all pixels in it to black. Each pixel is no longer identified with one number but two: X and Y. If we consider the width and height of the image as W and H, respectively, we will need to change X from 0 to W, and Y from 0 to H. We may try to write two separate loops:

```
For X=0 to X=W

    Set pixel (X,Y) to black

For Y=0 to Y=H

    Set pixel (X,Y) to black
```

For now, let's not concern ourselves with how we set the pixel color. The algorithm above is an abstraction that hides such details, which are dependent on what language and graphic library we use. But you can see that the algorithm doesn't work for two reasons: (1) what is the value of Y in the first loop or X in the second? (2) the total number of pixels that we set to black is W+H (first loop runs W times and the second H times), but it should be W*H (a 40-by-50-pixel image has 40x50=2000 pixels). The solution is that once we start the first loop, we have to run the second one for each of its iterations. That means the second loop has to be part of the first one. So, for X=0 we will try all Y values, then for X=1, etc.:

```
For X=0 to X=W

    For Y=0 to Y=H

            Set pixel (X,Y) to black
```

Examples 4.4p and 4.4c show how to do image manipulation based on nested loops in Python and C/C++. Some details are removed, and the full code can be found online at https://ali-arya.com/anyonecancode. The code is very versatile, and you can use it for all sort of operations on images. I will get back to it in Section 6.1.5.

### *Example 4.4p*

```
1.      window = GraphWin('Image', 600, 600) # give title and dimensions
2.      #get the image
3.      imgX = 200
4.      imgY = 200
5.      img = Image(Point(imgX, imgY), "prize.png")
6.      #image dimensions
7.      w = img.getWidth()
8.      h = img.getHeight()
9.      #color
10.     c = color_rgb(0,0,0)
11.     #loop
12.     for i in range(w):
13.         for j in range(h):
14.             img.setPixel(i,j,c)
15.     img.draw(window)    #draw
16.     window.getKey()     #wait
```

### *Example 4.4c*

```
1.      SDLX_Init("Image", 640, 480, true);
2.
3.      //get the image
4.      int imgX = 200;
5.      int imgY = 200;
6.      SDLX_Bitmap* img = SDLX_LoadBitmap("Prize.bmp");
7.
```

```
8.      //image dimensions
9.      int w = img->width;
10.     int h = img->height;
11.
12.     //color
13.     SDLX_Color c;
14.     c.r = c.g = c.b = 0;
15.
16.     //loop
17.     for(int i=0; i < w; i++)
18.         for (int j = 0; j < h; j++)
19.         {
20.             SDLX_PutPixel(img, i, j, &c);
21.         }
22.     //draw
23.     SDLX_DrawBitmap(img, imgX, imgY);
24.     //update the screen. Needed for SDL/SDLX
25.     SDLX_Render();
```

*Nested loops are a common construct when dealing with two variables, for example, XY coordinates for pixels in an image.*

*Write a program that has a set of first names and a set of last names, then shows all possible combinations.*

## 4.4. Modularization

Decomposition is in important problem-solving skill. It allows us to divide a complex task into smaller parts that are easier to solve. In software terminology, we usually refer to this concept as **modularization**. Defining a program in terms of a series of modules (components), or hierarchies of them, has multiple advantages, as it allows:

- working on smaller and simpler modules
- creating and using re-usable/generic modules
- collaborating and dividing task between different developers

Library functions, available in all programming languages, such as those for reading from keyboard and writing to screen, are examples of such pre-defined and re-usable modules that make programming easier and provide standard way of performing the related actions. Modules can be defined for both data and code. Arrays, stacks, and lists are examples of data modules (or data structures) that I will review in the next chapter. Functions are the most basic modules of code, and classes[1] are modules with both data and code. In Chapter 6, I will review algorithm design using classes. Here, let's see how we can improve our algorithmic thinking through the use of functions or modules of code.

Example 2.7 in Chapter 2 demonstrated the use of functions in algorithms. As a simple example, consider a calculator for performing four basic operations: addition, subtraction, multiplication, and division. The algorithm for such simplified calculator is easy to write by following data-centred approach. We have three main data items: the operation, the first number, and the second number. 3HQ guide us to write the algorithm as in Example 4.5.

***Example 4.5***

<u>Calculator</u>

```
1.  Ask for the operation
2.  Ask for the first number
3.  Ask for the second number
4.  If operation is to add
5.     Print first+second
6.  If operation is to subtract
7.     Print first-second
8.  If operation is to multiply
9.     Print first*second
10. If operation is to divide
```

---

[1] An instance of a class is called an Object. Classes and objects are like data types and variables or data items.

```
11.     Print first/second
```

Note that the first three lines get the information we need: what operation to perform and on what numbers. Then we perform different actions depending on the operation. Each of these two parts could be done differently. For example, we could get the data from a database or print the results in a different format. But the main two parts of the algorithm stay the same:

```
Get data

Perform the operation
```

By designing the algorithm in terms of these two modules first, we can have a basic structure for our program and then add details as needed. Examples 4.6p and 4.6c show the code in Python and C, respectively. The calculator is a good entry point to introduce another common design pattern that works well with modularization: **Command Processor**. In this pattern, a series of operations are defined, and each is performed based on the value of a command. The program is a loop consisting of (1) getting command, and (2) performing the related operation. Each operation and the command input can have their own modules. In Examples 4.6p and 4.6c the operation modules are very simple (possibly not worth making a module for) but the simplicity is only to make a point of clearly showing how the modules are made and related in a Command Processor.

### Example 4.6p

```
1.    #operation modules
2.    def Add(num1, num2):
3.         return num1+num2
4.    def Sub(num1, num2):
5.         return num1-num2
6.    def Mul(num1, num2):
7.         return num1*num2
8.    def Div(num1, num2):
9.        if num2 == 0:
10.            print("divide by zero!")
11.            return False
12.        else:
13.            return num1/num2
14.   #main code
15.   quit = False
16.   while  not quit :
```

```
17.      operation = int(input("Enter operation type.  1 for add, 2 for
    subtract, 3 for multiply, 4 for divide, 0 to exit: "))
18.      if operation>0 and operation<5:
19.          num1 = int(input("Enter the first number: "))
20.          num2 = int(input("Enter the second number: "))
21.      if operation == 0:
22.          quit = True
23.      elif operation == 1:
24.          print(Add(num1, num2))
25.      elif operation == 2:
26.          print(Sub(num1, num2))
27.      elif operation == 3:
28.          print(Mul(num1, num2))
29.      elif operation == 4:
30.          print(Div(num1, num2))
31.      else:
32.          print("Invalid operation!")
33.
```

### Example 4.6c

```
1.   //operation modules
2.   int Add(int num1, int num2)
3.   {
4.       return num1 + num2;
5.   }
6.   int Sub(int num1, int num2)
7.   {
8.       return num1 - num2;
9.   }
10.  int Mul(int num1, int num2)
11.  {
12.      return num1 * num2;
13.  }
14.  int Div(int num1, int num2)
15.  {
16.      if (num2 == 0)
17.      {
18.          cout << "divid eby zero!\n";
```

```
19.          return NULL;
20.      }
21.      else
22.          return num1 / num2;
23.  }
24.  //main code
25.  int main()
26.  {
27.      int operation, num1, num2;
28.      bool quit = false;
29.      while (!quit)
30.      {
31.          cout << "Enter operation type.  1 for add, 2 for subtract, 3
     for multiply, 4 for divide, 0 to exit: ";
32.          cin >> operation;
33.          if (operation > 0 && operation < 5)
34.          {
35.              cout << "Enter the first number: ";
36.              cin >> num1;
37.              cout << "Enter the second number: ";
38.              cin >> num2;
39.          }
40.          if (operation == 0)
41.              quit = true;
42.          else if (operation == 1)
43.              cout << Add(num1, num2);
44.          else if (operation == 2)
45.              cout << Sub(num1, num2);
46.          else if (operation == 3)
47.              cout << Mul(num1, num2);
48.          else if (operation == 4)
49.              cout << Div(num1, num2);
50.          else
51.              cout << "Invalid operation!";
52.          cout << "\n";
53.      }
54.      return 0;
55.  }
```

*Command Processor is a commonly used software design pattern where a series of commands are received and each processed separately, usually through independent modules and in a loop.*

*Can you think of the main loop in the simple game of Section 4.1.2 as a command processor?*

*Write a program to calculate restaurant bills as a command processor? What can the commands be?*

Game loops that I reviewed in Section 4.1 are also good candidates for using modularization. Instead of adding all the code together, we can define the basic structure of the game with high-level modules and then define each of them. Example 4.7p shows the highlights of a more modularized version of 4.2p, the simple graphics game.

### Example 4.7p

```
1.   #global variables
2.   playerMoveX = 0
3.   playerMoveY = 0
4.   enemyMoveX = 0
5.   enemyMoveY = 0
6.   #the rest of globals
7.
```

```
8.    #modules
9.    def prizeCheck():
10.       global playerX, playerY, prizeX, prizeY, prize, quit, win
11.       distanceX = abs(playerX - prizeX)
12.       distanceY = abs(playerY - prizeY)
13.       if distanceX<10 and distanceY<10:
14.           prize.move(1000,1000)
15.           quit = True
16.           win = True
17.   def enemyCheck():
18.       global playerX, playerY, enemyX, enemyY, quit
19.       distanceX = abs(playerX - enemyX)
20.       distanceY = abs(playerY - enemyY)
21.       if distanceX<10 and distanceY<10:
22.           quit = True
23.   def playerMove():
24.       global window, playerMoveX, playerMoveY, quit
25.       dx = 0
26.       dy = 0
27.       needsMove = False
28.       key = window.checkKey()
29.       if key=='Escape':
30.           quit = True
31.       if key=='Up':
32.           playerMoveY -= 1
33.       if key=='Down':
34.           playerMoveY += 1
35.       if key=='Right':
36.           playerMoveX += 1
37.       if key=='Left':
38.           playerMoveX -= 1
39.       playerX += playerMoveX
40.       playerY += playerMoveY
41.   def enemyMove():
42.       global enemyMoveX, enemyMoveY
43.       enemyMoveX = random.randint(-2,2)
44.       enemyMoveY = random.randint(-2,2)
45.   def update():
```

```
46.        prizeCheck()
47.        enemyCheck()
48.        playerMove()
49.        enemyMove()
50.  def draw():
51.        global playerMoveX, playerMoveY, playerX, playerY, enemyMoveX,
       enemyMoveY
52.        player.move(playerMoveX, playerMoveY)
53.        enemy.move(enemyMoveX,enemyMoveY)
54.  def init():
55.        #init code here
56.
57.  def mainloop():
58.        global quit, win, window
59.        while not quit:
60.            update()
61.            draw()
62.            time.sleep(0.05)
63.        window.close()
64.        if win == True:
65.            print("you win!")
66.        else:
67.            print("you lose!")
68.
69.  #game main loop is called after init
70.  mainloop()
```

In these examples, I demonstrated the concept of modularization by using functions. Modules are not limited to functions though. We also have modules of data and modules of code and data combined, which I will review in Chapters 5 and 6, respectively. Modularization of data helps manage groups of related data. For example, in 4.7p, you see that exchanging data between functions can be cumbersome as each object is using multiple pieces of data (x, y, shape, etc). To avoid complications, I have used global variables, which is not a recommended method. It makes it difficult to know which data is used where, and it ads to code dependencies. A single data module representing each of the game objects will be a lot more manageable as it can be easily passed as a parameter to a function or received as the result. Regardless of which type, modularization is a key element in software design and proper use of it will make our algorithmic thinking and algorithm design process more efficient and manageable.

## 4.5. Algorithmic Thinking Toolbox

There is no magic formula that allows us to think algorithmically. Algorithm design, like any other design work, requires practice, starting with simpler cases and advancing to more complex ones. But, so far, we have managed to collect a set of things that can help us in algorithmic thinking process. Think of this set as a toolbox: Having it won't make a person an expert, but experts have it and use it all the time. In our toolbox, we have four groups of tools:

- Five steps that form the data-centred design approach, including the golden rule #1. This is our basic method of designing algorithms.
- Selection, iteration, and modularization. These are programming constructs that we use as the building block of any program/algorithm.
- Decomposition, pattern recognition, and abstraction. These are problem-solving skills that we use to come up with our algorithms.
- Accumulation, search, sort, command processor, game loop, and many other patterns. These are the bases of many algorithms we commonly use.

As a final example of using all these tools together, imagine a Non-Player Character (NPC) in a game that acts as a guard in front of a prize object (or a door or anything else that the player needs to access). When there is no one around, the guard walks randomly. When the player comes shows up, the guard stops to watch the player. Finally, when the player comes close, the guard attacks. Let's see what happens in each step of the data centred approach:

Exhibit 4.5 shows the program visualization. Using the abstraction skill, we ignore some of the details that don't matter such as how the objects look like and in what setting the game is played. The important thing is to show the elements necessary for the main task, controlling the walk-watch-attack behaviour of the guard.

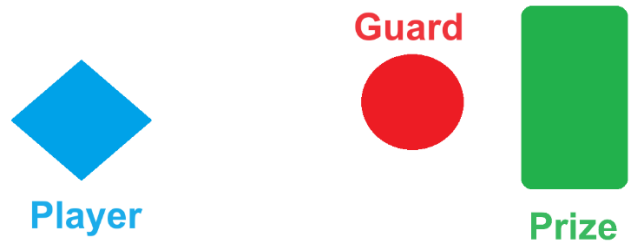

**Exhibit 4.5. Program Visualization for Guard NPC**

Looking at the visualization, it is fairly easy to identify three objects and their locations as the main data items.

- PlayerX, PlayerY, PlayerShape
- GuardX, GuardY, GuardShape
- PrizeX, PrizeY, PrizeShape

The 3HQ (how to initialize, how to use, how to change) will result in the following general answers:

- All locations will be initialized at the start with hard-coded values that are the game's starting positions.
- For simplicity (remember abstraction), shapes are initialized to some images and won't change. Effects such as walking and fighting animation could mean changing shapes, but they don't matter at this time as they won't affect the main algorithm.
- Player and Prize locations (their distance) will be used in every iteration of the game loop to control the Guard. The control takes the form of three distinct set of actions. Using our decomposition skill, we can define three separate modules for Walk, Watch, and Attack, each defined independently and called based on the distance between Player and Prize.
- Prize location doesn't change. Player location is changed by the player using arrow keys. Guard location is changed using our control algorithm and based on the distance between player and prize. These changes will happen in every iteration of the game loop.

Based on the above answers, the man algorithm can be written using the following modules:

- Initialize: Sets the initial values and is called at the start of the program.
- Mian Loop: Runs after the Initialize module and has two parts:

- Update: Changes the player position using arrow keys, calculates the player-score distance, and calls one of the three action modules to move the guard.
- Draw: Shows all objects at new locations.
- Walk: Called by Update.
- Watch: Called by Update.
- Attack: Called by Update. The algorithm was shown in Example 4.3, Section 4.2.1.
- GuardMove (optional): Called by Update to determine the type of movement for guard based on current state, and then call Walk/Watch/Attack

Finally, we add any extra details we need such as "glue code," putting all together, as shown in Example 4.7p. This program is based on our previous simple game (guard is the enemy) and demonstrates a common design pattern called **Finite State Machine (FSM)**. In an FSM, the behaviour of a system is defined as a set of distinct states. Each state is identified with two things: actions that happen at that state and possible transitions to other states. So, the main algorithm for an FSM-based system (the Update module in our example) performs two tasks: (1) detecting the state and do transitions if needed, and (2) performing the actions for current state through the related modules. Notice how the overall design is decomposed into modules and also uses previously discussed patterns such as following a target.

### Example 4.8p

```
1.   #code is based on 4.7
2.   #new global variables
3.   distance = 0     #player to prize
4.   state = "walk"   #guard/enemy state
5.   #new modules
6.   def walk():
7.       global enemyMoveX, enemyMoveY
8.       enemyMoveX = random.randint(-2,2)
9.       enemyMoveY = random.randint(-2,2)
10.  def watch():
11.      global enemyMoveX, enemyMoveY
12.      enemyMoveX = 0
13.      enemyMoveY = 0
14.  def attack():
15.      global enemyMoveX, enemyMoveY, enemyX, enemyY, playerX, playerY
```

```
16.    enemyMoveX = 0
17.    enemyMoveY = 0
18.    if playerX > enemyX:
19.        enemyMoveX = 1
20.    if playerX < enemyX:
21.        enemyMoveX = -1
22.    if playerY > enemyY:
23.        enemyMoveY = 1
24.    if playerY < enemyY:
25.        enemyMoveY = -1
26.    enemyX += enemyMoveX
27.    enemyY += enemyMoveY
28.  #modified modules
29.  def enemyMove():
30.    global state
31.    if state == "walk":
32.        walk()
33.    elif state == "watch":
34.        watch()
35.    else:
36.        attack()
37.  def update():
38.    global distance, state
39.    prizeCheck()
40.    enemyCheck()
41.    playerMove()
42.    #detect/transition state
43.    state = "walk"
44.    if distance > 200:
45.        state = "walk"
46.    elif distance > 100:
47.        state = "watch"
48.    else:
49.        state = "attack"
50.    #perform satate action
51.    enemyMove()
```

*Algorithmic thinking does not have a standard process but set of tools that help you design an algorithm: data-centred approach, basic programming constructs, problem-solving skills, and common software patterns.*

*Find out the algorithm and write the code for table-look-up path planning using the data-centred approach.*

## Highlights

- The basic structure of almost all games includes an initialization followed by a main/game loop where the main action happens in a repeated way: user input and results.
- Graphics games include three types of objects: player, stationary, and moved by the game. The main loop involves updating the information related to these and then drawing them.
- Data-centred algorithm design is an approach to algorithmic thinking where data is the key concept, and everything is organized around that concept.
- Data-centred design starts by identifying and visualizing data elements and then asking three how questions (3HQ).
- Nested loops are a common construct when dealing with two variables, for example, XY coordinates for pixels in an image.
- Command Processor is a commonly used software design pattern where a series of commands are received and each processed separately, usually through independent modules and in a loop.
- Algorithmic thinking does not have a standard process but set of tools that help you design an algorithm: data-centred approach, basic programming constructs, problem-solving skills, and common software patterns.

# Chapter 5: Data Structures

## Overview

In the previous chapter, I introduced the Data-Centred Design Approach that allows us to design an algorithm based on the data it processes. The examples were chosen to demonstrate this approach with simple data items such as single or at most an array of integer numbers. Dependency on data means that it makes sense to progress through examples as the data becomes more complicated. Just like the code, our data can be modularized for better and more efficient processing. Modules of data are commonly referred to as compound data or **Data Structures**. Most programming languages support simple (a.k.a. basic or atomic) data types such as integer, float, character/string, and Boolean. An array[1] is a good example of a data module that groups a set of data items of the same basic type[2]. Records[3], or User-Defined Types (UDT), are plain data structures that put together members of different types. More complex data structures can be defined using combinations of arrays and records, to establish a hierarchical modularization in which the modules become more complex as we go higher on the hierarchy; a record shows all the data for one item, an array of those records shows a collection of items, another record can have multiple such arrays and other data for a large organization, etc. Modules of data can have specific properties, purposes, and common operations. For example, a queue is a first-in-first-out structure used when the data is stored in an array and retrieved in the same order as it was stored (see Section 5.3). So, it has specific operations for adding and removing data. Such complex data structures are sometimes called Abstract Data Type (ADT). An ADT is a formal but abstract definition that can be implemented in different forms, for example an array and a set of global functions (non-object-oriented) vs. a class (object-oriented). In this chapter, I will discuss common algorithms that use data structures without the use of object-oriented classes. Chapter 6 will be focused on classes and object-oriented algorithms and programs.

---

[1] Lists and arrays may be considered the same or different. Most commonly, we use the term array when talking about a group of data items but we use list when there are defined operations such as insertion and deletion of members. C/C++ have standard arrays and the programmers can define lists using arrays. Python has standard lists with pre-defied functions like `append()`, which use arrays internally.

[2] For example, in C/C++ there is no standard string type and strings of text are made using an arrays of character types.

[3] In C, records are called structures or simply structs. C++ continues to use that name although C++ structs are really classes. I will discuss these in the next chapter.

## 5.1. Arrays

In many programs, we deal with a group of data items that have similar roles. The grades for different courses, price of all purchased items, heights of a team's players, and all the available names in a data base are examples of items that have the same type and are usually processed similarly in a program. Instead of assigning each of these items a separate name and dealing with them as separate variables, we collect them in one variable, an **array**, that has multiple members each identified with a number, called **index**. The common notation in programming languages for indexing arrays is the use of [ ], such as `grades[0], heights[1]`, etc, where the number inside [ ] is the index to the array named before the [ ]. Some important things to note about arrays are:

- While many languages such as C/C++ use the term array, others such as Python have **lists**. The terms array and list are sometimes used interchangeably but more commonly list is used for an array that has special abilities. For example, a Python list has functions to append and delete members, while a C array has no standard associated function, and it is up to programmer to implement different operations.

- Different programming languages have different ways of dealing with array sizes. Sometimes they are static (such as "`int grades[5]`" in C that creates an integer array of fixed size 5) and sometimes they are dynamic (such as "`grades=[]`" in Python that can be appended with new members).

- The index starts from zero to show the first member, so the last member of an array of size N is at index N-1.

- Index values can be constant numbers, variables, or any expression that calculates to a number. But they have to be within the range of array size.

- The name of the array with the index (e.g., `grades[2]`) refers to a single member of the array. The name of the array without the [ ] and index value can be used differently in different languages. In Python, the array name is an object with functions. In C/C++, the array name is simply a pointer to the start of the array.

*Arrays are simple modules of related data items that have the same type.*

*Create a circular array in C. Define an array of size N and functions that read and write its members at a given index. If the index value is greater or equal N, then the remainder of division by N is used. For example, if N is 5, the members at index 5 and 12 are actually at locations 0 and 2.*

### 5.1.1. Array Indexing

Indexing is the most basic operation performed on arrays, which allows us to access different members of the array. It can be done randomly for one member (as in a user selection) or sequentially for all members (as in a loop).

```
int Data = [1,2,3,4,5,6];

cout << "which member?";

cin >> i;

cout << Data[i];

for(i=0; i< 6; i++)

{

    cout << Data[i];

}
```

The loop example above shows the sequential indexing of an array that happens one after another. There are cases where we need to sequentially go through all members but not all in one loop, i.e., at different points in the program we need to advance to the next index. **Array indexing** is the act of keeping track of array members to be accessed using an index value, inside a single loop or distributed in the program. If inside a single loop, the index value is commonly the loop counter or a variation of it.

```
for(i=0; i< 6; i++)
```

```
{

    Cout << i * 2;  //print the first 6 even numbers

}
```

If indexing is not inside a single loop, then we follow the data-centred approach: we define a variable for any piece of information (in this case, current index) and we ask the three how questions to process that variable. For example, if we have an array of customers to be serviced, we can keep a variable called `currentCustomer` as the index to the array. This variable can be set to zero at the start and every time we are done serving a customer, it can be increased so at any time it is the index for the customer to be served.

Another example is animation. An animation is a series of images to be shown one after another. Assuming we have a class or structure called Image, the following code shows how to play an animation.

### Example 5.1c

```
1.   #define NUM_FRAMES     3
2.   //global variables
3.   SDLX_Bitmap* frames[NUM_FRAMES];
4.   int currentFrame = 0;
5.   int playerX, playerY;
6.   //global functions
7.   void LoadFrames()
8.   {
9.       frames[0] = SDLX_LoadBitmap("Player0.bmp");
10.      frames[1] = SDLX_LoadBitmap("Player1.bmp");
11.      frames[2] = SDLX_LoadBitmap("Player2.bmp");
12.  }
13.  void AdvanceAnimation()
14.  {
15.      currentFrame++;
16.      if(currentFrame >= NUM_FRAMES)
17.          currentFrame = 0;
18.  }
19.  void DrawAnimation()
20.  {
21.      SDLX_DrawBitmap(frames[currentFrame], playerX, playerY);
```

```
22.  }
23.
24.  //main function
25.  int main(int argc, char *argv[])
26.  {
27.      ////////////////////////
28.      // Initialize
29.      ////////////////////////
30.
31.      //init SDL library and graphics window
32.      SDLX_Init("Simple 2D Game", 640, 480, true);
33.
34.      //game object
35.      playerX = 0;
36.      playerY = 0;
37.      LoadFrames();
38.
39.      //loop variables
40.      SDLX_Event e;
41.      bool quit = false;
42.
43.      // Main Loop
44.      while (!quit)
45.      {
46.          ////////////////////////
47.          // Update
48.          ////////////////////////
49.
50.          //keyboard events
51.          if (SDLX_PollEvent(&e))
52.          {
53.              if (e.type == SDL_KEYDOWN)
54.              {
55.              //press escape to end the game
56.              if (e.keycode == SDLK_ESCAPE)
57.                  quit = true;
58.              //update player
59.              if (e.keycode == SDLK_RIGHT)
```

```
60.                playerX ++;
61.            if (e.keycode == SDLK_LEFT)
62.                playerX --;
63.            if (e.keycode == SDLK_UP)
64.                playerY --;
65.            if (e.keycode == SDLK_DOWN)
66.                playerY ++;
67.            }
68.        }
69.
70.        AdvanceAnimation();
71.
72.        ////////////////////////
73.        // Draw
74.        ////////////////////////
75.
76.        //First clear the renderer/buffer
77.        SDLX_Clear();
78.
79.        //Draw the objects (in order: background to foreground)
80.        DrawAnimation();
81.
82.        //Update the screen
83.        SDLX_Render();
84.
85.        //pause
86.        SDLX_Delay(1);
87.    }
88.
89.    //clean up
90.    SDLX_End();
91.
92.    return 0;
93. }
94.
```

The variable `currentFrame` is initialized at the start to point to the first member of the array. We have a function to change it (`AdvanceAnimation`) that is called any time we want to show the next frame of our animation. We also have another function (`DrawAnimation`) that is in charge of the actual drawing part using the current index. As we have seen before, any graphics program usually has a main loop that generates frames. These functions are called in that loop. Note that separating these two functions follows our approach in games (Section 4.1) to separate updating and drawing.

---

*Animation is an example of array indexing. We change the array index when we need to advance the animation and point to a new frame. Note that the animation frames for one object and the scene frames for the whole game are not the same.*

*In Example 5.1c, each frame of the game (each iteration of the main loop) advances the animation of the object. In practice, animated objects may not advance all the time. For example, a running or walking animation is played only when the character is moving. Modify the code to advance character (player) animation only when it is moving. Hint: Pay attention to arrow keys.*

---

## 5.1.2. Array Matching

Arrays are used by themselves and in combination with other code and data modules. The majority of the programs we work on have arrays as we commonly deal with groups of similar data items. As we saw in the previous chapters, arrays are almost always used together with different forms of iterations. Loops and array indexing allow us to repeat the same operation on all members of an array. I have already shown various design patterns such as accumulation, search, and sort that use arrays. Another common pattern that works with arrays is copy, i.e. setting the values of one array to another. Copy demonstrates the use of iteration very well.

### Example 5.2c

```
1.  int Data1 = {1,2,3,4,5,6};
```

```
2.    int Data2[6];
3.    for(int i=0; i< 6; i++)
4.    {
5.        Data2[i] = Data1[i];
6.    }
```

In Python, we could simply say `Data2 = Data1.copy()` as Python arrays have a built-in copy function[1]. Using `copy()` function is convenient but it doesn't help us understand and re-use the algorithm pattern I call **array matching**, which is forming a loop and going over corresponding members of two (or more) arrays in order to copy, compare, or otherwise process them pair-wise. For example, array matching can be used to count the number of equal members:

```
int count = 0;

for(int i=0; i< 6; i++)

{

      if(Data2[i] == Data1[i]

            count++;

}
```

*Array matching is the process of matching corresponding members of two arrays. Matching members may not have the same index, e.g., the member at index 5 of one array may be matched with a member at index 10 of another.*

---

[1] Note that saying `Data2=Data1` in Python doesn't copy create a new array that is a copy of Data1. It creates a new reference to the same data. So saying `Data2[0]=0` will set the first member of both `Data1` and `Data2` to zero, as they are the same array. Same concept applies to C/C++ where array names are really pointers to array data.

*Start with an array of random numbers. Create a second array and store the odd members of the first array. Hint: You can have a loop to go over the first array, but the loop index cannot be used as index for the second array. You need to define a counter.*

*In Python, you can add new members to an array without worrying about array size. Append() function simply extends the array when needed. C/C++ arrays, on the other hand, have a fixed size. So once "full" you can add more data unless you override existing ones. Assume there is no append() function. Write the algorithm, or C code, for creating an array with a certain size (e.g., [0,0,0,0,0]) and then asking the user to enter numbers. Keep adding the numbers to the array, starting from index 0 to the end of array (hint: you will need a counter). Every time you add a new member, check to see if you have reached the end. If so, create a new array with double size, copy the existing array, and the use the new array for storing data. Note that the array name is a reference. S you can always say old_array = new_array to keep using the old name but with reference to the new array.*

---

### 5.1.3. Array Shifting

A variation to copy operation is shift, which is, in fact, copying a part of an array over another part. For example, members from 5 to 10 will shift "up[1]" to become members from 6 to 11. That leaves member 5 empty[2] and overwrites member 10. Shift no longer follows the common array matching algorithm as we don't have two arrays processed with the same index. Python lists already have functions that take care of this by inserting a new member or deleting an existing one at any index, which will result in other members (from that index to the end) to be shifted. Languages like C/C++ that use plain arrays don't have such built-in functions and the programmers need to implement them. No matter what language we use though, the shift algorithm is a good one to practice as it improves our algorithmic thinking skills. Exhibit 5.1 visualizes the shift operation. Shifting up inserts a new empty member and shifting down deletes an existing member.

---

[1] Up and down mean towards the end of the array or its start, respectively.

[2] Technically, a memory location such as a member of array cannot be empty as it always has a value. We can consider an array location empty, when it has no intentional or useful value and so can be used without concern about data loss. Sometimes, we initialize such locations with a known value such as -1 or zero that is never used as valid value (for example, if the data is people's heights).

| Data to be inserted | Array (before) | Array (after) |
| --- | --- | --- |
| | [empty] | [empty] |
| | [empty] | 1 |
| | 1 | 90 |
| 23 | 90 | 23 |
| | 45 | 45 |
| | 12 | 12 |

(a)

| Data to be deleted | Array (before) | Array (after) |
| --- | --- | --- |
| | [empty] | [empty] |
| | 1 | [empty] |
| | 90 | 1 |
| | 23 | 90 |
| 45 | 45 | 23 |
| | 12 | 12 |

(b)

**Exhibit 5.1. Shift Operations: (a) shift up (towards the end of the array) by inserting, (b) shift down (towards the start of the array) by deleting.**

The main data items are the array and two index values that show the start and end of the portion of array to shift (in the example above, all data from index 5 to 10 will be shifted up). The shift operation, as illustrated, is a series of copying one member over another. So, we know we will have a loop going over all the locations that are to be shifted, and for each, we copy to the adjacent location (up or down, next or previous). The copy operation is how we use and change the data, following the data-centred approach:

```
For i=index1 to i=index2; i++

  Data[i+1] =Data[i]
```

The key to performing the operation correctly is to realize that when we copy a data over another, the destination loses its original value. So, if that value is needed, it has to be copied somewhere else first. We saw this point at the start of the book with the swap algorithm[1]. In the swap case, we used a temporary variable, but here have the array locations. All we need to remember is to start with the last thing that needs to be copied and loop in the opposite direction of the shift, as shown in Example 5.1. Note that the original data at members `Data[index2+1]` (for `ShiftUp`) and `Data[index1-1]` for `ShiftDown` are lost after the shift.

***Example 5.3a***

```
1.   ShiftUp(index1, index2)
2.      For i=index2; i>=index1; i--
3.         Data[i+1] =Data[i]
4.   ShiftDown(index1, index2)
5.      For i=index1; i<=index2; i++
6.         Data[i-1] =Data[i]
```

*Array shifting moves members of an array towards the start or end. It is used for inserting or deleting members.*

*How do we check for array length to make sure we don't go beyond array boundaries (i.e., index values are never negative or more than length-1?*

The full C code for `Insert()` and `Delete()` functions are shown in Example 5.3c. The functions use `ShiftUp()` and `ShiftDown()`, and receive the data array and index as parameters. `Insert()` also needs a value for the member being inserted at index. Note that (1) for inserting and deleting one member, we shift the whole array from that member to the end, and (2) C/C++ arrays do not identify their length, so any function that uses the entire array needs to have the length as parameter.

---

[1] Swapping `first` and `second` needed a temporary variable: `old_first = first; first=second; second=old_first`

*Example 5.3c*

```
1.   //shifting up the members from index1 to index2 (inclusive)
2.   void ShiftUp(int Data[], int index1, int index2)
3.   {
4.       for (int i = index2; i >= index1; i--)
5.       {
6.           Data[i + 1] = Data[i];
7.       }
8.   }
9.   //shifting down the members from index1 to index2 (inclusive)
10.  void ShiftDown(int Data[], int index1, int index2)
11.  {
12.      for (int i = index1; i <= index2; i++)
13.      {
14.          Data[i - 1] = Data[i];
15.      }
16.  }
17.  void Insert(int Data[], int len, int index, int value)
18.  {
19.      //move up all members from insertion point
20.      //this copies the member at index to index+1
21.      ShiftUp(Data, index, len - 1);
22.      //now add the new value
23.      Data[index] = value;
24.  }
25.  void Delete(int Data[], int len, int index)
26.  {
27.      //move down all members after insertion point
28.      //this overwrites the member at index by index+1
29.      ShiftDown(Data, index + 1, len - 1);
30.  }
```

Back in Section 3.3, I showed how to sort data. Using insert and delete operations, we can create sorted arrays by adding new items to the right place in the array, as shown in Example 5.4a and Exhibit 5.2, assuming that we want the array to be sorted in descending order.

| 44 | 36 | 25 | 12 | | |
|----|----|----|----|----|----|

**Add 39:**

| 44 | **39** | 36 | 25 | 12 | |
|----|----|----|----|----|----|

**Exhibit 5.2. Create Sorted List**

### Example 5.4a

```
1.   Create empty array
2.   While we need new data
3.       Num = new data
4.       For i=0; i < len; j++
5.           If Num > data[i]
6.               Index = i
7.               Break the loop (end search)
8.       Insert(data, i, Num)
```

If writing a program in Python, you can use the existing `append()` and `remove()` functions, while in C/C++ you need to use the new functions we wrote.

### Example 5.4c

```
1.   //create empty array
2.   int data[] = { 0,0,0,0,0,0,0 };
3.
4.   //create as many new members as we want (here assume 10)
5.   for (int j = 0; j < 10; j++)
6.   {
7.       //create new random number
```

```
8.      int num = rand() % 50 +1; //1 to 50
9.
10.     //find the right place assuming the array is descending order
11.     int index = 0;
12.     for (int i = 0; i < 7; i++)
13.         if (num > data[i])
14.         {
15.             index = i;
16.             break;
17.         }
18.     //write in the found location
19.     Insert(data, 7, index, num);
20. }
```

*Example 5.4p*

```
1.  data = [0,0,0,0,0,0,0]
2.  for j in range(10):
3.      num = random.randint(1,50)
4.      index = 0
5.      for i in range(0,len(data)):
6.          if num > data[i]:
7.              index = i
8.              break
9.      data.insert(index,num)
```

## 5.2. Records

Arrays help modularize data items when they have a similar role, i.e., the have the same type and perform the same action. For example, all the grades for a student are integer numbers between 0 and 100, and they are used similarly to calculate an average or determine a pass or fail. But if we think of all the information a school stores on a student, we will notice that they no longer have the same type and role. While grades are numeric, name and address are text. They are also used for identification and contact. On the other hand, student identification number (ID) is also an integer but has different range and role. It is possible to think of one array consisting of grades, IDs, and other integer numbers, but defining such modules is not common because managing the array will be difficult due to the difference in the nature of members. An alternative, supported in different forms in all programming languages, is a **record**, also called struct (following C/C++ terminology) or plain data structure. Members of a record are commonly referred to as **fields**.

---

*Records (or user-defined types) are modules of data consisting of members that can have different types (called fields). They encapsulate related data items that can have different type or role in the program into a single entity.*

---

### 5.2.1. Encapsulation

Records allow the use of user-defined types. They are a combination of different members, each with their own name and type, that together define a new, complex, data type. Here is an example of a C/C++ struct for student information[1]:

```
struct Student
{
    char name[50];
    int ID;
    int grades[5];
    int average;
}
Student s1;
s1.ID = 1000;
```

Student is a record/struct or user-defined type, while s1 is an **instance** of it, i.e., a variable of that type. In Python and many other object-oriented languages, records are implemented using classes, which I will discuss in the next chapter. The Python version of the above code uses the keyword `class`. For now, let's assume the Python class is a collection of related data as in records.

---

[1] Note that the Example 5.7c and other C/C++ struct examples in this book follow C++ syntax. In C, we will need to use `typedef` in line 1 or `struct` in line 6. See my notes in the Preface.

```
class Student:

    name = ""

    ID = 0

    grades = []

    average = 0

s1 = Student()

s1.ID = 1000
```

Note that in Python, unlike C/C++, we don't define members of a class without assigning values. The point of listing members in C/C++ `struct` is to show their name and type. Python identifiers (such as variable names), on the other hand, are dynamic and their type is set when we assign a value. In the above example, I have assigned a value so that we, as programmers, know what type of data each member should be assigned later. Alternatively, programmers can assign `None` to a placeholder name that is going to get a proper value later. Both C++ struct/class and Python class can have initialization functions to set initial values. Also, Python distinguishes between members of a class (`Student`) and members of an instance of that class (`s1`). The difference is somewhat, but not entirely, similar to static and regular class members in C/C++. I will discuss these details in the next chapter when I talk about objects and classes.

The most obvious use of user-defined types such as structs is to bundle related data. This is commonly called **encapsulation**. This term is used with different meanings though. The simplest one is "bundling related data" but a more complex one is "bundling related data and code" and yet another, even more complex, definition is "bundling related data and code, and restricting access to them." In this chapter, I stick to the first definition. The other two make more sense in the context of object-oriented programming. Example 5.5c shows how encapsulating student data in one record helps simplify our program.

***Example 5.5c***

```
1.   #define NUM_GRADES 5
2.   void StudentInfo1(char name[], int ID, int grades[], int average)
3.   {
4.       cout << name << "\n";
```

```
5.      cout << ID << "\n";
6.       for(int i=0; i < NUM_GRADES; i++)
7.           cout << grades[i] << "\n";
8.      cout << average << "\n";
9.  }
10. void StudentInfo2(Student s)
11. {
12.     cout << s.name << "\n";
13.     cout << s.ID << "\n";
14.      for (int i = 0; i < NUM_GRADES; i++)
15.          cout << s.grades[i] << "\n";
16.     cout << s.average << "\n";
17. }
```

`StudentInfo1` requires a long list of parameters to get all the information for one student. `StudentInfo2`, on the other hand, has only one parameter. In many cases, we don't even need to know what all the members of `Student` are. We simply pass the bundle with all its members and the function uses whichever member it needs.

The advantage of using records is more obvious when we have multiple data items of that type, as shown in the second part of Example 5.5c below with an array of records each with four members. Without records, we needed to have four separate arrays, and then match the right members to each other. This is possible but harder to manage the code. For one thing, we always have to remember which arrays are related.

### *Example 5.5c-continued*

```
1.  int main()
2.  {
3.      //create and use Student variable here
4.      Student year1[10];        //array of 10 students
5.      year1[0].ID = 1000;       //ID for the first student
6.      year1[0].grades[0] = 99; //first grade for the first student
7.      year1[0].average = 100;
8.      year1[1].ID = 1001;       //ID for the second student
9.      //continue individually for all students
10.     //or use a loop
11.      for (int i = 0; i < 10; i++)   //loop over all students
12.      {
```

```
13.          strcpy_s(year1[i].name, 50, "x");
14.          year1[i].ID = 1000 + i;
15.          for (int j = 0; j < NUM_GRADES; j++)
16.              year1[i].grades[j] = 0;
17.          year1[i].average = 0;
18.      }
19.
20.      //using a loop to access all Student array members
21.      for(int i=0; i<10; i++)  //loop over all students
22.          StudentInfo2(year1[i]);
23. }
```

*Write functions to initialize Student and calculate average.*

## 5.2.2. Spawn

A good example to show the use of records is spawning characters in games, or creating and recreating any complex data items at different stages of a program runtime. The entities in a game are commonly referred to as **game objects**. They can play different roles (player character, non-player character, environment items, etc.) and have different types of action (moved by the player, moved by the game, and stationary). But generally, they share some basic attributes such as location and shape. In a 2D game, these can be defined through a record:

```
struct GameObject
{
    int x;

    int y;

    Image shape; //assuming we have a user-defined type called Image
};
```

The code assumes the shape is stored in a struct called `Image`. Standard C/C++ has no graphics support, as I mentioned before. All 3<sup>rd</sup> party libraries for graphics have a user-defined type (struct or class) for images. Id using SDL/SDLX as in Examples in Section 4.1, we will have SDLX_Bitmap*[1].

*Modify the code in Example 4.2 to use the GameObject struct. See Example 5.6c for a solution.*

Each character has many attributes that need to be initialized when they come alive (spawn) at the start of the game or after they are killed (respawn). One of the most important things to remember when spawning characters is that they go to a certain default state even if previously they had changed their state. For example, once an enemy is killed and needs to respawn, they will start from a default location no matter how much movement they had prior to getting killed and where they were killed. This can apply to player character too, except that usually there are no arrays of player characters.

Let's focus on only one aspect of spawning, i.e., placing the character at the right coordinates. Considering multiple levels of the game, spawn can be done using a record that has current coordinates, and also the default coordinates for each level. If the character needs to be spawned or respawned at any level, the coordinates will be set to the default for that level. Example 5.6c shows spawning the enemy in a game with five levels, based on our simple SDL/SDLX game (Example 4.2c)[2].

### Example 5.6c

```
1.    struct GameObject
2.    {
3.        int x;
```

---

[1] Note that SDL and many other graphics libraries use a pointer to access the objects, for various memory management and data exchange reasons.

[2] In the next chapter, we will see how to define a class/struct based on another, for example, Enemy as a child for GameObject.

```
4.      int y;
5.      SDLX_Bitmap* shape;
6.  };
7.  struct Enemy
8.  {
9.      int x;
10.     int y;
11.     int defaultX[5];
12.     int defaultY[5];
13.     SDLX_Bitmap* shape;
14. };
15. void Spawn(int level, Enemy* e)
16. {
17.     e->x = e->defaultX[level];
18.     e->y = e->defaultY[level];
19. }
20. //in main() - Init part
21.     //player
22.     GameObject player;
23.     player.shape = SDLX_LoadBitmap("Player.bmp");
24.     player.x = 0;
25.     player.y = 0;
26.     //enemy
27.     Enemy enemy;
28.     enemy.shape = SDLX_LoadBitmap("Enemy.bmp");
29.     enemy.defaultX[0] = 100;
30.     enemy.defaultX[1] = 120;
31.     enemy.defaultX[2] = 140;
32.     enemy.defaultX[3] = 160;
33.     enemy.defaultX[4] = 180;
34.     enemy.defaultY[0] = 100;
35.     enemy.defaultY[1] = 120;
36.     enemy.defaultY[2] = 140;
37.     enemy.defaultY[3] = 160;
38.     enemy.defaultY[4] = 180;
39.     Spawn(0, &enemy);
40.     //prize
41.     GameObject prize;
```

```
42.      prize.shape = SDLX_LoadBitmap("Prize.bmp");
43.      prize.x = 200;
44.      prize.y = 200;
45.      int level = 0;
46.  // in main() - Update part
47.             //press space bar to advance level
48.             if (e.keycode == SDLK_SPACE)
49.             {
50.                 level++;
51.                 if (level == 5)
52.                     level = 0;
53.                 Spawn(level, &enemy);
54.             }
```

The above example is simplified to make a point. It advances the game level, simply by pressing the space bar. In a real game, this will happen through normal game progression. Also, other parts of the game object may be reset, and not just the location. A similar algorithm can be used in a banking system if a transaction is going through multiple stages and each stage may be canceled before it is finished, or an image processing program that applies multiple filters to an input image.

*Write a function that uses the structures as parameter for Example 5.6 to simplify drawing objects. What functions can we add to the code in Example 5.6c to simplify the use of structures?*

*Write a Clone function that create a copy of an object.*

### 5.2.3. Tables

An array of records is basically a **table**, as the output of the code in Example 5.5c shows (see Exhibit 5.3). An alternative for creating tables is a two-dimensional array. `Data[50][10]` is a 50x10 table but all its members have the same type and they don't get to have meaningful different names. So, an array of records is a more understandable and flexible way to define tables of data.

```
Student data before sorting:
Name    ID      g1      g2      g3      g4      g5      average
──────  ──────  ──────  ──────  ──────  ──────  ──────  ──────
x       141     67      34      0       69      24      38
x       778     58      62      64      5       45      46
x       881     27      61      91      95      42      63
x       427     36      91      4       2       53      37
x       392     82      21      16      18      95      46

Student data after sorting by ID:
Name    ID      g1      g2      g3      g4      g5      average
──────  ──────  ──────  ──────  ──────  ──────  ──────  ──────
x       881     27      61      91      95      42      63
x       778     58      62      64      5       45      46
x       427     36      91      4       2       53      37
x       392     82      21      16      18      95      46
x       141     67      34      0       69      24      38
```

**Exhibit 5.3. Table as an Array of Records**

There are many operations (and their related algorithms) that are based on tables and records. If you open Microsoft Excel, Apple Numbers, Google Sheets, or any other spreadsheet program, you will see a series of such operations. Take a simple one, for example, sorting. As I showed in Section 3.3, there are multiple ways for sorting a single array of data. When working with table, sorting happens based on one column as the one array that is being sorted. But the reordering needs to apply to all other columns so that all members of a record stay together (on the same row). In other words, comparisons to find the order are done with one member of the record but moving things around is done for the whole record. Example 5.7c demonstrates sorting by ID for a table of student information (results shown in Exhibit 5.3). Note how sorting is done by comparing ID values but the whole struct is being switched.

*Example 5.7c*

```
1.      //sort by ID
2.      for (int i = 0; i < NUM_STUDENTS; i++)
3.          for (int j = i; j < NUM_STUDENTS; j++)
4.          {
5.              if (s[i].ID < s[j].ID)
6.              {
7.                  Student temp;
```

```
8.                  temp = s[i];
9.                  s[i] = s[j];
10.                 s[j] = temp;
11.             }
12.         }
```

*Optimization using a formula to combine max and min of two columns.*

## 5.3. Queues

Being able to insert and remove items in and out of any location of an array is very helpful, but there are cases that we don't want to allow such a "random access." Imagine staying in a coffeeshop line to purchase your morning drink before going to work or school. You are probably in a rush and need to get to a meeting or the start of a lecture. Suddenly, someone shows up and starts chatting with a friend who is ahead of you in the line, and staying there and ordering a coffee. Obviously, you or many others waiting in line will not appreciate this behaviour and prefer newcomers to go to the end of the line and the barista to first serve those who got in line earlier. This method of dealing with people (or anything else, for that matter) is called **first-come-first-serve** or **first-in-first-out** and the group of items managed this way is a **queue**. As an Abstract Data Type (ADT), queues are data structures defined using a group of data items and some specific operations that are supported[1]. The queue data is commonly defined as an array or list, but you could imagine other ways. The operations allowed on a queue are similar to append/insert and remove/delete but with a tricky note: We can only add new data to the end of the queue and remove from the start of it. We sometimes call these two operations **enqueue** and **dequeue**, respectively, to distinguish them from regular append/insert and remove/delete which can be done at any location. When using arrays for a queue, these operations can be implemented using regular append/insert and remove/delete but with the restriction on the location.

---

[1] You don't need to use the term ADT to work with queues. You could just design your data as an array and define some functions for it. Or you could design the whole thing as a class. The beauty of ADTs is that you can define the behaviour of the data structure in a way that is independent of its implementation and then anyone can comply with that definition while using a different way to implement it.

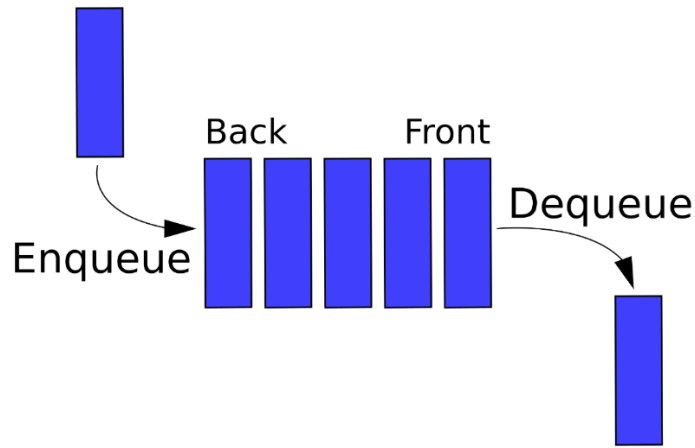

**Exhibit 5.4. Queue. Front and End can be fixed or moving.**

Of course, when you are waiting in line, you can always take a look to see who else is there without adding and removing people. So, queues allow a third, read-only, operation to get the content of any existing location. We call this operation **peek**. It can be implemented by simply reading the array from an index. But keep in mind that ADTs can be implemented in different ways, so we may not even have a simple array and in that case the peek function will be different.

*Queues are first-in-first-out ordered data structures.*

### 5.3.1. Simple Queue Implementation

Example 5.8c shows the implementation of queue in C/C++. It follows our data-centred design approach, starting with the visualization in Exhibit 5.4. The main data elements are the array, the location where the new data item is to be added (back), the location where data is to be extracted (front), and the number of data items (count) which is the difference between front and back. The array can be empty to start. Back and front can be managed in different ways:

- We can add new data to the end of the array. So, the oldest data (front) is always zero, and the number of data items in the queue (count) identifies the next available location to add (back). When removing a data item, we use the index 0 and then delete that data, shifting other down so again the oldest data will be at 0.
- We can insert new data at the start of the array, shifting existing data up. So, the oldest data (front) is at the location identified by the number of data items (count), and the location to add new data is always at index 0.
- We can also have variable back and front, which no longer need any shifting up or down. In this method, we don't insert or delete any data but identify that location as empty by increasing the value of back and using the newly emptied location when needed. This method can be a little confusing since we need to keep track of the front and back both changing and potentially cycling and starting from 0 again to use the emptied locations. On the other hand, it doesn't need any shifting of data, so it can be faster and more efficient when working with large amounts of data.

In Example 5.c, I use the first method. So, front is always zero and back is the count of data in the array. We can have two variables for the array and its length as in Example 5.3c, but now that we have discussed records, we can define a struct to bundle the related data.

```
struct Queue
{
    int data[]; //created dynamically or statically
    int count;  //number of data items in the queue
    int max;    //array size
};
```

The only important actions are inserting the data at the end/back of the array, and extracting from the start/front. In this simple example, we assume each data item is an integer number (imagine the customer number to be served in a busy restaurant lineup). All we do with the extracted data is to print it, which in real-world situations can be replaced with more complicated tasks (see the next Section).

### Example 5.8c

```
1.  bool DeQueue(Queue* q, int* result)
2.  {
```

```
3.        if (q->count != 0)
4.        {
5.            *result = q->data[0];
6.            ShiftDown(q->data, 1, q->count);
7.            q->count--;
8.            return true;
9.        }
10.       else
11.       {
12.           printf("Could not remove data. Queue is empty.\n");
13.           return false;
14.       }
15.   }
16.   bool EnQueue(Queue* q, int newdata)
17.   {
18.       if (q->count != q->max)
19.       {
20.           q->data[q->count] = newdata;
21.           q->count++;
22.           return true;
23.       }
24.       else
25.       {
26.           printf("Could not add data. Queue is full.\n");
27.           return false;
28.       }
29.   }
30.   int main()
31.   {
32.       Queue q;
33.       InitQueue(&q);
34.
35.       bool quit = false;
36.       while (quit == false)
37.       {
38.           int command;
39.           cout << "\n >> 0-exit 1-add 2-remove 3-show: ";
40.           cin >> command;
```

```
41.        if (command == 0)
42.            quit = true;
43.        else if (command == 1)
44.        {
45.            cout << "enter new data: ";
46.            cin >> command;
47.            EnQueue(&q,command);
48.            Print(q.data, q.count);
49.        }
50.        else if (command == 2)
51.        {
52.            DeQueue(&q,&command);
53.            Print(q.data, q.count);
54.        }
55.        else if (command == 3)
56.        {
57.            Print(q.data,q.count);
58.        }
59.    }
60. }
```

In the above code, note that the functions `DeQueue` and `EnQueue` return false/true to show if they were successful. If we knew that the data is limited to a certain range (for example, only positive), then we could return a value outside that range to show failed operation (queue empty when reading or full when writing). When there is no limit to what our data can be, it is common to return an error code (false/true or even an integer number showing the type of error). Since the return value is no longer the data, the `DeQueue` function receives an extra parameter that is a pointer to the data we removed from the queue. Both this parameter and the queue need to be passed as pointer, so the function can modify their original value. See Section 5.5.1 for a refresher on pointers and references. For now, it suffices to say that they are the address of a variable used to access the original data.

*Write the queue code in Python.*

### 5.3.2. Transaction Processing

In the previous example, we had a simple integer number to store and process. In many cases, the items to be added and removed from the queue are more complicated and involve multiple data. For example, as I am typing this text, I can click my mouse on different parts of the screen. The word processor program receives a series of mouse events from the operating system corresponding to these clicks and it has to process them in a first-in-first-out fashion. The related que cannot have just one number though, because the mouse pointer had X and Y coordinates and mouse click can be using left, right, or middle button. A queue of mouse-click events needs to have at least three pieces of information for each item: X, Y, and the button. Any other command processing queue has a series of related data for each queued item. As I discussed in Section 5.2, encapsulation of related data is commonly done using records.

In computing, a bundle of actions that have to happen together and have a clear result is called a transaction. Transactions are commonly defined for databases, such as adding, removing, modifying, and listing items. But they can be defined for other applications too. **Transaction processing** is a queue-based algorithm pattern, as transactions commonly need to be performed in a first-in-first-out fashion. If you initiate a deposit transaction for your bank account and then ask for a report, you expect the report to show the new balance. So, the banking system has to perform your transactions (deposit and report) in the same order as they are received. In Exhibit 5.4, you can imagine each item is a record consisting of all information required for the transaction, while enqueue and dequeue are creating and processing the transaction, correspondingly.

To demonstrate the concept of transaction processing, and the combination of queues and records, imagine the simple banking system of Exhibit 5.5. The main database is a list of accounts (account number, customer information such as names, and balance). The transactions include an account number, a type (withdraw, deposit, and report), and an amount (not needed for report). Of course, in a real banking program, things are a lot more complicated, but this gives you a general idea.

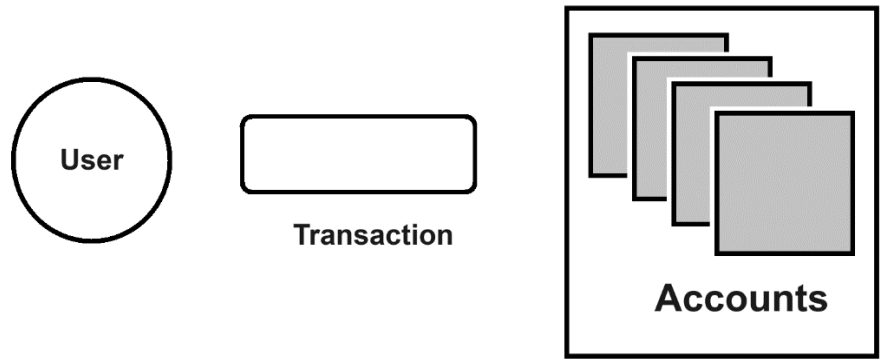

**Exhibit 5.5. Banking System**

Following the data-centred design approach, and based on the visualization in Exhibit 5.5, our main data items are as followed:

- Database is an array of accounts. For the sake of simplicity, we assume the account number is the index to the array. We also assume the accounts are already created and there is no need to deal with adding and removing accounts.
- Account is a record with name and balance.
- Queue is an array of transactions.
- Transaction is a record with number, type, and amount.
- NewTransaction is what we create to be added to the Queue.
- CurrentTransaction is what we extract from the queue to be performed.

The only thing that needs to be initialized is the Database. NewTransaction and CurrentTransaction change when we add and remove from the Queue. In our simple program, these will be done by the same user, but in real banking systems we have the customer and the banking agent (person or automated) as separate users for creating and processing transactions. So, the first thing our program needs to do is to provide options (menu items) for creating and processing transactions. Finally, CurrentTransaction will be used to modify the Database.

Example 5.9p shows a simple banking program written in Python.

***Example 5.9p***

```
1.   #Record types
2.   class Transaction:
3.       """A simple transaction class/record"""
```

```
4.        account = 0
5.        type = " "
6.        amount = 0
7.  class Account:
8.        """A simple account class/record"""
9.        number = 0
10.       name = " "
11.       balance = 0
12. #Global variables
13. accounts = []
14. transactions = []
15. trtypes = ["Report", "Deposit", "Withdraw"]
16. #Functions
17. def GetAccount(i):
18.       acc = Account()
19.       acc.number = 1000+i #start from 1000
20.       acc.name = input ("Enter account holder's name: ")
21.       acc.balance = 0
22.       return acc
23. def PrintAccount(acc):
24.       print("Account:")
25.       print("Account #: ", acc.number)
26.       print("Name: ", acc.name)
27.       print("Balance: ", acc.balance)
28. def GetTransaction():
29.       tr = Transaction()
30.       tr.account = int(input ("Enter account number: "))
31.       tr.type = int(input ("Enter transaction type (1-Report, 2-Deposit,
      3-Withdraw): "))
32.       tr.amount = int(input ("Enter transaction amount: "))
33.       return tr
34. def PrintTransaction(tr):
35.       print("Transaction:")
36.       print("Account #: ",tr.account)
37.       print("Type: ",trtypes[tr.type-1])
38.       print("Amount: ", tr.amount)
39. def RunTransaction(tr):
40.       PrintTransaction(tr)
```

```
41.        if tr.type == 1:
42.            PrintAccount(accounts[tr.account-1000])
43.        elif tr.type == 2:
44.            accounts[tr.account-1000].balance += tr.amount
45.            PrintAccount(accounts[tr.account-1000])
46.        elif tr.type == 2:
47.            accounts[tr.account-1000].balance -= tr.amount
48.            PrintAccount(accounts[tr.account-1000])
49.        else:
50.            print("Invalid transaction")
51. def Init():
52.        global accounts
53.        for i in range(3):
54.            accounts.append(GetAccount(i))
55.        for i in range(3):
56.            PrintAccount(accounts[i])
57. def Loop():
58.        global accounts, transactions
59.
60.        quit = False
61.        while not quit:
62.            command = int(input("0-Exit, 1-Create Transaction, 2-Run
    Transaction, 3-List Transactions: "))
63.            if command == 0:
64.                quit = True
65.            elif command == 1:
66.                newTransaction = GetTransaction()
67.                transactions.append(newTransaction)
68.            elif command == 2:
69.                currentTransaction = transactions[0]
70.                RunTransaction(currentTransaction)
71.                transactions.remove(currentTransaction)
72.            elif command == 3:
73.                for i in range(len(transactions)):
74.                    PrintTransaction(transactions[i])
75.            else:
76.                print("invalid command")
77.
```

```
78.  #Main code
79.  Init()
80.  Loop()
```

The code has two main global variables `accounts` (Database) and `transactions` (Queue). These are arrays of `Account` and `Transaction` types. Each type has its own Get (to create an instance) and Print (to report an instance) functions. Transaction also has a Run function. Two high-level functions, `Init()` and `Loop()`, initialize accounts and run the main loop to get commands for creating a new transaction (adding it to the queue) and running a transaction (removing it from the queue). Note how creating a new transaction acts as Enqueue by appending a new item to the end of queue, and running a transaction acts as Dequeue by removing the oldest item, as in Example 5.8c.

*Add error checking to Example 5.9. (For example, to handle incorrect user input)*

*Add options to create a new account or delete one.*

*Add account number. Hint: run a search to find the index, instead of using the account number as index*

*Write the code in C/C++ (no class)*

## 5.4. Stacks

The first-in-first-out system may be an intuitive way to deal with list items but there are cases when we don't follow it. For example, if you create a pile or stack of clothes to iron or fold on your laundry day (if you do any ironing or folding), there is no reason to work on them in the same order they were added to the pile. In fact, we generally pick an item more conveniently from the top of the pile, which is the reverse order, i.e., **last-in-first-out**. A data structure that represents such a mechanism is called a **stack**, as shown in Exhibit 5.6.

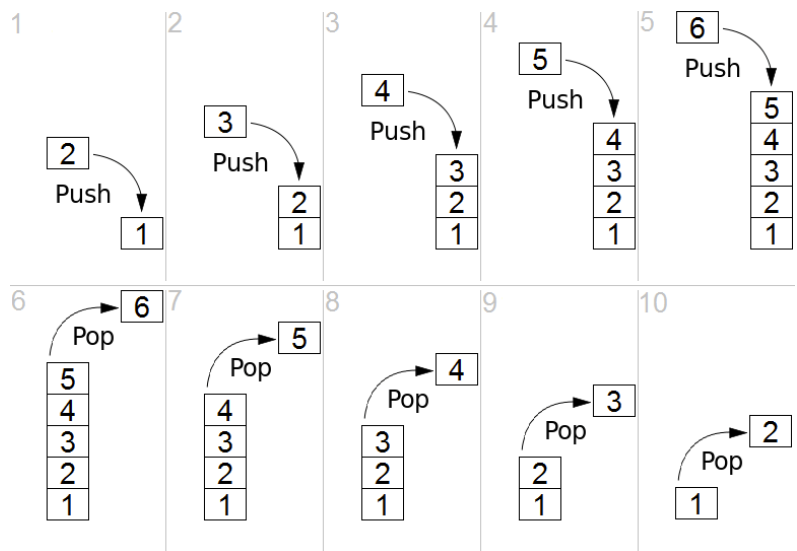

**Exhibit 5.6. Stack**

The most common use of this data structure in computing happens for function calls. Programs to be executed by a computer reside in its memory. Each instruction has its own memory address, and the processor has an internal register, called program counter, that holds the address of the current instruction. In normal sequential execution, this register increments after each instruction to point to the next one. When a program calls a function, the program counter changes to the starting address of that function. The computer needs to know where to return at the end of the called function, so the caller address will be stored. If another function call happens within the called function, another return address has to be saved, and so on for all nested function calls. Once the last function is finished without any other calls, the computer needs to return back in reverse order. This means the last saved address is the first one to be returned to. The return addresses are stored in a stack. Storing in stack and retrieving form it are the **push** and **pop** operations, respectively. A **peek** operation, similar to queues, can give us the value of a location without affecting the stack.

Example 5.10c shows the implementation of stack in C. In this example, stack items are simple integer variables. Similar to queues, stack scan be implemented in different ways. One of the simplest methods is to add new data to the next available location in the array (count) while the oldest data is at location 0. Note that with this implementation, there is no need for shifting the data.

***Example 5.10c***

```
1.   struct Stack
2.   {
3.       int data[MAX];
4.       int count;
5.       int max;
6.   };
7.   void InitStack(Stack* s)
8.   {
9.       s->max = MAX;
10.      s->count = 0;
11.      for (int i = 0; i < MAX; i++)
12.          s->data[i] = -1; //not really necessary but let's assume -1
     means unused
13.  }
14.  bool PeekStack(Stack* s, int* result)
15.  {
16.      if (s->count > 0)
17.      {
18.          *result = s->data[s->count - 1];
19.          return true;
20.      }
21.      else
22.      {
23.          printf("Stack is empty.\n");
24.          return false;
25.      }
26.  }
27.  bool PopStack(Stack* s, int* result)
28.  {
29.      if (s->count != 0)
30.      {
31.          *result = s->data[s->count-1];
32.          s->count--;
33.          return true;
34.      }
35.      else
36.      {
37.          printf("Could not remove data. Stack is empty.\n");
```

```
38.        return false;
39.     }
40.  }
41.  bool PushStack(Stack* s, int newdata)
42.  {
43.     if (s->count != s->max)
44.     {
45.        s->data[s->count] = newdata;
46.        s->count++;
47.        return true;
48.     }
49.     else
50.     {
51.        printf("Could not add data. Stack is full.\n");
52.        return false;
53.     }
54.  }
```

*Stacks are last-in-first-out ordered data structures.*

As we saw for queue, stacks can be combined with records to make more interesting applications. Many programs use the last-in-first-out mechanism and a stack. Among them, I discuss undo operation and a game called the Tower of Hanoi.

### 5.4.1. Undo

Undo is a common feature in many programs that allows uses to remove the effect of what they have done. This feature has to happen in a last-in-first-out fashion as each operation affects the data resulted from its previous ones. For example, consider in a game, you make a change in location A and then go to location B. Undoing the change first will result in a state where you are in B but A has already changed. Such a state never existed in the actual history of your gameplay. While undoing requires a stack structure, the actual information stored in the stack can be different. You can think of two possibilities: saving the information on what has been done or saving the information on what changed (actions vs. items they affected, verbs vs. nouns). Either way, the data in the stack are records containing information for each undoable action. Different programs may also consider different actions as undoable. For example, using arrow keys in a text processor is generally not undoable but in a game it may be. The copy operation (Control-C or Command-C) is usually not undoable but paste (Control-V or Command-V) is.

Let's consider a simple example, a plain text processor such as Notepad on a Windows PC. For simplicity, imagine that we only undo typing a character, and that typing inserts a character at current location or replaces the selected text if any exists. Moving the cursor using arrow keys or mouse is not an undoable action and we don't concern ourselves with commands such as copy/paste. Following our data-centred approach, we first need to identify the data items. Each undoable action requires the following information:

- The entered character
- The location
- The replaced text, if any

Our undo database is a stack of the above items (undoable action) and typing a character and undo command initiate push and pop operations. The C structures for a simple stack are shown below. Note that a full Notepad-like program requires other elements including user interface (menu, mouse support, etc.), but the push and pop operations are fairly similar to those in Example 5.10, except that we will use the `UndoElement` instead of an `int` data.

```
struct UndoElement
{
    char c;
    int row;
    int column;
    char text[MAX_TEXT];
};
struct UndoStack
{
```

```
    UndoElement data[MAX];

    int count;

    int max;

};
```

### 5.4.2. Tower of Hanoi

The Tower of Hanoi is a mathematical puzzle game made of three bars and a series of disks with different sizes that can be placed onto the bars as shown in Exhibit 5.7. The point of the game is to move all disks from a source bar (A) to a destination bar (C) using the third bar (B) as a temporary place, and following these rules:

- Disks can be moved one at a time.
- For each move, we can only take the top disk from a bar and put it on the top of another.
- Bars have to hold disks in increasing size order from the top (looking like a cone).

The solution to this puzzle requires the proper use of the temporary bars[1] and is not the subject of this book. However, the implementation of the puzzle as a game so players can try their solution provides a good example of using stacks. Each bar is used in a last-in-first out fashion, so we need to have three stacks.

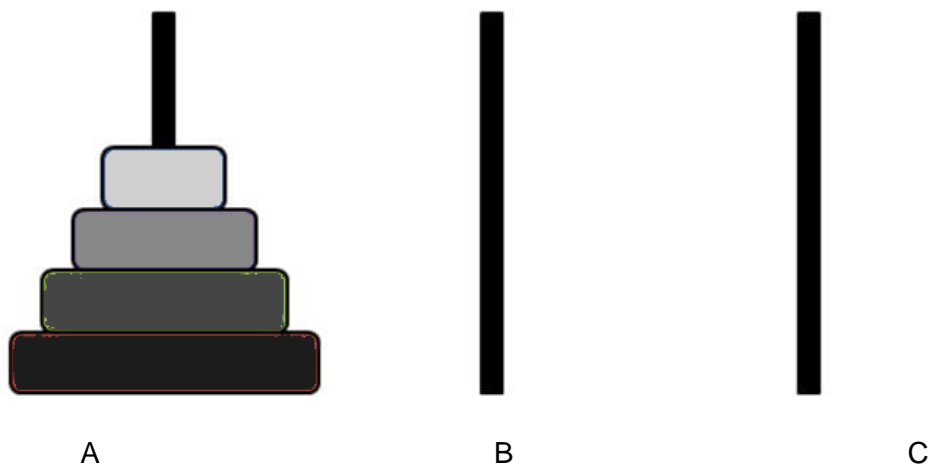

A                          B                          C

[1] https://en.wikipedia.org/wiki/Tower_of_Hanoi#/media/File:Iterative_algorithm_solving_a_6_disks_Tower_of_Hanoi.gif

**Exhibit 5.7. Tower of Hanoi**

Example 5.11c shows the full implementation of a text-based version of the puzzle. Instead of drawing the bars and disks, we simply print the content of the bars with integer numbers representing the disks (1 meaning a disk of size 1, etc.). The stack functions are similar to Example 5.10c. The code creates and initializes three stacks corresponding to bars A, B, and C in Exhibit 5.7. It then stays in a loop and keep asking the user to pop an item from a bar and push it to another.

*Example 5.11c*

```
1.   int main()
2.   {
3.       Stack sa, sb, sc;
4.       InitStack(&sa);
5.       PushStack(&sa, 4);
6.       PushStack(&sa, 3);
7.       PushStack(&sa, 2);
8.       PushStack(&sa, 1);
9.       InitStack(&sb);
10.      InitStack(&sc);
11.
12.      bool quit = false;
13.      while (quit == false)
14.      {
15.          //show three bars
16.          cout << "A: ";
17.          Print(sa.data, sa.count);
18.          cout << "B: ";
19.          Print(sb.data, sb.count);
20.          cout << "C: ";
21.          Print(sc.data, sc.count);
22.
23.          int command;
24.          int data = -1;
25.          bool success;
26.
27.          //pop an item
```

```
28.          cout << "\n >> 0-exit 1-pop from A, 2-pop from B, 3-pop from C:
     ";
29.          cin >> command;
30.          switch (command)
31.          {
32.             case 0:
33.             quit = true;
34.             break;
35.             case 1:
36.             success = PopStack(&sa, &data);
37.             break;
38.             case 2:
39.             success = PopStack(&sb, &data);
40.             break;
41.             case 3:
42.             success = PopStack(&sc, &data);
43.             break;
44.          }
45.          if (quit)
46.             break; //end game
47.          if (!success)
48.             continue; //go to the next iteration
49.
50.          //push the item
51.          cout << "\n >> 0-exit 1-push to A, 2-push to B, 3-push to C: ";
52.          cin >> command;
53.          switch (command)
54.          {
55.          case 0:
56.             quit = true;
57.             break;
58.          case 1:
59.             success = PushStack(&sa, data);
60.             break;
61.          case 2:
62.             success = PushStack(&sb, data);
63.             break;
64.          case 3:
```

```
65.              success = PushStack(&sc, data);
66.              break;
67.          }
68.      }
69.  }
```

*There is no error checking or determining if the game is over, in Example 5.11. Add the necessary code.*

## 5.5. Linked Lists

So far, I have discussed examples of data structures that can be implemented using arrays, as a very intuitive way of making collections of similar items. Arrays are very efficient due to the fact that they represent a set with clear length and members that can be addressed using an index (#1, #2, etc). Although there are operations such as append and remove that dynamically change the organization and length of an array (or list), these data structures are not designed to be dynamic and easily changed. For example, the insert and delete operations require shifting (potentially a large number of) array members. Even if a list function is doing this for the programmer and writing the code is not needed, the shifting is still happening behind the scenes and requires time and system resources. **Linked lists** are an easily configurable collection of items that are designed to be dynamic. A linked list is not implemented using arrays and there is no index value to refer to a member. Instead, each member, in addition to its regular data, has a reference to the next one (**single-linked**) or to both next and previous members (**double-linked**)[1], as shown in Exhibit 5.8[2].

---

[1] Some refer to these two types as "singly linked" and "doubly linked."

[2] There are also the circular and multiple-linked options. Circular-linked list is similar to the circular array we saw in Section 5.1. Multiple-linked list is similar to a tree that I discuss in Section 5.5.4.

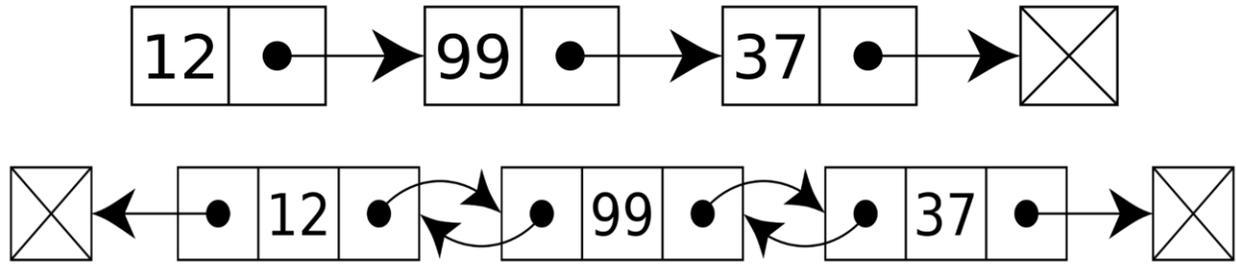

**Exhibit 5.8. Linked List. Top: Single, Bottom: Double**

Linked lists can be standalone sets of items, or they can be used, instead of arrays, to implement other data structures such as queues and stacks. Let's consider a simple example of using linked lists. Imagine a program that is keeping track of a series of players joining a multiplayer online game. Each player is identified using a unique ID, so our database is a growing set of ID numbers. As in any other similar case, we can have three possible solutions:

- We can use a fixed-length integer array that can hold the maximum number of players we may have. This solution is simple and easily done in languages, such as C/C++, that don't have dynamic ways to add/remove members in arrays. But it can be quite wasteful if the array is too large or restricting if it is too small.

- We can start with a small array and use append and remove functions to modify the array. Such functions exist in Python and can be written by the programmer in C/C++. As I discussed earlier though, they need a potentially large number of shifting or copying data.

- We can use a linked list and add and remove members by changing the references in only one (for single-linked) or two (for double-linked) members. When dealing with potentially large set of data, this is a more efficient solution. In this example, each member of the list will be a record (struct or class) with these members:

  - ID
  - Next (reference to the next item)
  - Previous (if double-linked)

The key to understanding how linked lists work is the concept of reference to a data item. Each member of a linked list includes a reference to the next one, not the next one itself. So, before moving forward, it makes sense to pause and review references (or pointers, as they called in C/C++).

---

*Linked lists are data structures similar to arrays but use a pointer/reference to the next and/or previous items instead of indexing. They are used when we have a dynamic set of items and need to easily add and remove.*

---

### 5.5.1. References and Pointers

Each data item (variable) is commonly identified by a name. It also has a value and a type. The following C/C++ code defines a new variable with the name x, type int, and value 0;

```
int x = 0;
```

In Python, the type is not defined explicitly but it exists and is set by the runtime environment when a value is assigned, which can be an int, a string, or a user-defined type. C/C++ have a specific type called pointer, which is an integer that holds the address of a data. So in the following line, `p` is a pointer (identified by `*`) and hold the address of (`&`) the variable `x`.

```
int* p = &x;
```

Just like the value of `x` can change in the program, the value of `p` can change making it point to another data.

```
int y = 1;

p = &y;
```

In C/C++, if we need to have a reference to a data item, we simply define a pointer of that type and set it to the address. Python and some other languages such as C#, on the other hand, don't have a pointer type. In these languages, the program automatically decides which variables are the actual data and which are a reference. Usually, variables representing objects (instances of classes and structs) are always reference to the actual object, which is similar to a C/C++ pointer. Variables representing basic types such as integer and float hold the actual data. We commonly refer to these as **reference types** and **value types**. The following Python code demonstrates how these two types work. In this example, `a1` is an integer data. When you create `a2` as equal to `a1`, it will be a new data item and changing it won't affect `a1`. But `x1` is an instance of a class. If you set `x2` to be equal to `x1`, they are both pointing to the same object. When you change `x2` (one of its attributes/members), you can see that change in `x1`. In other words, `x1` and `x2` are not the actual data items but references to them. In other words, their value is not the data but its address.

```
#value type
```

```
a1 = 0          #a1 is 0

a2 = a1         #a2 is equal to a1 but a separate data

a2 = 1

print(a1)       #prints 0 because changing a2 has on effect on a1
```

**#reference type**

```
class X:

  data = 0

x1 = X()        #x1 is a reference to newly created object

print(x1.data)  #prints 0

x2 = x1         #x2 is a reference to the same object

x2.data = 1

print(x1.data)  #prints 1 because x1 and x2 refer to the same object
```

Some other languages such as C# are similar to Python in the sense that they automatically decide if a variable is value or reference type. In C/C++, we achieve the same result using pointers. Two pointers can point to the same data, which can be changed through either of them. When passing data to a function, it can be done **by value**:

```
void Test1(int data);  //example call: int x=0; Test1(x);
```

Function `Test1()` has a local variable called `data` that is initialized by the value of the data passed as the parameter. Changing this local variable has no effect in the original variable passed to the function (`x`). On the other hand, we can pass data **by reference** (pointer):

```
void Test2(int* pdata); //example call: int x=0; Test2(&x);
```

Function `Test2()` has a local variable called `pdata` that is a pointer initialized by the address that was passed (address of `x`). Using this address, `Test2()` can modify the original variable (`x`).

In linked lists, each member is an instance of a struct or class with Next and Previous members that are references/pointers to the corresponding other instances.

## 5.5.2. Simple Linked List

Let's look at a simple single-linked list example, where we assume that the list is tracking the players in the multiplayer online game I mentioned earlier. Note that we only deal with tracking players and not the actual game. Each member of the list is a record with two data: ID and Next. In C/C++, we implement the record using a struct (or a class).

```
struct Player
{
    int ID;
    Player* Next;
};
```

In Python, we use a class to implement our player record.

```
class Player:
    ID = None
    Next = None
```

In both languages, we need functions for adding a new player and removing one. The simplest action is to add a new member to the end of the list. So, let's start with that. Following our data-centred approach, all we need is the existing list (empty at the start) and the new item. A linked list is not an array so there is no data item representing the whole list. As such, we identify a list using the first and the last items[1]. For an empty list, all these can be initialized to NULL. A function to create a new player and one to add it are the places where we change these data items, respectively.

Adding, of course, uses both last and new item. Adding a new member simply means (1) setting the Next member of the last item to point to this new one, and (2) setting the new one to be the last item.

The following code demonstrates these functions in C/C++.

```
//global data for list.
Player* firstPlayer = NULL;
```

---

[1] Sometimes called head and tail.

```
Player* lastPlayer = NULL;

Player* newPlayer = NULL;

//functions

void CreatePlayer()

{

    newPlayer = new Player(); //or malloc(sizeof(Player))

    newPlayer->Next = NULL;

    //init ID and other player data

}

void AddPlayer()

{

    lastPlayer->Next = newPlayer;

    lastPlayer = newPlayer;

    newPlayer = NULL; //to show the new player is already added

}
```

There are a couple of things to note:

- I am using global data to simplify the code in this example. It is generally not recommended to use global data as they make the code hard to understand and manage, especially in more complex cases. Here, our focus is only on demonstrating the use of linked lists.

- New players have to be created dynamically. In C++, this can be done using `new` operator for structs and classes. In C and C++, we can use `malloc()` function to dynamically create arrays and structures with no member function. The variable `newPlayer` will point to a new item every time we create a new player, and then `lastPlayer` will be pointing to the one that used to be newest.

  - When creating a data item dynamically, it is the responsibility of the programmer to release the memory when the data is no longer needed. This can be done using `delete` operator or `free()` function, as I show later.

*Add error checking for add (what if newPlayer or lastPlayer is NULL).*

A new player can be added at the end of the list, but we can also insert in the middle. Imagine we want the list members to be organized in ascending order of IDs. The reason for such a decision can be the ease of searching for a particular player[1]. Such a sorted list requires new players to be inserted in the right position. Looking for this position in the list or a particular item and many other operations such as reporting the current players or deleting all players are examples of a general operation called **linked list traversing**, that allows us to go through the members one by one to print, compare, or otherwise process them. Example 5.12c expands the previous example by adding Insert, Delete, and Report (as an example of full list traversing). The algorithm for each of these three functions can be designed through our data-centred approach. See 5.13p in the next section for Python code.

***Example 5.12c***

```
1.    #include <stdio.h>
2.    #include <iostream>
3.    using namespace std;
4.
5.    ////////////////////////////////
6.    // linked list item type
7.
8.    struct Player
9.    {
10.       int ID;
```

---

[1] Remember binary search in Section 3.2.2.

```
11.      Player* Next;
12.  };
13.
14.  //global data for list.
15.  Player* firstPlayer = NULL;
16.  Player* lastPlayer = NULL;
17.  Player* newPlayer = NULL;
18.  int numPlayers = 0;  //number of players on the list
19.
20.  /////////////////////////////////
21.  // linked list functions
22.
23.  //create a new player item
24.  void CreatePlayer()
25.  {
26.      newPlayer = new Player(); //or malloc(sizeof(Player))
27.      newPlayer->Next = NULL;
28.      //initialize player data
29.      //this is just an example: setting ID randomly from 1000 to 1100
30.      newPlayer->ID = 1000 + rand()%100;
31.  }
32.  //add the new player item to the end
33.  //we could just call InsertPlayer with numPlayers (end of the list) as
     parameter
34.  void AddPlayer()
35.  {
36.      //error checking: don't add if there is no new player
37.      if (newPlayer == NULL)
38.          return;
39.      //add the new item
40.      if (lastPlayer == NULL) //no existing item
41.          firstPlayer = lastPlayer = newPlayer;
42.      else //at least one item exists
43.      {
44.          lastPlayer->Next = newPlayer;
45.          lastPlayer = newPlayer;
46.      }
47.      numPlayers++;
```

```
48.      newPlayer = NULL;  //to show the new player is already added
49.  }
50.  void InsertPlayer(int i)
51.  {
52.      //error checking: don't add if there is no new player
53.      if (newPlayer == NULL)
54.          return;
55.      //add the new item
56.      if (lastPlayer == NULL) //no existing item
57.          firstPlayer = lastPlayer = newPlayer;
58.      else if (i >= numPlayers)
59.      {
60.          lastPlayer->Next = newPlayer;
61.          lastPlayer = newPlayer;
62.      }
63.      else
64.      {
65.          int n = 0;
66.          Player* tempPlayer = firstPlayer;
67.          Player* prevPlayer = NULL;
68.          while (tempPlayer != NULL)
69.          {
70.              if (n == i - 1) //point previous to new
71.              {
72.                  prevPlayer = tempPlayer;
73.              }
74.              if (n == i) //insert here
75.              {
76.                  newPlayer->Next = tempPlayer;
77.                  if(prevPlayer != NULL)
78.                      prevPlayer->Next = newPlayer;
79.                  if (i == 0)
80.                      firstPlayer = newPlayer;
81.                  break; //get out of the loop (optional. otherwise the
    loop finishes doing nothing more)
82.              }
83.              else //go to the next item
84.              {
```

```
85.              n++;
86.              tempPlayer = tempPlayer->Next;
87.          }
88.      }
89.  }
90.  numPlayers++;
91.  newPlayer = NULL;
92. }
93. void DeletePlayer(int i)
94. {
95.      int n = 0;
96.      Player* tempPlayer = firstPlayer;
97.      Player* prevPlayer = NULL;
98.      while (tempPlayer != NULL)
99.      {
100.         if (n == i - 1) //save the previous to fix its reference later
101.         {
102.             prevPlayer = tempPlayer;
103.         }
104.         if (n == i) //remove this item
105.         {
106.             if (prevPlayer != NULL)
107.                 prevPlayer->Next = tempPlayer->Next;
108.             if (tempPlayer->Next == NULL) //last
109.                 lastPlayer = prevPlayer;
110.             if(i == 0)
111.                 firstPlayer = tempPlayer->Next;
112.             delete tempPlayer;
113.             numPlayers--;
114.             break; //get out of the loop (optional)
115.         }
116.         else //go to the next item
117.         {
118.             n++;
119.             tempPlayer = tempPlayer->Next;
120.         }
121.     }
122.}
```

```
123.void ReportList()
124.{
125.    Player* tempPlayer = firstPlayer;
126.    while (tempPlayer != NULL)
127.    {
128.        cout << tempPlayer->ID;
129.        cout << "\n";
130.        tempPlayer = tempPlayer->Next;
131.    }
132.}
133. ///////////////////////////////////////////////////////////
134. //main function
135.
136. int main()
137. {
138.
139.    bool quit = false;
140.     while (quit == false)
141.     {
142.        int command;
143.        cout << "\n >> 0 - exit, 1 - add, 2 - insert, 3 - delete, 4 -
    report : ";
144.        cin >> command;
145.        switch (command)
146.        {
147.        case 0:
148.            quit = true;
149.            break;
150.        case 1:
151.            CreatePlayer();
152.            AddPlayer();
153.            break;
154.        case 2:
155.            cout << "enter index: ";
156.            cin >> command;
157.            CreatePlayer();
158.            InsertPlayer(command);
159.            break;
```

```
160.        case 3:
161.            cout << "enter index: ";
162.            cin >> command;
163.            DeletePlayer(command);
164.            break;
165.        case 4:
166.            ReportList();
167.            break;
168.        default:
169.            cout << "\n enter valid command 0-4\n";
170.            break;
171.        }
172.    }
173. }
```

*Can we remove but not delete the item?*

*Use link traversing to delete all members. Hint: once you delete a list member, you can no longer use its Next pointer. If you are using Next to traverse, at each item you need to save the Next value then delete the current item. Alternatively, you can use a double-linked list.*

*\*\*\**

*Design a queue using linked lists.*

### 5.5.3. Game Player List

When defining linked lists, as shown in Exhibit 5.8, each member of the list has two parts: the regular data (which itself can be a group of items) and the reference(s). In the previous section, I showed a simple example where the regular data was only an ID. I put all data for each list member together in a record called `Player`. In many cases, we may have more regular data. For example, we may need to keep a player's name and other information. We can address this need by simply adding more items to the `Player` record. On the other hand, it is possible that we need to use the player information in other parts of our program, or even other programs, that are not using the linked list. In other terms, the referencing data and the regular ones are not really of the same type or always related. It is more logical and flexible to have a record that has all the regular `Player` data for any purpose, and then another, specifically for linked lists, that has the first record plus referencing data.

- `Player:`
  - `ID`
  - `Name`
  - `Score`
- `PlayerListItem:`
  - `Player`
  - `Next`
  - `Previous`

This approach follows the modularization idea that is one of the key elements of algorithmic thinking. `PlayerListItem` is a higher-level module that has all the information for a member of our list. `Player`, on the other hand, is a lower-level module that only has player-specific data, and nothing related to the list or how players are used.

Example 5.13p demonstrates this approach applied to the previous example.

***Example 5.13p***

```
1.   import random
2.
3.   #Record types
4.   class Player:
```

```
5.      """A simple Player class/record"""
6.      ID = 0
7.      Name = " "
8.      Score = 0
9.
10. class PlayerListItem:
11.     """A class/record for player list items"""
12.     player = None
13.     next = None
14.     previous = None
15.
16. #Global variables
17. firstPlayerItem = None
18. lastPlayerItem = None
19. newPlayerItem = None
20. numPlayers = 0
21.
22. #list functions
23. def CreatePlayer():
24.     global firstPlayerItem
25.     global lastPlayerItem
26.     global newPlayerItem
27.     global numPlayers
28.     p = Player()
29.     p.ID = 1000 + random.randint(0,100)
30.     newPlayerItem = PlayerListItem()
31.     newPlayerItem.player = p
32.     newPlayerItem.next = None
33.     newPlayerItem.previous = None
34.
35. def AddPlayer():
36.     global firstPlayerItem
37.     global lastPlayerItem
38.     global newPlayerItem
39.     global numPlayers
40.
41.     if newPlayerItem == None:
42.         return
```

```
43.     if firstPlayerItem == None:
44.         firstPlayerItem = newPlayerItem
45.         lastPlayerItem = newPlayerItem
46.     else:
47.         lastPlayerItem.next = newPlayerItem
48.         newPlayerItem.previous = lastPlayerItem
49.         lastPlayerItem = newPlayerItem
50.     numPlayers += 1
51.     newPlayerItem = None
52.
53. def InsertPlayer(i):
54.     global firstPlayerItem
55.     global lastPlayerItem
56.     global newPlayerItem
57.     global numPlayers
58.
59.     if newPlayerItem == None:
60.         return
61.     if lastPlayerItem == None:
62.         firstPlayerItem = newPlayerItem
63.         lastPlayerItem = newPlayerItem
64.     elif i >= numPlayers:
65.         lastPlayerItem.next = newPlayerItem
66.         newPlayerItem.previous = lastPlayerItem
67.         lastPlayerItem = newPlayerItem
68.     else:
69.         n = 0
70.         tempPlayerItem = firstPlayerItem
71.         prevPlayerItem = None
72.         while tempPlayerItem != None:
73.             if n == i-1:
74.                 prevPlayerItem = tempPlayerItem
75.             if n == i:
76.                 newPlayerItem.next = tempPlayerItem
77.                 tempPlayerItem.previous = newPlayerItem
78.                 if prevPlayerItem != None:
79.                     prevPlayerItem.next = newPlayerItem
80.                 if i == 0:
```

```
81.                    firstPlayerItem = newPlayerItem
82.                break
83.            else:
84.                n += 1
85.                tempPlayerItem = tempPlayerItem.next
86.        numPlayers += 1
87.        newPlayerItem = None
88.
89. def DeletePlayer(i):
90.     global firstPlayerItem
91.     global lastPlayerItem
92.     global newPlayerItem
93.     global numPlayers
94.     n = 0
95.     tempPlayerItem = firstPlayerItem
96.     prevPlayerItem = None
97.     while tempPlayerItem != None:
98.        if n == i-1:
99.            prevPlayerItem = tempPlayerItem
100.       if n == i:
101.           if prevPlayerItem != None:
102.               prevPlayerItem.next = tempPlayerItem.next
103.           if tempPlayerItem.next == None:
104.               lastPlayerItem = prevPlayerItem
105.           if i == 0:
106.               firstPlayerItem = tempPlayerItem.next
107.           #no explicit delete needed
108.           numPlayers -= 1
109.           break
110.       else:
111.           n += 1
112.           tempPlayerItem = tempPlayerItem.next
113.
114. def ReportList():
115.     global firstPlayerItem
116.     global lastPlayerItem
117.     global newPlayerItem
118.     global numPlayers
```

```
119.    tempPlayerItem = firstPlayerItem
120.    while tempPlayerItem != None:
121.        print(tempPlayerItem.player.ID)
122.        tempPlayerItem = tempPlayerItem.next
123.
124. #main code
125. #sample operations
126. CreatePlayer()
127. AddPlayer()
128. CreatePlayer()
129. AddPlayer()
130. CreatePlayer()
131. AddPlayer()
132. ReportList()
133.
134. input("insert at 0")
135. CreatePlayer()
136. InsertPlayer(0)
137. ReportList()
138.
139. input("insert at 1")
140. CreatePlayer()
141. InsertPlayer(1)
142. ReportList()
143.
144. input("insert at 10")
145. CreatePlayer()
146. InsertPlayer(10)
147. ReportList()
148.
149. input("delete 0")
150. CreatePlayer()
151. DeletePlayer(0)
152. ReportList()
153.
154. input("delete 1")
155. CreatePlayer()
156. DeletePlayer(1)
```

*Add proper UI to the code in Example 5.13p.*

### 5.5.4. Trees and Graphs

Linked lists are flexible structures that can change dynamically. This flexibility distinguishes them from arrays, but these two types of modules share a linear nature: all members are orders along one path, as we move from one to the next. It is not hard to imagine the possibility of forking these paths to create more than one, or merging multiple paths to turn into one. This idea results in new data structures that are based on a concept similar to linked lists, i.e., each item has references to what comes next and what was before. Tree and Graph are two common examples of these multi-path structures, as shown in Exhibit 5.9.

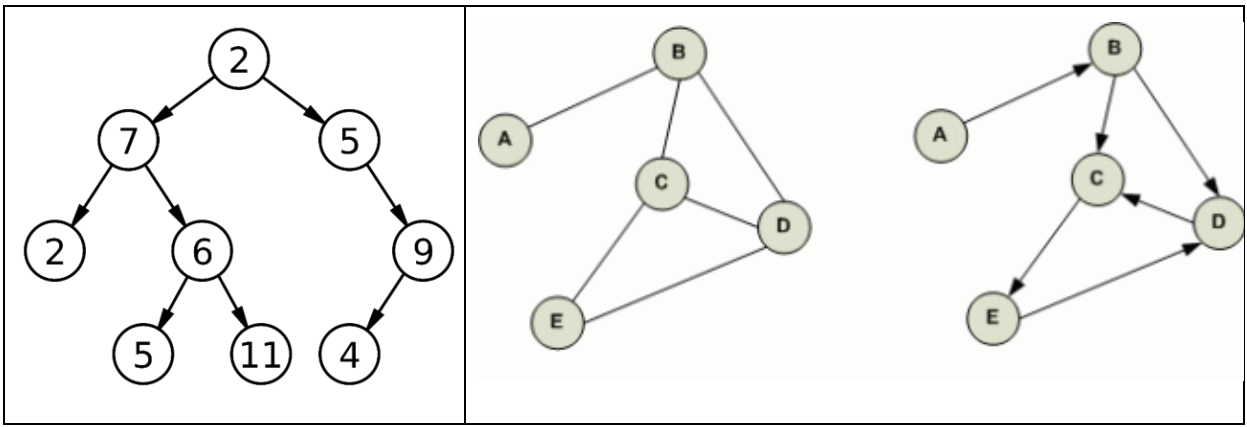

**Exhibit 5.9. Trees (left) and Graphs (right, directed and undirected)**

A **tree** is a linked list where each member can lead to more than one member. The structure starts with a **root** (or parent) item (similar to the first/head in linked lists) and then grows as each member points to more than one **branch** (or child) following it. Trees are the basic form of hierarchies. In object-oriented programming that I discuss in Chapter 6, classes can be defined based on each other to establish a class hierarchy represented by a tree. Note that hierarchies allow multiple parents that happens when multiple paths merge together, as in a flow chart. Some people still call such a structure a tree and some consider multi-parent hierarchies as different structures.

Single or multi-parent, hierarchies can have multiple paths but still a single direction, as branches cannot go back to the root level to form a circle. A **graph**, on the other hand, is made of a series of nodes and each can have incoming or outgoing connection to any other one.

A detailed discussion of trees and graphs and algorithms based on them is beyond the scope of this book.

## 5.6. Simple Game: Revisited

As I mentioned earlier, linked lists are suitable data structures when the data is a collection of items that frequently and dynamically change. A good example is shooting bullets in a 2D game such as Space Invaders or River Raid, shown in Exhibit 5.10. In these games, the player can fire bullets at any time and, sometimes, with no limit.

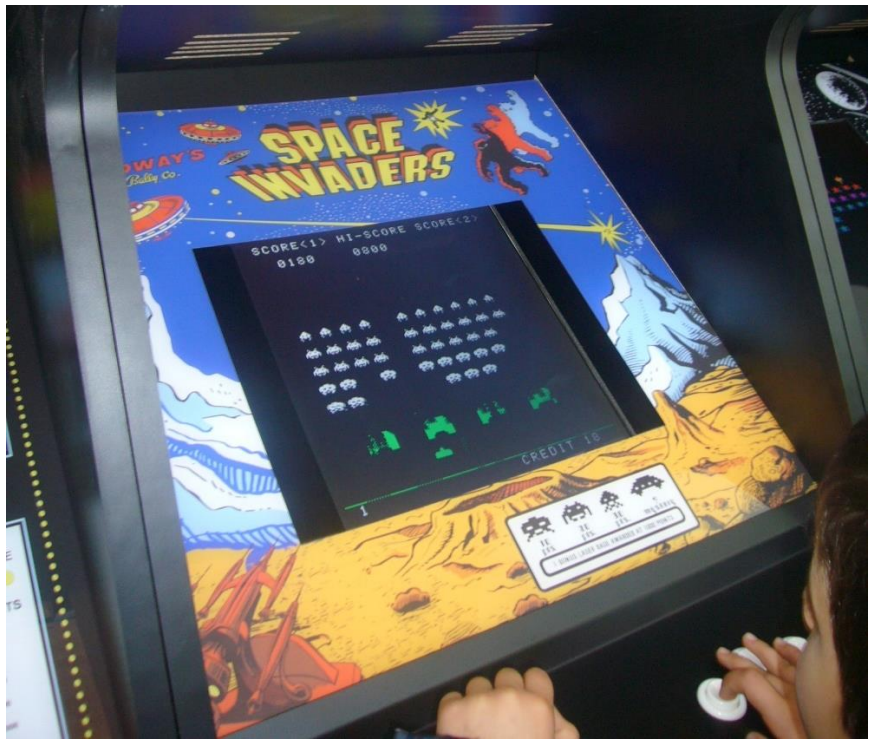

**Exhibit 5.10. 2D Games with Shooting**

https://en.wikipedia.org/wiki/Shooter_game#/media/File:Space_Invaders_-_Midway's.JPG

Let's start by defining the requirements of our game:

- The game has one player character that is displayed at the bottom of the screen and can be moved left and right using arrow keys.
- The game has one or more enemies that move left and right (or move down from the top of the screen).
- Player can shoot bullets to destroy the enemies by pressing space bar. Player bullets go straight up from where they were fired.
- Enemies periodically shoot bullets, which go straight down from where they were fired. Enemies can only have one active bullet at a time.
- The goal is to destroy all enemies before they shoot the player (or reach the bottom of the screen if they move down).
- Player, enemies, and bullets are represented using distinct images.

This game is a very simplified version of Space Invaders. Later, we can add more features to make them more complex. Following the data-centred approach, here are the main data items for this program:

- Bullets with coordinates, image, and direction (up or down).
- Player with coordinates, image, and a set of bullets (array or linked list).
- Enemies with coordinates, image, and one bullet (that can be re-fired once it is out of the screen).

Player bullets can be an array, but they will be fired and destroyed at random times and can have unlimited number. So, a linked list is the preferred option.

You notice that these data items are best encapsulated as a record. They share coordinates and image as attributes/members, as we saw in the GameObject struct earlier. But each has some extra information. When dealing with such cases, we have three common solutions:

- Define three data types.
- Define one data type that has all of the required information by three groups, and then each group can use only the members it needs.
- Define a base/parent type and then use inheritance to derive new types as children.

The third solution is probably the most efficient one, but I leave it to the next chapter when we discuss classes.

The first option is pretty straight forward and is probably what we think of first. But it has the disadvantage that there is no single type that has common aspects. This can be a problem if, for example, we have a function that draws objects. In C/C++, we will need three different functions to deal with three types, even though drawing only uses the common members (location and image). In languages such as Python where functions don't specify the type of parameters, it will be less of a problem.

The second solution does offer a common type, but it lacks efficiency as each object includes unused members. In a simple case like our example, having a couple of small unused data may not be a big deal. But in large programs with many items, such an approach can lead to inefficient use of system resources such as memory.

The direction of movement is something that all objects may use in a later version, and we can always have a set of bullets for the enemies but limit the firing to one at a time. So, the player and enemies can indeed use the same type. Considering these two points, the second solution above is probably appropriate for now. After defining all major date items as GameObject, we will ask the three how questions to structure our program.

I use the SDL/SDLX library to implement this game. First, I define our data types for game objects and the linked lists:

```
struct GameObject
{
    int x;
    int y;
    SDLX_Bitmap* shape;
    int vx;
    int vy;
    bool visible;
};
struct ListItem
```

```
{
    GameObject* obj;

    ListItem* next;

};

struct PlayerBullets

{

    ListItem* first;   //first on the linked list

    ListItem* last;           //last on the linked list

    SDLX_Bitmap* shape;      //common shape for all bullets

};
```

The `GameObject` structure represents player, enemy, and bullets. The `vx` and `vy` variables control the velocity (speed) in both directions while the `visible` one allows us to remove an object from screen. The `ListItem` structure is for the linked list as shown in 5.13. Finally, I define a structure to encapsulate all the info for the list: references to the first and last item plus a common shape for all bullets (instead of each one loading the same image separately). Sharing a single resource among a series of objects is referred to as **flyweight** and is a common software design pattern.

Next, I modularize the code by defining a series of functions needed for the game objects, starting with the generic `GameObject` operations: initializing, moving, and drawing, plus a helper function to detect collision of two objects.

```
//init object with a filename to load for shape

void InitObject(GameObject* obj, char* fname, int x, int y, int vx, int
vy, bool v)

{

    obj->shape = SDLX_LoadBitmap(fname);

    obj->x = x;

    obj->y = y;

    obj->vx = vx;

    obj->vy = vy;
```

```
      obj->visible = v;

   }

   //init object with a pre-loaded shape

   //this helps save resources when multiple objects share teh same image

   void InitObject(GameObject* obj, SDLX_Bitmap* bmp, int x, int y, int vx,
int vy, bool v)

   {

      obj->shape = bmp;

      obj->x = x;

      obj->y = y;

      obj->vx = vx;

      obj->vy = vy;

      obj->visible = v;

   }

   void MoveObject(GameObject* obj)

   {

      obj->x += obj->vx;

      obj->y += obj->vy;

   }

   void DrawObject(GameObject* obj)

   {

      if(obj->visible == true)

            SDLX_DrawBitmap(obj->shape, obj->x, obj->y);

   }

   bool Collision(GameObject* o1, GameObject* o2)

   {

      int distanceX = abs(o1->x - o2->x);

      int distanceY = abs(o1->y - o2->y);
```

```
    if (distanceX < 10 && distanceY < 10)

        return true;

    else

        return false;

}
```

Moving, drawing, and collision detection are fairly straight-forward. I have provided two versions of initialization: one that uses a filename and another that uses the shared pre-loaded image (flyweight).

Next, we have the bullet-specific modules for the player. The enemy bullet is much simpler as we have only one (we'll see it later). The player bullets though need to be created dynamically and added to the list. In this case, compared to 5.13, we have a simpler task as the bullets are always added to the end (as in `AddItem()`) and the earliest one (first) is the one that can hit the enemy (assuming that enemy doesn't move vertically, the earliest bullet is the one that reaches enemy level first). Moving and drawing bullets follows the logic of `ReportList()` (list traversing), while `CheckBullets()` only checks the first bullet to see if it has hit the enemy or reached the top of the screen. Bullets that go off screen need to be deleted.

**//Fire a new bullet**

**//Note that if this is our first bullet (i.e., if last is NULL), it has to be set for first and last item on the list**

**void FireBullet(GameObject\* p, PlayerBullets\* b)**

```
{

    GameObject* new_bullet = new GameObject();

    InitObject(new_bullet, b->shape, p->x, p->y, 0, -1*BULLET_SPEED, true);

    ListItem* li = new ListItem();

    li->obj = new_bullet;

    li->next = NULL;

    if (b->last == NULL)

    {

        b->last = b->first = li;

    }
```

```
    else

    {

            b->last->next = li;

            b->last = li;

    }

}

void MoveBullets(PlayerBullets* b)

{

    ListItem* temp = b->first;

    while (temp != NULL)

    {

            MoveObject(temp->obj);

            temp = temp->next;

    }

}

void DrawBullets(PlayerBullets* b)

{

    ListItem* temp = b->first;

    while (temp != NULL)

    {

            DrawObject(temp->obj);

            temp = temp->next;

    }

}

bool CheckBullets(PlayerBullets* b, GameObject* e)

{

    //check if any hits the enemy

    bool result = false;
```

```
        if (b->first == NULL)

                return false;

        if (Collision(b->first->obj, e))

                result = true;

        if (b->first->obj->y < 0) //remove when gets to the top

        {

                ListItem* temp = b->first;

                b->first = b->first->next;

                if (b->first == NULL)

                        b->last = NULL;

                delete temp->obj;

                delete temp;

        }

        return result;

}
```

Example 5.14c shows the main body of the game code. It has three parts (init, update, and draw) and it is written in response to our three main questions (how to initialize, how to change, and how to use data items). Here are some quick notes:

- All main data items are initialized in INIT section of the code outside the main loop.

  - See how the list for player bullets is created and uses the shared image (`bullet_shape`, loaded only once.

  - Four Boolean flags are used to determine if the loop continues, if we have won or lost, and if arrow keys are pressed.

- The first part of the UPDATE section detects keyboard events and changes some of the variables immediately. After that, and depending on the arrow keys, we move the player and also automatically move the enemy and any bullets.

***Example 5.14c***

```
1.   int main(int argc, char* argv[])
2.   {
```

```
3.      /////////////////////////////////////////////////////////////
4.      // INIT part of the game
5.      // things that are done only once at the start (initialization)
6.      /////////////////////////////////////////////////////////////
7.
8.      //init SDL library and graphics window
9.      SDLX_Init("Simple 2D Game", SCREEN_W, SCREEN_H, true);
10.
11.     //game map (background)
12.     SDLX_Bitmap* map = SDLX_LoadBitmap("map.bmp", false, NULL); //do
   not use any mask for transparent color
13.
14.     //objects
15.     SDLX_Bitmap* bullet_shape = SDLX_LoadBitmap("Bullet.bmp");
16.     GameObject player;
17.     InitObject(&player, (char*) "Player.bmp", 300, 450, 0, 0, true);
18.     PlayerBullets player_bullets;
19.     player_bullets.last = player_bullets.first = NULL;
20.     player_bullets.shape = bullet_shape;
21.     GameObject enemy;
22.     InitObject(&enemy, (char*)"Enemy.bmp", 300, 0, ENEMY_SPEED, 0,
   true);
23.     GameObject e_bullet;
24.     InitObject(&e_bullet, bullet_shape, 300, 0, 0, BULLET_SPEED, true);
25.
26.
27.     //main loop and winning
28.     bool quit = false;
29.     bool win = false;
30.
31.     //left and right movement for player
32.     //true when key is pressed down. false when key released
33.     bool right = false;
34.     bool left = false;
35.
36.
37.     /////////////////////////////////////////////////////////////
38.     //--------Main game Loop---------- -
```

```
39.     // things that are done repeatedly
40.     ///////////////////////////////////////////////////////////////
41.     while (!quit)
42.     {
43.         //////////////////////////////////////////////
44.         // UPDATE
45.         //////////////////////////////////////////////
46.
47.         //check if there is any movement
48.         //player controlled by keyboard
49.         SDLX_Event e;
50.         //keyboard events
51.         while (SDLX_PollEvent(&e))
52.         {
53.             if (e.type == SDL_KEYDOWN)
54.             {
55.             //press escape to end the game
56.             if (e.keycode == SDLK_ESCAPE)
57.                 quit = true;
58.             //press space bar to fire
59.             if (e.keycode == SDLK_SPACE)
60.             {
61.                     FireBullet(&player, &player_bullets);
62.             }
63.
64.             //update player  movement
65.             //we could just move the player here, but this way we can
    process multiple key presses better
66.             if (e.keycode == SDLK_RIGHT)
67.                 right = true;
68.             if (e.keycode == SDLK_LEFT)
69.                 left = true;
70.             }
71.             //this is needed only if we are using left and right flags
    for movement
72.             if (e.type == SDL_KEYUP)
73.             {
74.             if (e.keycode == SDLK_RIGHT)
```

```
75.                right = false;
76.            if (e.keycode == SDLK_LEFT)
77.                left = false;
78.            }
79.        }
80.
81.        //update enemy
82.        MoveObject(&enemy);
83.        if (enemy.x < 0 || enemy.x > SCREEN_W)
84.            enemy.vx *= -1;
85.        if (e_bullet.visible == false && abs(enemy.x - player.x)<5)
    //fire enemy bullet
86.        {
87.            e_bullet.x = enemy.x;
88.            e_bullet.y = enemy.y;
89.            e_bullet.visible = true;
90.        }
91.        MoveObject(&e_bullet);
92.        if (Collision(&player, &e_bullet))
93.        {
94.            quit = true;
95.            win = false;
96.            player.visible = false;
97.        }
98.        //check if enemy bullet out of screen
99.        if (e_bullet.y > SCREEN_H)
100.           e_bullet.visible = false;
101.
102.       //update player
103.       if (right)
104.           player.x += 2;
105.       if (left)
106.           player.x -= 2;
107.       MoveBullets(&player_bullets);
108.       if (CheckBullets(&player_bullets, &enemy))
109.       {
110.           enemy.visible = false;
111.           quit = true;
```

```
112.          win = true;
113.        }
114.
115.        //////////////////////////////////////////
116.        // DRAW
117.        //////////////////////////////////////////
118.        //note the order (last one will be drawn on top)
119.        SDLX_DrawBitmap(map, 0, 0);
120.        DrawObject(&e_bullet);
121.        DrawObject(&enemy);
122.        DrawObject(&player);
123.        DrawBullets(&player_bullets);
124.
125.        //Update the screen
126.        SDLX_Render();
127.
128.
129.        //now wait for loop timing
130.        SDLX_Delay(1);
131.    }
132.
133.    SDLX_Delay(3000);
134.
135.    return 0;
136. }
```

## Highlights

- Arrays are simple modules of related data items that have the same type.
- Animation is an example of array indexing. We change the array index when we need to advance the animation and point to a new frame. Note that the animation frames for one object and the scene frames for the whole game are not the same.
- Array matching is the process of matching corresponding members of two arrays. Matching members may not have the same index, e.g., the member at index 5 of one array may be matched with a member at index 10 of another.

- Array shifting moves members of an array towards the start or end. It is used for inserting or deleting members.
- Records (or user-defined types) are modules of data consisting of members that can have different types (called fields). They encapsulate related data items that can have different type or role in the program into a single entity.
- Queues are first-in-first-out ordered data structures.
- Stacks are last-in-first-out ordered data structures.
- Linked lists are data structures similar to arrays but use a pointer/reference to the next and/or previous items instead of indexing. They are used when we have a dynamic set of items and need to easily add and remove.

# Chapter 6: Objects and Classes

## Overview

In the previous chapter, I discussed records (a.k.a, user-defined types or structures, or simply structs, as called in C/C++). I explained that records have the ability to bundle various related data, what we call **encapsulation**, which can also mean "bundling related data and code" and "bundling related data and code, and restricting access to them." These two, more complex, definitions of encapsulation are based on not the simple records as group of data, but a more complex data module called a **class**. In this chapter, I review how to encapsulate code and data using classes, how to create instances of them as objects, and what are some of most common algorithms that are based on classes and objects. Using objects in software design is referred to as object-oriented programming (OOP). It is a distinct paradigm compared to other software design approaches such as structured, functional, or aspect-oriented programming. OOP is based on three principles: encapsulation, **inheritance**, and **polymorphism**. Designing algorithms that use objects requires an understanding of these principles and how we can use them in defining the structure and behaviour of our algorithm.

## 6.1. Encapsulation: Revisited

Earlier in the book, I introduced modularization as a key concept in algorithm and software design. Defining modules of code, such as functions, and modules of data, such as arrays and records, allows us to organize and reuse our program better. In many of previous examples, we saw that modules of data are commonly processed with modules of code that are created specifically to process that data. For example, consider the following code:

***Example 6.1c***

```
1.   int numStudents = 0; //global variables holding the number of students
2.   struct Student
3.   {
4.       int ID;
5.       int Grades[20];
6.       int GPA;
7.   };
8.
```

```
9.    Student InitStudent()
10.   {
11.       Student s;
12.       s.ID = 1000 + numStudents;        //start from 1000
13.       numStudents++;
14.       for (int i = 0; i < 20; i++)
15.       {
16.           s.Grades[i] = -1;             //use -1 to mean not-defined-yet
17.       }
18.       s.GPA = -1;
19.        return s;
20.   }
21.   void PrintStudent(Student s)
22.   {
23.       printf("ID = % d\n", s.ID);
24.       for (int i = 0; i < 20; i++)
25.       {
26.           printf("Grade - % d = % d\n", i, s.Grades[i]);
27.       }
28.       printf("GPA = % d\n", s.GPA);
29.   }
30.
31.   int main()
32.   {
33.       Student s1 = InitStudent();
34.       PrintStudent(s1);
35.       Student s2 = InitStudent();
36.       PrintStudent(s2);
37.   }
```

In this example, the functions `InitStudent()` and `PrintStudent()` are in charge of initializing and using the data module `Student`. These code modules cannot be used without the related data, and the data cannot be used without these functions (or something similar). While it is possible to keep the data and code modules separate like this, increasing the number of such modules will make it hard to keep track of dependencies and reuse the modules as we need to remember which modules are related to each other. Also, having separate functions means that there is no guaranty that processing data will happen in a standard way. For example, if we require the unused grades to be -1 (to show the variable doesn't hold a valid grade), then a program may use the `Student` struct but make the mistake of not initializing it correctly. In other terms, when using the `Student` struct, a program needs to implement related functions and be aware of how the struct data should be initialized, reported, etc.

Defining modules that include both data and its related code facilitates a more efficient use of both data and code. It helps managing and reusing the program as the module becomes self-sufficient and can be more easily reused without knowing how things are done inside the module. Such a combined module has all the functionality it needs within the module. We refer to modules of combined code and data as **class** and variables based on them as their instances or **object**. Programming languages that support the use of classes and objects are called object-oriented (OO) languages.

In Python, there is no distinction between records/structs and classes, and the keyword `class` defines a module that can hold just data or data and code. In C, a non-OO language, we can only have modules of data using `struct` keyword. In C++, an OO extension of C, there is the new keyword `class` for OO programming. The keyword `struct` is kept for compatibility with C, but it can be used for classes too, i.e., C++ `struct` can have functions. The only difference between C++ structs and classes is that the former has public members by default and the latter has private members by default. Private members of a struct/class can only be accessed by functions that are members of the class/struct. Public members, on the other hand, can be accessed by any code in the program.  Each member can be defined as public or private, or they will have the default status. Examples 6.2c and 6.2p show how to define classes and objects in C++ and Python. These classes encapsulate the same data as 6.1 with two required functions into a single module. Being members of the same module, the functions no longer need to have the data as parameters. They access the data members as if they are local variables. Note in `main()` how we create two variables of `Student` type and each has its own data. When calling `s1.Init()`, for example, the function operates on s1 data.

### *Example 6.2c*

```
1.   int numStudents = 0; //global variables holding the number of students
2.   class Student
3.   {
4.       int ID;
5.       int Grades[20];
```

```
6.       int GPA;
7.   public:
8.       void Init()
9.       {
10.          ID = 1000 + numStudents;   //start from 1000
11.          numStudents++;
12.          for (int i = 0; i < 20; i++)
13.          {
14.             Grades[i] = -1;        //use -1 to mean not-defined-yet
15.          }
16.          GPA = -1;
17.       }
18.      void Print()
19.       {
20.          printf("ID = % d\n", ID);
21.          for (int i = 0; i < 20; i++)
22.          {
23.             printf("Grade - % d = % d\n", i, Grades[i]);
24.          }
25.          printf("GPA = % d\n", GPA);
26.       }
27.  };
28.
29.  int main()
30.   {
31.       Student s1;
32.       s1.Print();
33.       Student s2;
34.       s2.Print();
35.   }
```

C++ allows member functions of a class to be defined inside the class (as in 6.2c) or outside. In the latter case, only a declaration is used inside the class to show the member exists while the full definition is added elsewhere with the name of the class added before the name of the function:

```
class Student
{
```

```
    void Init(); //only a function declaration here

    //other members here

};

void Student::Init()

{

    //code here

}
```

The Python code is structurally similar although the syntax looks a little different. Python requires the keyword `self` (also added as a parameter to any member function) before using any data members to distinguish between a local variable and a class member (see line 8, for example). This and declaration of global variables inside a function are needed because Python doesn't use any keywords or types when defining a new variable. Without them, the first time we use `numStudents` or `ID` in a member function, it wouldn't be clear if we intend to create a new local variable or use the previously defined variables, and Python would simply create new local variables with those names.

***Example 6.2p***

```
1.   class Student:
2.       """A simple Student class"""
3.       ID = 0
4.       Grades = []
5.       GPA = 0
6.       def Init(self):
7.           global numStudents
8.           self.ID = 1000 + numStudents
9.           numStudents += 1
10.          for i in range(0,20):
11.              self.Grades.append(-1)
12.          self.GPA = -1
13.      def Print(self):
14.          for i in range(0,20):
15.              print(self.Grades[i])
16.          print(self.GPA)
17.
```

```
18.  s1 = Student();
19.  s1.Init()
20.  s1.Print()
```

The Python code for creating and initializing array could be simpler by saying `Grades = [-1] * 20`, which doesn't need the loop, but I kept it similar to the C++ code for ease of comparison. Also, instead of using the `Init()` initialization functions, both languages support another method (constructors) which I will discuss shortly.

*Encapsulation can be used to define objects and classes, which are modules that combine related data and code.*

Various algorithms and algorithmic patterns use objects and classes. As part of the data-centred approach, I discussed three how questions that deal with how to initialize, use, and change data. In the following sub-sections, I use these questions as the basis for two categories of tasks in object-oriented algorithm design and algorithmic thinking, initializing and using/changing data. Then I discuss a particular aspect of encapsulation that is controlling access to data, i.e., private vs. public members. But first, let's revisit simple 2D games and see how they look like using objects.

**Note on Multiple Files**

In object-oriented programming, the common and recommended practice is to have separate files for each class. Ease of change and reuse are among the reasons for this practice. Different programming languages provide different methods for a code in file A to use something in file B. In Python, we use the `import` keyword, as we have seen when using external libraries (e.g., Example 4.1p). Java and C# allow the compiler to look inside all project files for a module that we use in a source file. This makes it easy to have a separate file for each class. In C++, on the other hand, each file has to include at least a declaration of what it is using. The header files are made for that purpose. For example, "stdio.h" has declaration (not definition with the actual code) for all functions used in that library. Including "stdio.h" allows us to compile a source file that is using a stdio function. The actual code is then linked to the program, in this case from a pre-compiled library. Other header files can be used for code that is defined in a different source file of our project.

```cpp
// in File A.cpp

//variable

int bar;

//function definition

void foo()

{

    //actual code here

}

// in File A.h

//external variables

extern int bar; //this doesn't create a new variable

//function declaration

void foo(); //no actual code

// in File B.cpp

#include "A.h" // use < > for header files in standard folder and " " for current folder

//somewhere in the code

foo(); //call the external function

bar = 0;
```

For C++ classes, we create a header and a cpp file. The code for the `Student` class of Example 6.2c can be written as follows:

```cpp
//Student.h only lists the members of the class

class Student

{

    int ID;
```

```
    int Grades[20];

    int GPA;

public:

    void Init();

    void Print()

};
```

**//Student.cpp provides the implementation of the member functions**

```
#include "Student.h" //any file that uses the class needs this

int numStudents = 0; //global variables holding the number of students
```

**void Student::Init()**

```
{

    ID = 1000 + numStudents;      //start from 1000

    numStudents++;

    for (int i = 0; i < 20; i++)

    {

        Grades[i] = -1;           //use -1 to mean not-defined-yet

    }

    GPA = -1;

}
```

**void Student::Print()**

```
{

    printf("ID = % d\n", ID);

    for (int i = 0; i < 20; i++)

    {

        printf("Grade - % d = % d\n", i, Grades[i]);

    }

    printf("GPA = % d\n", GPA);

}
```

For simplicity, in the rest of the book, I assume all code for a program is in one file.

## 6.1.1 Simple Object-Oriented Game

If we are following the data-centred approach, after identifying the main data items and operations that need to be performed on them, we can simply bundle these together to form our objects. Later, in Section 6.4, I formalize this and other things we learn about designing algorithms with objects into an extension of data-centred design that I refer to as object-oriented design process.

Imagine that our simple game involves a player that is controlled by arrow keys, a series of enemies that are moving around screen and have to be avoided, and a series of scores that need to be picked up, as in Example 4.2c. A C++ object-oriented version of this game is shown in Example 6.3c. The logic is exactly what we have seen before. The only difference is the use of encapsulation to define self-sufficient `GameObject` class. As you can see, each object now includes functions that process its data. Note that member functions deal with data members of their class as global variables, i.e., no local declaration is needed.

*A common class in games is GameObject that encapsulates data and code for a basic game entity.*

**Example 6.3c**

```
1.  class GameObject
2.  {
3.  public:
4.      int x;
5.      int y;
6.      SDLX_Bitmap* shape;
7.      bool visible;
8.
9.      void Init(char* fname, int xx, int yy)
10.     {
11.         shape = SDLX_LoadBitmap(fname);
```

```
12.        x = xx;
13.        y = yy;
14.        visible = true;
15.      }
16.      void Move()
17.      {
18.        x += rand() % 5 - 2;
19.        y += rand() % 5 - 2;
20.      }
21.      void Draw()
22.      {
23.        if(visible)
24.            SDLX_DrawBitmap(shape, x, y);
25.      }
26.      bool Collision(GameObject* obj)
27.      {
28.        int distanceX = abs(x - obj->x);
29.        int distanceY = abs(y - obj->y);
30.        if (distanceX < 10 && distanceY < 10)
31.            return true;
32.        else
33.            return false;
34.      }
35.  };
36.  int main(int argc, char* argv[])
37.  {
38.      ////////////////////////////////////////////////////////////
39.      // INIT part of the game
40.      // things that are done only once at the start (initialization)
41.      ////////////////////////////////////////////////////////////
42.
43.      //init SDL library and graphics window
44.      SDLX_Init("Simple 2D Game", 640, 480, true);
45.
46.      //game map (background)
47.      SDLX_Bitmap* map = SDLX_LoadBitmap("map.bmp", false, NULL); //do
    not use any mask for transparent color
48.
```

```
49.     //game objects
50.     GameObject player;
51.     player.Init((char*)"Player.bmp", 0, 0);
52.     GameObject enemy;
53.     enemy.Init((char*)"Enemy.bmp", 100, 180);
54.     GameObject prize;
55.     prize.Init((char*)"Prize.bmp", 200, 200);
56.
57.     bool quit = false;
58.     bool win = false;
59.
60.     bool right = false;
61.     bool left = false;
62.     bool up = false;
63.     bool down = false;
64.
65.     ////////////////////////////////////////////////////////////////
66.     //--------Main game Loop---------- -
67.     // things that are done repeatedly
68.     ////////////////////////////////////////////////////////////////
69.     while (!quit)
70.     {
71.         ////////////////////////////////////////////
72.         // UPDATE
73.         ////////////////////////////////////////////
74.
75.         //check if we hit the prize
76.         if(player.Collision(&prize))
77.         {
78.             quit = true;
79.             win = true;
80.             prize.visible = false;
81.         }
82.         //check if we hit the enemy
83.         if (player.Collision(&enemy))
84.         {
85.             player.visible = false;
86.             quit = true;
```

```
87.              win = false;
88.          }
89.
90.          //check if there is any movement
91.          //player controlled by keyboard
92.          SDLX_Event e;
93.          //keyboard events
94.          if (SDLX_PollEvent(&e))
95.          {
96.              if (e.type == SDL_KEYDOWN)
97.              {
98.              if (e.keycode == SDLK_ESCAPE)
99.                  quit = true;
100.             //update player
101.             if (e.keycode == SDLK_RIGHT)
102.                 right = true;
103.             if (e.keycode == SDLK_LEFT)
104.                 left = true;
105.             if (e.keycode == SDLK_UP)
106.                 up = true;
107.             if (e.keycode == SDLK_DOWN)
108.                 down = true;
109.             }
110.             if (e.type == SDL_KEYUP)
111.             {
112.             if (e.keycode == SDLK_RIGHT)
113.                 right = false;
114.             if (e.keycode == SDLK_LEFT)
115.                 left = false;
116.             if (e.keycode == SDLK_UP)
117.                 up = false;
118.             if (e.keycode == SDLK_DOWN)
119.                 down = false;
120.             }
121.         }
122.
123.         //update player
124.         if (right)
```

```
125.              player.x += 3;
126.          if (left)
127.              player.x -= 3;
128.          if (up)
129.              player.y -= 3;
130.          if (down)
131.              player.y += 3;
132.
133.          //update enemy
134.          enemy.Move();
135.
136.          /////////////////////////////////////////
137.          // DRAW
138.          /////////////////////////////////////////
139.          //note the order (last one will be drawn on top)
140.          SDLX_DrawBitmap(map, 0, 0);
141.          prize.Draw();
142.          enemy.Draw();
143.          player.Draw();
144.
145.          //Update the screen
146.          SDLX_Render();
147.
148.
149.          //now wait for loop timing
150.          SDLX_Delay(1);
151.      }
152.
153.      SDLX_Delay(3000);
154.
155.      return 0;
156. }
```

Compared to 4.2c, Example 6.3c modifies the logic of arrow keys to detect both key-down and key-up events, as we saw in 5.14c. Similar to 5.14, we have initialize, move, draw and collision detection functions which are now part pf the GameObject module. Note that all members of the class are public and can be accessed from outside the class (e.g., line 125). I will discuss the access rights later, but next let's see how object-oriented languages make the process of initializing data more streamlined.

## 6.1.2. Initializing Data

As I have mentioned multiple times before, initializing data is a very important part of data processing. Giving proper initial value to our data can have a significant effect on how the program works. Consider, for example, the old case of adding a series of data in the Python code below:

```
Sum = 0

for i in range(5):

   Sum += data[i]
```

In this case, if we don't initialize the variable `Sum` to `0`, the program won't work properly. If we decide to multiply all those data items, then 0 is not going to be the proper initial value and we need to use 1. 0 is the neutral value in addition and 1 the neutral value in multiplication. The initial value of our data in both cases has to be the neutral value (i.e., before anything happens) for the operation we are performing:

```
M = 1

for i in range (5):

   M *= data[i]
```

Note that in the above examples, the first time we use Sum or M, they need to have a value (`Sum+=data[i]` means `Sum=Sum+data[i]`). If the first time we use a data item, we give it a value, then we don't need to initialize it as the initial value is never used. See the C code below as an example where we are using a temporary variable to help switch the values of two other variables entered by user. There is no point initializing `temp`, as it gets its value as part of the switching algorithm.

```
int temp;

cin >> x;

cin >> y;

temp = x;

x = y;
```

```
y = temp;
```

When we create a new object, its data members need to be similarly initialized. The most straight-forward way is to do that by the same part of the code that creates the object:

```
Student s;

s.ID = 1000;
```

While this approach seems right and simple, it defeats the main point of object-oriented programming: the object's data is being processed by external code not one of its own functions. Alternatively, we can have a function that initializes the object:

```
//part of the class definition

Student::Init()

{

    ID = 1000+numStudents;

    numStudents++;

}

//somewhere else in the code

Student s;

s.Init();
```

The `Init()` function guaranties that the data is initialized correctly. It is written when the class is designed and by the people who are aware of that design. As such, its code uses the initial values that are appropriate (`numStudents+1000`). It can also perform extra operations that are necessary after we have a new object (`numStudents++`). While all these could have been done outside the class, it would make it more likely for a programmer using the class to make mistakes and would result in the class being less reusable and self-sufficient.

Having an initialization function that can be called is reasonable, but one more step towards a fully reusable and self-sufficient class is to have an initialization function that is called automatically when we create an object. Most object-oriented languages provide member functions that are called automatically when an instance of the class is created or destroyed. These functions are called **constructor** and **destructor**, respectively. Examples 6.4c and 6.4p demonstrate the syntax for these functions in C++ and Python.

*The most common way of initializing an object is in a special member function of the class called constructor that is called automatically when the object is created.*

### Example 6.4c

```
1.   int numStudents = 0;
2.   int nextID = 1000;
3.
4.   class Student
5.   {
6.       int ID;
7.       int Grades[20];
8.       int GPA;
9.   public:
10.      Student()
11.      {
12.          ID = nextID;
13.          numStudents++;
14.          nextID++;
15.          Init();
16.      }
17.      ~Student()
18.      {
19.          numStudents--;
20.      }
21.      void Init()
22.      {
23.          for (int i = 0; i < 20; i++)
24.          {
25.              Grades[i] = -1;        //use -1 to mean not-defined-yet
26.          }
27.          GPA = -1;
28.      }
```

```
29.      void Print()
30.      {
31.          printf("ID = % d\n", ID);
32.          for (int i = 0; i < 20; i++)
33.          {
34.              printf("Grade - % d = % d\n", i, Grades[i]);
35.          }
36.          printf("GPA = % d\n", GPA);
37.      }
38.  };
39.  int main()
40.  {
41.      Student* s1 = new Student();
42.      printf("%d students\n", numStudents);
43.      Student* s2 = new Student();
44.      printf("%d students\n", numStudents);
45.
46.      delete s2;
47.      printf("%d students\n", numStudents);
48.
49.      Student* s3 = new Student();
50.      printf("%d students\n", numStudents);
51.      s3->Print();
52.
53.      delete s1;
54.      delete s3;
55.      return 0;
56.  }
```

In 6.4c, I am using dynamic object creation, which means student object scan be deleted throughout the program. Once an object is deleted, the number of students has to be reduced but to not repeat the same ID, I have defined two variables: `numStudents` (the number of current students, which can go up and down) and `nextID` (which always goes up). A constructor is called automatically every time an object is created (dynamic or static). A destructor is called every time the object is deleted explicitly or the statically created variable reaches the end of its scope. In this case, constructor is increasing both `numStudents` and `nextID`, but destructor is only decreasing `numStudents`.

Having a constructor doesn't mean we won't need a function like `Init()`. Commonly, there are parts of initialization that can be done right after the object is created and parts that need to be done later, or cases like re-spawn for game characters when we need to reset an object and initialize it again. In 6.4c, the Init function resets the grades.

*An alternative way of initializing an object is an explicit function, as in Init() in Example 6.4c. Try an Init() function for the GameObject class.*

**Example 6.4p**

```
1.  class Student:
2.      """A simple Student class"""
3.      ID = 0
4.      Grades = None
5.      GPA = 0
6.
7.      #constructor
8.      def __init__(self):
9.          global numStudents
10.         self.ID = 1000 + numStudents
11.         numStudents += 1
12.         self.Grades = [-1] * 20
13.         self.GPA = -1
14.
15.      #destructor
16.      def __del__(self):
17.          print("object deleted")
18.
19.      #other functions
20.      def Print(self):
21.          for i in range(0,20):
```

```
22.              print(self.Grades[i])
23.          print(self.GPA)
```

There are various ways to initialize data members of an object, but they fall into two common patterns. The first one is using some logic internal to the class, as we saw in Examples 6.4c and 6.4p. The class uses a global scheme to assign IDs. The second method allows the external code that creates the object to pass parameters to the constructor to be used to initialize the object. These can be individual values corresponding to different members or a full object of the same type whose members will be used to initialize the new object's members. This latter method is commonly referred to as **cloning**. Sometimes classes have a `clone()` function so this action can be done at any time.

To demonstrate everything that we have discussed so far about initializing objects, let's consider adding some difficulties to the sample game of Example 6.3c. Instead of a fixed number of enemies, we start with only one enemy but allow enemies to clone themselves, i.e., each enemy can spawn a copy of itself. This spawn happens after a certain time. To design the algorithm for this game, we follow the data-centred approach, ignoring the player and scores for new, as they are the same as previous examples:

- The main data elements are enemies. Each enemy has XY coordinates and a shape.
- Each enemy needs to spawn a clone after a certain time, so we need to keep track of "age" information for each enemy. This can be a new variable associated with the enemies in addition to coordinates and shape. Time can be measured using the system timer but in programs that use a main loop with fixed cycle time, the number of iterations of the main loop (number of frames) can be a good proxy for time.
- We need a set of enemies that is constantly growing. A linked list will be a good choice.
- Asking three how questions, we can see that constructors are where we initialize the enemies, `Update()` is where we change them, and `Draw()` is where we use them. Adding players will result in a new use in collision detection and ending the game. Changing enemies involves moving them in random direction and cloning if age reaches a certain value.
- Moving enemies requires a new piece of information which is their direction. This direction is fixed but randomly set when we create the object. As per the first rule of programming, we will need a new variable for direction which will become a new member of the class. This direction information can be combined with speed as two `vx` and `vy` variables. Different combinations of speed in X and Y will cause different direction of movement.

- Note that encapsulation means we bundle all the data and functions that act on an object within its class, including a `Move()` and a `Draw()` function. These functions allow the object to be updated and drawn by any logic that is controlled by the class itself. The outside code doesn't need to know how to move or draw the object, and simply calls these functions.

Example 6.5c shows the modified game with only the enemies. Using the concepts of decomposition/modularization and abstraction (Section 3.4), we can divide the bigger task of managing enemies into a hierarchy of modules at different levels:

- The `Enemy` class is the lowest level one and implements the basic functionalities of an enemy object. It is not involved with any linked list actions, as we may want to use it in another case without a linked list.

  - It has three constructors: the **default constructor** has no parameters and uses default values for members, and the alternative constructors have different combination of parameters (only the clone is used in the example). Note that in C++, we can have functions with the same name but different parameters.
  - Similar to `GameObject`, it has move, draw, and collision functions, but `Move()` is now also in charge of creating a clone after a certain time. Later on, I'll show how we can use inheritance to make `GameObject` functions used by `Enemy` without any repetition.

- The `EnemyListItem` class is what we saw in Example 5.13 and simply holds the actual object plus linked list references (next, and if needed previous). A list item class allows us to extend a basic class to linked list operations without modifying the basic class itself. As a layer of abstraction on top of Enemy, this `EnemyListClass` adds higher level functionality (linked list) without adding complications to `Enemy` class, so it can be used in non-linked list cases.

- The `EnemyList` class is in charge of a linked list of enemies. It has data members that hold reference to the first, last, and common shape of the enemies. The member functions initialize the list with a starting enemy, and traverse the list to move, draw and detect collision for all enemies. These functions don't implement the actual operations and simply call the appropriate function from the `Enemy` class.

You can easily add the player and scores from Example 6.3. The `main()` function is the same as 6.3, except that instead of `GameObject enemy`, we have `EnemyList enemies`. Player and prize continue to be instances of `GameObject` class.

**Example 6.5c**

```
1.   class Enemy
2.   {
3.   public:
4.       int x;
5.       int y;
6.       SDLX_Bitmap* shape;
7.       bool visible;
8.       int vx;
9.       int vy;
10.      int age;
11.
12.      Enemy()
13.      {
14.          shape = SDLX_LoadBitmap("Enemy.bmp");
15.          x = 200;
16.          y = 200;
17.          visible = true;
18.          vx = rand() % 5 - 2;
19.          vy = rand() % 5 - 2;
20.          age = 0;
21.      }
22.      Enemy(char* fname, int xx, int yy)
23.      {
24.          shape = SDLX_LoadBitmap(fname);
25.          x = xx;
26.          y = yy;
27.          visible = true;
28.          vx = rand() % 5 - 2;
29.          vy = rand() % 5 - 2;
30.          age = 0;
31.      }
32.      Enemy(SDLX_Bitmap* bmp, int xx, int yy)
33.      {
34.          shape = bmp;
35.          x = xx;
36.          y = yy;
```

```
37.        visible = true;
38.        vx = rand() % 5 - 2;
39.        vy = rand() % 5 - 2;
40.        age = 0;
41.     }
42.     Enemy(Enemy* e) //clone
43.     {
44.        shape = e->shape;
45.        x = e->x;
46.        y = e->y;
47.        visible = e->visible;
48.        vx = rand() % 5 - 2;
49.        vy = rand() % 5 - 2;
50.        age = 0;
51.     }
52.     Enemy* Move()
53.     {
54.        Enemy* e = NULL;
55.        //move
56.        x += vx;
57.        y += vy;
58.        //bounce
59.        if (x < 0 || x > SCREEN_W)
60.            vx *= -1;
61.        if (y < 0 || y > SCREEN_H)
62.            vy *= -1;
63.        //clone
64.        if (age != -1)
65.        {
66.            age++;
67.            if (age == 50)
68.            {
69.            e = new Enemy(this);
70.            age = -1; //don't clone again
71.            }
72.        }
73.        return e;
74.     }
```

```
75.    void Draw()
76.    {
77.        if (visible)
78.            SDLX_DrawBitmap(shape, x, y);
79.    }
80.    bool Collision(GameObject* obj)
81.    {
82.        int distanceX = abs(x - obj->x);
83.        int distanceY = abs(y - obj->y);
84.        if (distanceX < 10 && distanceY < 10)
85.            return true;
86.        else
87.            return false;
88.    }
89. };
90. class EnemyListItem
91. {
92. public:
93.    Enemy* e;
94.    EnemyListItem* next;
95.    EnemyListItem()
96.    {
97.        e = NULL;
98.        next = NULL;
99.    }
100.    EnemyListItem(SDLX_Bitmap* shape, int x, int y)
101.    {
102.        e = new Enemy(shape, x, y);
103.        next = NULL;
104.    }
105. };
106. class EnemyList
107. {
108. public:
109.    EnemyListItem* first;    //first on the linked list
110.    EnemyListItem* last;             //last on the linked list
111.    SDLX_Bitmap* shape; //common shape for all enemies
112.
```

```
113.    EnemyList()
114.    {
115.        shape = SDLX_LoadBitmap("Enemy.bmp");
116.        first = last = new EnemyListItem(shape, 200, 200);
117.    }
118.    void Delete()
119.    {
120.        //remove an item from the list and delete it
121.    }
122.    void Move()
123.    {
124.        EnemyListItem* temp = first;
125.        while (temp != NULL)
126.        {
127.            Enemy* e = temp->e->Move();
128.            if (e != NULL)
129.            {
130.            EnemyListItem* li = new EnemyListItem();
131.            li->e = e;
132.            last->next = li;
133.            last = li;
134.            }
135.            temp = temp->next;
136.        }
137.    }
138.    void Draw()
139.    {
140.        EnemyListItem* temp = first;
141.        while (temp != NULL)
142.        {
143.            temp->e->Draw();
144.            temp = temp->next;
145.        }
146.    }
147.    bool Collision(GameObject* obj)
148.    {
149.        bool result = false;
150.        EnemyListItem* temp = first;
```

```
151.        while (temp != NULL)
152.        {
153.            result = temp->e->Collision(obj);
154.            if (result)
155.            break;
156.            temp = temp->next;
157.        }
158.        return result;
159.    }
160. };
```

*Add Delete function to EnemyList and use it to remove an enemy after a certain time.*

### 6.1.3. Using and Changing Data

The simple 2D game I discussed earlier shows the game world with a top view and allows the player to move in four directions. Another group of 2D games are side-view games, where the player moves in one dimension. The movement in side-view games is usually horizontal, on the ground. Vertical movement can happen as the result of jumping and falling, which is an important game mechanics in platformers but also in other forms of side-view games[1]. Some side-view games have a fixed screen, where the player goes left and right. Some others scroll the screen as the player moves to keep the player at the centre but show new parts of the game world. This corresponds to a fixed virtual camera vs. a camera that moves with the player, as shown in Exhibit 6.1. Fixed screen games have smaller scenes (maps) that fit into one screen. Bigger environments (game worlds/maps) can be managed by multiple levels and rooms that the player visits through game progression. Side-scrolling games (a.k.a. **side-scrollers**) offer another method of displaying bigger scenes that don't fit into the screen where player/camera movement allows us to see different parts.

---

[1] Not all platformers are side-view or even 2D, and not all side-view games are platformer.

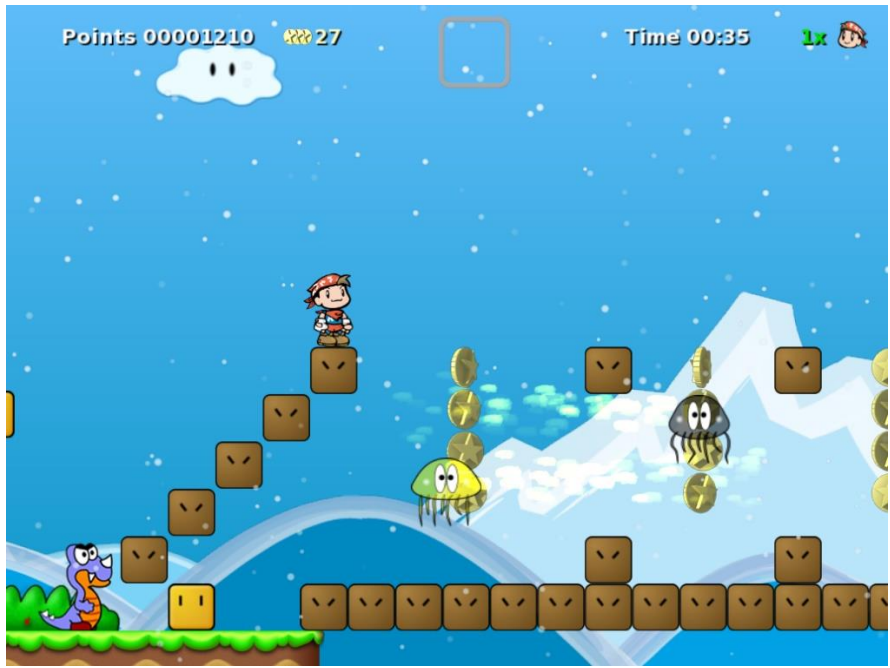

**Exhibit 6.1. Side-scrolling Game**

https://en.wikipedia.org/wiki/Side-scrolling_video_game#/media/File:Secretmaryochronicles.png

To demonstrate the use of data in objects, let's design a simple side-scroller where the player moves and jumps over obstacles. This example will help us understand how to use object data through two tasks, each of them has its own algorithmic thinking: (1) how to scroll the screen, and (2) how to use Physics.

**Scrolling Screen and Coordinate Systems**

The key to understanding how scrolling works is remembering that the game world is larger than what we see on screen, which is the view from a virtual camera attached to the player. The game world corresponds to the background image and the screen shows only a portion of it. Most graphics libraries allow drawing images with an offset, i.e., starting to draw not from the top-left but somewhere in the middle. The following code is an example of how to draw a background image with scrolling. The result is shown in Exhibit 6.2.

```
SDLX_DrawBitmap(map, -300, 0);  //a 300-pixel offset in X direction
```

The notion of a game world that is more than what we see on screen leads to another important fact, that is the difference between the coordinates of any point relative to the screen vs. its coordinate relative to the world. We refer to these two ways of defining coordinates as screen coordinate system and world coordinate system. There is also an object coordinate system that defines the location of any point relative to the object to which it belongs. These coordinate systems are demonstrated in Exhibit 6.2, where the screen has an offset of 300 and a green object is placed at 300 from the left side of the screen.

- The left side of the green object is at x=0, x=300, and x=600 in the object, screen, and world coordinate systems, respectively.
- The left side of the screen is at x=300 in the world coordinate system.
- The left side of the map is at x=-300 in the screen coordinate system.

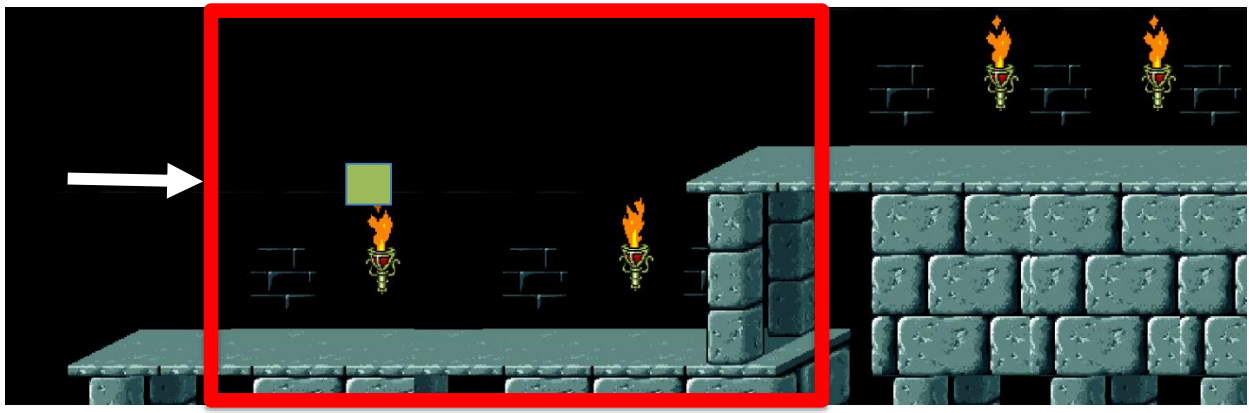

**Exhibit 6.2. Coordinate Systems. The bordered rectangle shows the screen, and the bigger image is the game world/map. The arrow represents the offset or scrolling amount.**

There are simple rules to translate the coordinates from each system to another:

- The reference point (a.k.a. centre) of a coordinate system can be anywhere but we commonly use top-left or the actual centre. For the rest of this book, I use top-left.
- Virtual camera is located at the centre of player character, so it's world coordinates are the player's world coordinate plus half of the player's size.
- Each object's world coordinates are equal to its screen coordinates plus the camera's world coordinates.
- Each point's world coordinates are equal to its object coordinates plus the object's world coordinates.
- All items that store location use the world coordinate system by default as it is the most universal value in the game.

Imagine that at the start of our simple game, the camera is at far left of the game world so that the world and screen coordinates are the same. We draw the background image on screen without any offset which means the left-most part of the image will be drawn and the rest will stay off screen. Using the above rules, and to have the player at the centre of screen, we will need to set the player's world coordinates to:

playerX = screenWidth/2 – playerWidth/2

Player's Y value can be anything depending on where the ground level is. We don't have vertical camera movement in this game.

As the player starts moving to the right, the camera moves with it and the difference between screen and world coordinate systems grows. This won't affect where on the screen the player is drawn, even though the player's X coordinate changes[1]. What changes is which part of the game world (in this case, the background image) we draw on screen, which we saw earlier. The amount of scrolling at any time is equal to the total movement of the player in X direction compared to the starting position. We can calculate this at any frame to find out how much scrolling we need, or we can follow the programming rule #1 and define a variable for it called `offset`:

- At the start: `int offset = 0;`
- Change at any player move: `offset += playerSpeed;`[2]
- Use `offset` when drawing by converting world coordinates to screen coordinates for drawing all objects.

Following the data-centred approach, we can identify the following data items for our simple scrolling game:

- Player with coordinates and image
- A collection of high-level data including scrolling, background image, and ground level (to set the vertical coordinates of the player). All these can be encapsulated into a new item called `Game`, which can also have code modules for updating and drawing.

---

[1] Remember that we are storing the world coordinates in the GameObject members.

[2] This is the amount of movement we have from one frame to another if an arrow key is pressed.

Example 6.6.c shows the new scrolling game based on Example 6.3. Objects are now placed on ground level and player only moves horizontally. As the game is getting more complicated, we have defined a new class called Game to hold all the game-related information (background, scrolling, and player) and functions (initialize, update, and draw). Defining the game (or app) as a main class is a common algorithmic pattern that helps us encapsulate all high-level elements of the program. Exhibit 6.3 shows our objects in what we sometimes call **object aggregation model**, a representation of what objects we have in the program and which ones are members of the others, i.e., how bigger objects include smaller ones.

The `GameObject` class from previous examples is used here with no change for player, enemy, and prize. The `Game` class has functions for Update and Draw and a function called `Run()` to implement the main loop. Initialization is done in the `Game` constructor. The `main()` function only creates a Game object and calls its `Run()`.

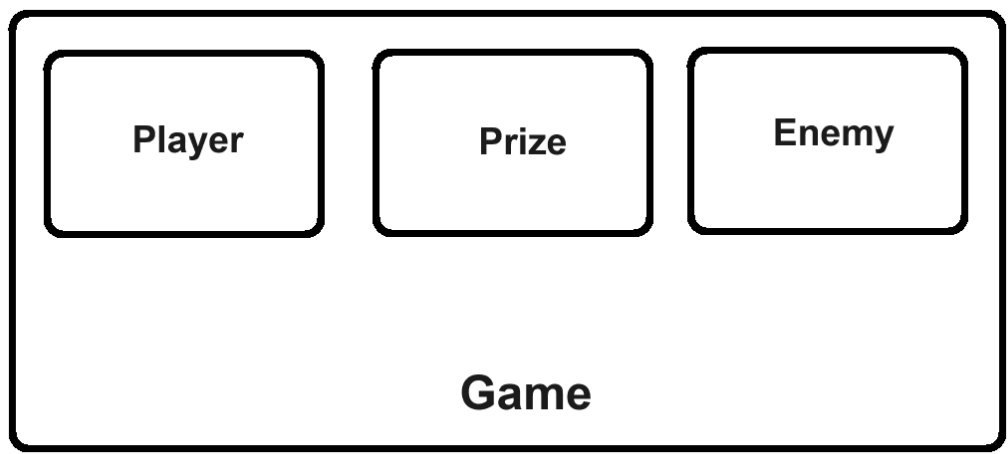

**Exhibit 6.3. Object Aggregation Model**

*Example 6.6c*

```
1.    class Game
2.    {
3.    public:
4.        GameObject player;
5.        GameObject prize;
6.        GameObject enemy;
7.        int offset;
```

```
8.      SDLX_Bitmap* map;

9.      int end_wait;

10.     bool quit;

11.     bool win;

12.     bool left;

13.     bool right;

14.

15.     Game() //same as Init()

16.     {

17.         //init SDL library and graphics window

18.         SDLX_Init("Simple 2D Game", 640, 480, true);

19.

20.         //game map (background)

21.         map = SDLX_LoadBitmap("map.bmp", false, NULL); //do not use any
    mask for transparent color

22.

23.         //control variables

24.         quit = false;

25.         win = false;

26.         right = false;

27.         left = false;

28.         offset = 0;

29.         end_wait = 0;

30.

31.         //game objects

32.         player.Init((char*)"Player.bmp", SCREEN_W/2-PLAYER_W/2);
        //middle of screen

33.         enemy.Init((char*)"Enemy.bmp", 0.75*SCREEN_W);
        //between player and end of screen

34.         prize.Init((char*)"Prize.bmp", 1.5*SCREEN_W);
        //off screen

35.     }

36.     void Update()

37.     {

38.         //check if we hit the prize

39.         if (player.Collision(&prize))

40.         {

41.             quit = true;
```

```
42.          win = true;
43.          prize.visible = false;
44.          end_wait = 3000;
45.      }
46.      //check if we hit the enemy
47.      if (player.Collision(&enemy))
48.      {
49.          player.visible = false;
50.          quit = true;
51.          win = false;
52.          end_wait = 3000;
53.      }
54.
55.      //check if there is any movement
56.      //player controlled by keyboard
57.      SDLX_Event e;
58.      //keyboard events
59.      if (SDLX_PollEvent(&e))
60.      {
61.          if (e.type == SDL_KEYDOWN)
62.          {
63.          if (e.keycode == SDLK_ESCAPE)
64.              quit = true;
65.          //update player
66.          if (e.keycode == SDLK_RIGHT)
67.              right = true;
68.          if (e.keycode == SDLK_LEFT)
69.              left = true;
70.          }
71.          if (e.type == SDL_KEYUP)
72.          {
73.          if (e.keycode == SDLK_RIGHT)
74.              right = false;
75.          if (e.keycode == SDLK_LEFT)
76.              left = false;
77.          }
78.      }
79.
```

```
80.        //update player
81.        if (right)
82.            player.x += 3;
83.        if (left)
84.            player.x -= 3;
85.
86.        //update enemy
87.        enemy.Move();
88.
89.        offset = player.x - (SCREEN_W / 2 - PLAYER_W / 2); //movement
    from original X
90.    }
91.    void Draw()
92.    {
93.        //note the order (last one will be drawn on top)
94.        SDLX_DrawBitmap(map, -1*offset, 0);
95.        prize.Draw(offset);
96.        enemy.Draw(offset);
97.        player.Draw(offset);
98.
99.        //Update the screen
100.        SDLX_Render();
101.    }
102.    void Run()
103.    {
104.        while (!quit)
105.        {
106.            Update();
107.            Draw();
108.            //now wait for loop timing
109.            SDLX_Delay(1);
110.        }
111.        //done
112.        SDLX_Delay(end_wait);
113.    }
114. };
```

The code for initializing, updating, and drawing is almost entirely the same as Example 6.3, with these differences:

- Class members are used instead of local variables.

- The new variable `end_wait` is used in `Run()` to control the amount of wait after the game is over. The default value is zero (when we press Escape to end the game) but it is set to 3000 (3 seconds) if the game ends by winning or losing.

- `Update()` calculates the new offset value as the amount of player movement from original position.

- `Draw()` passes offset to all drawing functions. `GameObject`'s `Draw()` is modified to accept an offset value and calculates screen coordinates using offset and world coordinates.

```
void Draw(int offset=0) //GameObject
{
    if (visible)
        SDLX_DrawBitmap(shape, x-offset, y);
}
```

*Control/limit the offset value to prevent going beyond the game map.*

**Physics of Jump and Fall**

While scrolling games usually have only horizontal movement, they do allow some vertical movement through jumping and falling, which are essential game mechanics in platformer games. Jumping can be to get on platforms or over obstacles. Falling happens when the player gets off a cliff or platform. Let's add a simple jump up/down ability to our scrolling game.

In real world, jumping and falling are accelerated movement controlled by gravity. But to start, let's make things simpler. It is possible to implement jumping and falling using a fixed speed vertical movement. The jump speed simply changes direction to cause falling or coming back down. The main data items that we need are:

- When we start the jump, for example pressing the space bar on keyboard.
- How fast is the jump/fall movement (V)
- How far we go up. With fixed speed, this is proportional to the amount of time, which itself can be measured in actual time units or the number of frames.

If we define new variables for speed and time as members of `GameObject`, we will have a simple algorithm shown below[1]:

```
Initialize:

Vertical_speed = 0

Jump-time = 0

Update:

If spacebar pressed:

    Vertical_speed = V         //set speed

    Jump_time = 0              //reset timer

If Vertical_speed != 0         //if in jump, move and advance time

    Y += Vertical_speed

    Jump_time ++

If Jump_time == Max_time   //if time is up

    If Vertical_speed == V  //if at the end of jumping up, start falling

        Vertical_speed *= -1

        Jump_time = 0

    If Vertical_speed = -V  //if at the end of falling, then stop

        Vertical_speed = 0

Draw:
```

---

[1] Note that these are information about the player, so the variables have to be members of the player object.

```
Draw object at XY as usual
```

*Try the code for simple jump.*

This simple jump works, but it may not look very realistic. The laws of physics define an accelerated movement as $Y = Y_0 + VT + 0.5GT^2$, where $Y_0$ is the initial Y value and G is the gravity acceleration (9.8 m/s$^2$ on Earth). V is the initial speed for jump (0 for free fall)[1] and T is the `Jump_time` variable above. Since the speed is no longer fixed, we can define a new variable to show that we are "jumping" and modify the algorithm:

```
If spacebar pressed:

   Jumping = True

   Jump_time = 0

If Vertical_speed != 0      //if in jump, move and advance time

   Y += Vertical_speed

   Jump_time ++

If Jump_time == Max_time   //if time is up

   If Vertical_speed == V  //if at the end of jumping up, start falling

         Vertical_speed *= -1

         Jump_time = 0

   If Vertical_speed = -V  //if at the end of falling, then stop

         Vertical_speed = 0
```

Example 6.7c shows the implementation of this new algorithm in C++, within the `GameObject` class. Note that the class has a `Move()` function that deals with all sort of movements for the object. It is called in `Game::Update()` on every frame. The default behaviour is to move the object linearly based on `vx` and `vy` amounts (default zero). For the player, `vx` changes when we press arrow keys. If there is jumping, then `Move()` applies the gravity acceleration formula to calculate new Y values and ends jump when the player reaches the ground level.

---

[1] Speed is no longer fixed.

### *Example 6.7.c*

```
1.  #define GROUND_LEVEL 380       //based on the map
2.  #define GRAVITY_ACCELERATION 1 //not the real 9.8. Proportional to
    frame rate
3.  class GameObject
4.  {
5.  public:
6.      int x;
7.      int y;
8.      SDLX_Bitmap* shape;
9.      bool visible;
10.     int vy;
11.     int vx;
12.     //for jumping
13.     int y0;        //Y at the start of the jump
14.     int v0;        //VY at the start of the jump
15.     int t;         //jumping time
16.     bool jumping;
17.
18.     GameObject()
19.     {
20.         shape = NULL;
21.         x = y = vx = vy = 0;
22.         visible = false;
23.     }
24.     void Init(char* fname, int xx, int yy)
25.     {
26.         shape = SDLX_LoadBitmap(fname);
27.         x = xx;
28.         y = yy;
29.         visible = true;
```

```
30.        }
31.      void Move()
32.      {
33.          x += vx;
34.          y += vy;
35.
36.          int gorund = GROUND_LEVEL;  //can change along the map
37.
38.          if (jumping)
39.          {
40.              int newy = y0 + v0 * t + GRAVITY_ACCELERATION * t * t / 2;
     //in jumping sy is the "initial" velocity
41.
42.              if (newy > GROUND_LEVEL)
43.              {
44.              y = GROUND_LEVEL;
45.              jumping = false;
46.              }
47.              else
48.              {
49.              y = newy;
50.              }
51.              t++;
52.          }
53.      }
54.  //Draw() and Collision are the same as before
55.  };
56.  //Game class is the same as Example 6.6 except for Update()
57.        //changes are in this part
58.        if (e.type == SDL_KEYDOWN)
59.        {
60.              //this part is added
61.              if (e.keycode == SDLK_SPACE && player.jumping==false)
62.              {
63.                  player.jumping = true;
64.                  player.y0 = player.y;
65.                  player.t = 0;
66.                  player.v0 = -25;
```

```
67.            }
68.            // the rest is the same
69.        }
70.
71.        //update player is changed as well
72.        player.vx = 0;
73.        if (right)
74.            player.vx += 3;
75.        if (left)
76.            player.vx -= 3;
77.        player.Move();
```

The logic for implementing jump/fall in GameObject is encapsulation of all data and function related to the object in one place. On the other hand, one could imagine a scenario where objects could be moved with different systems, for example, with gravity, with fixed speed, or with other imaginary methods, all based on the game settings. In that case, the algorithm for movement is no longer a property of each object but the game. It could be even a module on its own that can be re-used in different games for implementing different forms of movement applied to any object. Similarly, drawing (a.k.a. **rendering**) an object can be done in different ways, for example, with various styles or added visual effects. Such approaches are common for things such as movement or rendering, where the same object can be drawn in different styles. When using a stand-alone `Physics` or `Render` class, they will provide the `Move()` and `Draw()` functions for an object. The `GameObject` class may keep its default `Move()` and `Draw()` functions, or not. It can be as complicated or simple as the game needs as long as it has the coordinates and image which will be used in `Physics` and `Render` classes.

---

*Create a Physics class that deals with movement of all objects.*

*\*\*\**

### 6.1.4. Controlling the Access to Class Members

Recall that initially I defined encapsulation as "bundling related data." Earlier in this chapter, I showed how that definition can extend to "bundling related data and code." Now, let's discuss the third definition which is "bundling related data and code, and restricting access to them." In the previous example, we assumed all class members were public. This makes programming easier at first, as we don't need to worry about what is public and what is not, and how to access private members. But it can cause complications, especially in more complex programs written by multiple people or classes reused in multiple programs.

Private members can be needed for three important reasons: (1) there are certain valid values that class data members can have, (2) there certain actions that need to be performed when a data member changes, and (3) the implementation of data may change in future versions. When outside code can access (especially write) data members, it is more likely that the programmer sets wrong values (because they don't know what is valid or forget to check) or not perform the required follow-up action. For example, imagine a case when every time data is changed, we need to set a flag that object has to be saved. When we are writing the code for the class, we are aware of this and implement it in any class function that changes the data. Later on, when we or someone else uses that class, we may not remember. Another example is when a set of data is implemented in our class as an array but in a later version, we change it to a linked list. If this class is used by another part of the program or in multiple programs, and they have accessed the data directly, then all their code needs to change to maintain compatibility. On the other hand, if the class has a function to return or change data, that function can still be called with no change. Only the implementation of that function in the class needs to be updated.

This feature of encapsulation (access control) results in a very important outcome that we refer to as **information hiding**[1]. Also called abstraction, information hiding creates black boxes and hides details about implementation of the object. Those who use the object don't need to know how things happen inside the object, but only what public functions to call in order to read or write or otherwise process the data. The set of these public functions is usually called the **interface**, as it allows outside code to connect to the object[2].

---

*Information hiding can involve making some members of the class private, so their definition and implementation is not accessible to outside code, which will become independent of how these members are created.*

---

To demonstrate the use of private members and the notion of information hiding, let's look at another example that is commonly used in games and other graphics applications: particle systems. In computer graphics, particles are small and simple objects that can be grouped together to form complex visual effects that are very fluid and hard to design as one piece. For example, smoke, fire, and water are not solid objects that can be easily visualized. But using a set of particles and a logic that binds them together can simulate them fairly reasonably. The trick is to have a large enough number of particles and a good algorithm that controls their appearance and movement. Figure 6.3 shows a simple particle system simulating smoke coming out of an airplane.

The main data items for the particle system of Figure 6.3 are:

- The individual particles with coordinates and shape, similar to the `GameObject` we saw earlier.

---

[1] Some people consider information hiding a separate property of object-oriented programming, but some (like myself) prefer to see it as an outcome of encapsulation, inheritance, and polymorphism.

[2] Abstraction is commonly defined as adding a higher layer with less details between an object and outside world. Interfaces play the role of that layer of abstraction. Some software systems include multiple layers of abstraction, each dedicated to a certain aspect of the system, for example application layer, operating system layer, and hardware layer.

- A set of particles that we call `ParticleSystem` that can be implemented as an array or linked list
- The source object that can be another simple `GameObject`. This object may or may not be needed in different applications and also may be considered part of the particle system itself.
- The optional Game/App object that encapsulates everything

What makes the design of this example particularly different from the previous ones is that a particle system is very sensitive to the proper working of its element. The way the particles are organized and the logic that controls them are implemented within the `ParticleSystem` class, and it is important that outside code does not interfere with it (for example, moving a particle or adding one that can disturb the visual effect). It is also possible that we want this logic to change without affecting the outside code (for example, making the smoke look different). For this reason, we want to not assume such interference won't happen and instead, establish mechanisms that prevent them. In object-oriented languages, the notion of encapsulation as "bundling related data and code, and restricting access to them." is generally done by **access modifiers**, such as the keywords `public`, `private` and `protected` in C++:

```
Class X
{
    private:    //no need to say this as it is default
    int data1;
    public:    //from here everything is public
    int data2;
    X()
    {
        //some operations to initialize
    }
    void SetData1(int d)
    {
        if (d > 0)
```

```
            data1 = d;

    }

    int GetData1()

    {

            return data1;

    }

    private:     //we can always switch back to private

    int data3;

    void ProcessData3(); //only called by another function in class X

    protected:  //accessible to child classes

    int data4;

    void ProcessData4()

    {

            //some operations

    }

};
```

Some particular things that are worth noting about the access modifiers and the above code are:

- In C++, the default access is private for a class and public for a struct.
- In C++, access modifiers are not for a single member, but they apply to all members in the class listing until the next modifier, if any. They can be used as many times as we want.
- Using a pair of public Get/Set functions is common for private/protected data members. It allows outside code to access these data members in a controlled manner.
- Protected members are accessible within the class and its child classes (see Section 6.2).
- In Java and C#, the same keywords are used with the same meaning but defined per member. In Python, any member that has a name starting with a single _ is protected and anyone with a double underscore __ is private (such as __init__() that is the

constructor[1]). Other members are public. These access modifiers can apply to data members or functions to control who can access them.

Examples 6.8c and 6.8p show this particle system in C++ and Python. The `ParticleSystem` class is the key part of this program. For simplicity, I assume we have a fixed size array of particles. In reality, new smoke particles are created continuously as the old ones spread around and disappear in the air. We can simulate this effect through the use of particles that are all invisible. They start one-by-one at a certain rate (for example, one particle per frame) with full visibility and then gradually becoming transparent. Once a particle has disappeared, it can re-appear as a new one. A constant random movement can give a more realistic look to the particles[2]. Such an **array re-use** algorithmic pattern can be used for other cases such as bullets in Chapter 5, where each bullet can re-spawn once it hits its target or gets off screen. The only point to remember is that we will have a limited number of items, unlike the linked list option or a Python list which can have unlimited members.

### Example 6.8.c

```
1.    #define NUM_PARTICLES 20
2.    class GameObject
3.    {
4.        int x;
5.        int y;
6.        SDLX_Bitmap* shape;
7.        int visible;
8.        int vy;
9.        int vx;
10. public:
11.       GameObject()
12.       {
13.           shape = NULL;
14.           x = y = vx = vy = 0;
15.           visible = 100;
```

---

[1] In all these languages, constructor is called automatically when the object is created and cannot be called directly. Python considers constructor private, but C++ considers it public. Despite that difference, the actual behaviour is similar.

[2] Note that the smoke particles don't move with their source object once they are created. They stay in the air and a trail of smoke particles is formed as new ones are coming out of the moving source.

```
16.     }
17.     void Init(int xx, int yy)
18.     {
19.         x = xx;
20.         y = yy;
21.         visible = 100;
22.     }
23.     void Init(SDLX_Bitmap* bmp, int xx, int yy, int v)
24.     {
25.         shape = bmp;
26.         x = xx;
27.         y = yy;
28.         if (v >= 0 && v <= 100)
29.             visible = v;
30.     }
31.     void Move(bool random=false)
32.     {
33.         if (random)
34.         {
35.             x += rand() % 5 - 2;
36.             y += rand() % 5 - 2;
37.         }
38.         else
39.         {
40.             x += vx;
41.             y += vy;
42.         }
43.     }
44.     void Move(int dx, int dy)
45.     {
46.         x += dx;
47.         y += dy;
48.     }
49.     void Draw()
50.     {
51.         SDLX_DrawBitmap(shape, x, y, visible);
52.     }
53.     int GetVisible()
```

```
54.        {
55.            return visible;
56.        }
57.        void SetVisible(int v)
58.        {
59.            if (v >= 0 && v <= 100)
60.                visible = v;
61.        }
62.        int GetX()
63.        {
64.            return x;
65.        }
66.        int GetY()
67.        {
68.            return y;
69.        }
70.    };
71.    class ParticleSystem
72.    {
73.        GameObject particles[NUM_PARTICLES];
74.        GameObject source;
75.        SDLX_Bitmap* source_shape;
76.        SDLX_Bitmap* particle_shape;
77.    public:
78.        ParticleSystem()
79.        {
80.            //we can't do any SDL/SDLX functions yet
81.            //this constructor is called before Game constructor, so
       SDLX_Init() is not called yet
82.            source_shape = NULL;
83.            particle_shape = NULL;
84.        }
85.        void Init()
86.        {
87.            //second parameter asks ot use a mask to not show certain color
       (background)
88.            source_shape = SDLX_LoadBitmap("Player.bmp", true);
89.            //default color mask is white
```

```
90.         //if we want to use a different one, we can pass hat to the
      LoadBitmap function
91.         SDLX_Color mask_color;
92.         mask_color.r = mask_color.g = mask_color.b = mask_color.a =
      255;
93.         particle_shape = SDLX_LoadBitmap("Prize.bmp", true,
      &mask_color);
94.         //show source in the middle of screen
95.         source.Init(source_shape, SCREEN_W / 2, SCREEN_H / 2, 100);
96.         //prepare particles but don't show yet
97.         for (int i = 0; i < NUM_PARTICLES; i++)
98.         {
99.             particles[i].Init(particle_shape, SCREEN_W / 2, SCREEN_H /
      2, rand()%100);
100.        }
101.    }
102.    void Update(int dx, int dy)
103.    {
104.        //move source
105.        source.Move(dx,dy);
106.        //update particles
107.        for (int i = 0; i < NUM_PARTICLES; i++)
108.        {
109.            particles[i].Move(true); //random move
110.            int v = particles[i].GetVisible();
111.            v--;
112.            particles[i].SetVisible(v);
113.            if (v <= 0)
114.            {
115.                particles[i].Init(source.GetX(), source.GetY());
116.                particles[i].SetVisible(100);
117.            }
118.        }
119.    }
120.    void Draw()
121.    {
122.        for (int i = 0; i < NUM_PARTICLES; i++)
123.            particles[i].Draw();
```

```
124.         source.Draw();
125.     }
126. };
127. class Game
128. {
129.     SDLX_Bitmap* map;
130.     ParticleSystem ps;
131.     bool quit;
132.     bool left;
133.     bool right;
134.     bool up;
135.     bool down;
136. public:
137.     Game() //same as Init()
138.     {
139.         //init SDL library and graphics window
140.         SDLX_Init("Simple 2D Game", 640, 480, true);
141.
142.         //background
143.         map = SDLX_LoadBitmap("map.bmp");
144.
145.         //particle system
146.         ps.Init();
147.
148.         //control variables
149.         quit = false;
150.         down = up = right = left = false;
151.     }
152.     void Update()
153.     {
154.         SDLX_Event e;
155.         if (SDLX_PollEvent(&e))
156.         {
157.             if (e.type == SDL_KEYDOWN)
158.             {
159.             if (e.keycode == SDLK_ESCAPE)
160.                 quit = true;
161.             //update source
```

```
162.            if (e.keycode == SDLK_RIGHT)
163.                    right = true;
164.            if (e.keycode == SDLK_LEFT)
165.                    left = true;
166.            if (e.keycode == SDLK_UP)
167.                    up = true;
168.            if (e.keycode == SDLK_DOWN)
169.                    down = true;
170.            }
171.            if (e.type == SDL_KEYUP)
172.            {
173.            if (e.keycode == SDLK_RIGHT)
174.                    right = false;
175.            if (e.keycode == SDLK_LEFT)
176.                    left = false;
177.            if (e.keycode == SDLK_UP)
178.                    up = false;
179.            if (e.keycode == SDLK_DOWN)
180.                    down = false;
181.            }
182.        }
183.
184.        //update particle system
185.        int dx = 0;
186.        if (right)
187.            dx += 3;
188.        if (left)
189.            dx -= 3;
190.        int dy = 0;
191.        if (up)
192.            dy -= 3;
193.        if (down)
194.            dy += 3;
195.        ps.Update(dx, dy);
196.    }
197.    void Draw()
198.    {
199.        SDLX_DrawBitmap(map,0,0);
```

```
200.        ps.Draw();
201.        SDLX_Render();
202.    }
203.    void Run()
204.    {
205.        while (!quit)
206.        {
207.            Update();
208.            Draw();
209.            SDLX_Delay(1);
210.        }
211.    }
212. };
213.
214. int main(int argc, char* argv[])
215. {
216.    Game game;
217.    game.Run();
218.    return 0;
219. }
```

Python code (6.8.p) is fairly similar, but I should mention that it has some differences compared to the C++ one. These are mainly due to the use of Graphics.py library and its simplified nature, and don't affect the general algorithm and program design:

- Graphics.py has a `move()` function for graphics objects such as `Image` that does the drawing too, so there is no need for drawing every frame. As such, the Python `Game` class has no `Draw()` and the `Draw()` function in `ParticleSystems` is only called at the start.

- Graphics.py `Image` class holds the current location (x and y) so there is no need to have separate variables for them in `GameObject`.

- Graphics.py doesn't offer the ability to control the transparency of images when drawing. So, the particles don't gradually disappear. The `visible` variable is used as a timer and once it reaches zero, the particle is moved to the current location of source to appear as a newly emitted particle. On the other hand, it does support PNG format with built-in transparency.

- Graphics.py `Image` class doesn't support flyweight (sharing a single image for multiple objects), so the code is simpler as there is no need to keep the shared shape, but it is less efficient.

- Graphics.py doesn't generate different events for key_down and key_up, and only shows if a key is down at the time of calling `checkkey()`. So, the keyboard handling part in `Game::Update()` is simpler for the Python program, although less efficient due to the use polling.

### *Example 6.8.p*

```
1.    class GameObject:
2.        # when using Graphics.py, x and y are included in the Image object
3.        __image = None
4.        __visible = 0
5.        def __init__(self, fname, x, y, v):
6.            self.__image = Image(Point(x,y),fname)
7.            if v>=0 and v<=100:
8.                self.__visible = v
9.        def Move(self, dx, dy):
10.            self.__image.move(dx,dy)
11.        def Draw(self,window):
12.            self.__image.draw(window)
13.        def GetV(self):
14.            return self.__visible
15.        def SetV(self, v):
16.            self.__visible = v
17.        def GetX(self):
18.            return self.__image.anchor.x
19.        def GetY(self):
20.            return self.__image.anchor.y
21.
22.    class ParticleSystem:
23.        __source = None
24.        __particles = []
25.        def __init__(self):
26.            for i in range(20):
27.                p = GameObject("prize.png",320,240,random.randint(0,100))
28.                self.__particles.append(p)
```

```
29.          self.__source =
     GameObject("player.png",320,240,random.randint(0,100))
30.      def Update(self,dx,dy):
31.          self.__source.Move(dx,dy)
32.          for i in range(20):
33.              self.__particles[i].Move(random.randint(-2,2),
     random.randint(-2,2))
34.              v = self.__particles[i].GetV()
35.              v -= 1
36.              self.__particles[i].SetV(v)
37.              if v == 0:
38.                  dx = self.__source.GetX() - self.__particles[i].GetX()
39.                  dy = self.__source.GetY() - self.__particles[i].GetY()
40.                  self.__particles[i].Move(dx,dy)
41.                  self.__particles[i].SetV(100)
42.
43.      def Draw(self, window):
44.          for i in range(20):
45.              self.__particles[i].Draw(window)
46.          self.__source.Draw(window)
47.
48.  class Game:
49.      ps = None
50.      window = None
51.      map = None
52.      quit = False
53.
54.      def __init__(self):
55.          self.window = GraphWin('Game', 640, 480)
56.          self.map = Image(Point(320,240),"map.png")
57.          self.map.draw(self.window)
58.          self.ps = ParticleSystem()
59.          self.ps.Draw(self.window)
60.      def Update(self):
61.          dx = 0
62.          dy = 0
63.          key = self.window.checkKey()
64.          if key=='Escape':
```

```
65.                self.quit = True
66.          if key=='Up':
67.                dy -= 3
68.          if key=='Down':
69.                dy += 3
70.          if key=='Right':
71.                dx += 3
72.          if key=='Left':
73.                dx -= 3
74.          self.ps.Update(dx,dy)
75.     def Run(self):
76.          while self.quit == False:
77.                self.Update()
78.                time.sleep(0.05)
79.
80.  #main code
81.  game = Game()
82.  game.Run()
```

## 6.1.5. Sharing Data between Instances of a Class

Classes provide a template for all their instances, i.e., a set of members they all have. Once an object is created as an instance of a class, it has all those members, but an object's data are different variables from the same members in another instance of that class. Changing the value of a member data in one object does not change the value of that same member in another object. Even if they were initialized similarly, they are separate entities. This is a helpful feature because members are similarly structured but have different identities (for example, students in a class or particles in a smoke effect). On the other hand, there are cases that a certain piece of data needs to be the same for all instances of a class. Consider the simple student example that we have seen multiple times.

```
class Student
{
   public:
      int ID;
```

```
int grades[20];

int GPA;

Student()

{

    for(int i=0; i < 20; i++)

    {

        cout << "enter grade: ";

        cin >> grades[i];

    }

}

void Report()

{

    for(int i=0; i < 20; i++)

    {

        cout << "\n";

        cout << grades[i];

    }

}

};
```

To avoid repeating the constant 20, we can have a `#define` (in C/C++), a global variable, or a new class member to hold that value. A #define is simply a label and cannot be changed but it has its own advantages and is recommended every time we use a constant[1]. A global variable is a more generic option, but it means relying on information outside the class and is not generally recommended. Having a member in the class is the best option as it offers flexibility in array size too.

_______________________

[1] It is meaningful and can be changed by the programmer once instead of going and changing everywhere the value if used in the code.

In previous examples, we assumed a fixed-size array for grades. To make it more realistic, imagine we need to define the number of grades each student needs to have, so make the array dynamic. We can use dynamic memory allocation in C/C++ and create an array of the desired size.

```
class Student
{
    public:
    int ID;
    int grades[];
    int numGrades;
    int GPA;
    Student(int n)
    {
        numGrades = n;
        grades = (int*)malloc(numGrades *sizeof(int));
        for(int i=0; i < numGrades; i++)
        {
            cout << "enter grade: ";
            cin >> grades[i];
        }
    }
    ~Student()
    {
        free(grades);
    }
};
```

Using `numGrade` member is a good solution but has the problem that each instance can potentially end up with different number of grades. In some cases, this may be desired but, in many cases, it is not[1]. To allow all instances of a class to share a data item, most object-oriented languages define two types of members: class members (vs. instance members) in Python and static members (vs. regular members) in C++ are good examples of this shared data. In C++, the Student class with static number of grades will look like this:

```
class Student
{
    public:
    int ID;
    int grades[];
    static int numGrades;
    static ChangeNumGrades(int n)
    {
        if (n > 0)
            numGrades = n;
            //reallocate the array
    }
    int GPA;
    Student()
    {
        grades = (int*)malloc(numGrades*sizeof(int));
        for(int i=0; i < numGrades; i++)
        {
            cout << "enter grade: ";
            cin >> grades[i];
        }
```

---

[1] All students need to take the same number of courses; all particles should disappear after the same number of frames; all falling objects should have the same gravity acceleration, etc.

```
    }

    ~Student()

    {

        free(grades);

    }

};

Student::numGrades = 20;

Student::SetnumGrades(20); //static functions can only use static data
```

Note that static members are accessed through the class name (and ::) not the object (and .), as they don't belong to a single object. Also keep in mind that static functions can only use static data members.

*To allow all instances of a class to share a data item, most object-oriented languages define two types of members: class members (vs. instance members) in Python and static members (vs. regular members) in C++ are examples of shared data between class instances.*

## 6.1.6. Using Functions

In our data-centred approach, we commonly define classes based on their core data and the function needed for processing them. But there are cases where a class includes functions that are used to process data outside the class itself. Such classes are generally defined not as a collection of data and its related functions, but as a collection of a set of related functions. An example is a math class that provides various helper functions to perform math operations. Physics and Render classes that I talked about earlier are other examples of classes that are not encapsulating data and code, but mainly defined as a provider of certain functions.

As another example, let's consider a filter (visual effect) class that supports different image processing filters. An image is basically a 2D array of pixel values. Processing images is a good example of nested loops that I discussed in Section 4.3. Example 6.9c shows a class called `Filter` that applies multiple filters to any input image:

- Gray: Converts the image to gray scale buy changing red, green, and blue components into an equal value
- Blur: Makes the image smoother by changing any pixel to the average of its adjacent ones
- Bright: Increases or decreases the brightness of the image

*Example 6.9.c*

```
1.    class Filter
2.    {
3.    public:
4.        void Gray(GameObject* obj)
5.        {
6.            //image to apply filter
7.            SDLX_Bitmap* img = obj->GetShape();
8.
9.            //image dimensions
10.           int w = img->width;
11.           int h = img->height;
12.
13.           //color
14.           SDLX_Color c;
15.
16.           //loop
17.           for (int i = 0; i < w; i++)
18.              for (int j = 0; j < h; j++)
19.              {
20.                 SDLX_GetPixel(img, i, j, &c);
21.                 int g = (c.b + c.g + c.r) / 3; //a simplified algorithm for
      changing to gray scale
22.                 c.r = c.g = c.b = g;
23.                 SDLX_PutPixel(img, i, j, &c);
24.              }
25.        }
26.        void Blur(GameObject* obj)
```

```
27.        {
28.            //image to apply filter
29.            SDLX_Bitmap* img = obj->GetShape();
30.
31.            //image dimensions
32.            int w = img->width;
33.            int h = img->height;
34.
35.            //color
36.            SDLX_Color c;
37.            SDLX_Color cr;
38.            SDLX_Color cl;
39.
40.            //loop
41.            //this code only considers the neighbouring pixels on the left
    and right
42.            for (int i = 1; i < w-1; i++) //the pixels at the edge don't
    have two neighbours
43.                for (int j = 0; j < h; j++)
44.                {
45.                    SDLX_GetPixel(img, i, j, &c);
46.                    //the following if statements are to make sure we don't
    conisder masked(transparent) pixles
47.                    if ((c.r != color_mask.r || c.g != color_mask.g || c.b !=
    color_mask.b) && c.a != 0)
48.                    {
49.                        int n = 1;
50.                        SDLX_GetPixel(img, i + 1, j, &cr);
51.                        if ((cr.r != color_mask.r || cr.g != color_mask.g ||
    cr.b != color_mask.b) && cr.a != 0)
52.                        {
53.                            n++;
54.                            c.r += cr.r;
55.                            c.g += cr.g;
56.                            c.b += cr.b;
57.                            SDLX_GetPixel(img, i - 1, j, &cl);
58.                            if ((cl.r != color_mask.r || cl.g !=
    color_mask.g || cl.b != color_mask.b) && cl.a != 0)
```

```
59.                     {
60.                         n++;
61.                         c.r += cl.r;
62.                         c.g += cl.g;
63.                         c.b += cl.b;
64.                     }
65.                 }
66.                 c.r /= n;
67.                 c.g /= n;
68.                 c.b /= n;
69.                 SDLX_PutPixel(img, i, j, &c);
70.             }
71.         }
72.     }
73.     void BrightUp(GameObject* obj)
74.     {
75.         //change RGB levels. Make sure we don't exceed 0-255 range
76.     }
77.     void BrightDown(GameObject* obj)
78.     {
79.         //change RGB levels. Make sure we don't exceed 0-255 range
80.     }
81. };
```

The `Filter` class can be used in application. For example, the `Game` class of Example 6.8c can be modified to apply filters by calling filter functions when a certain key is pressed:

```
//apply filter in Game::Update

if (e.keycode == SDLK_1)

        filter.Gray(&player);

if (e.keycode == SDLK_2)

        filter.Blur(&player);
```

*Implement full Blur and Bright. Hint:*

*Make filter functions static so we don't need a Filter object in Game*

*How can we Undo? Apply to a clone.*

*Why not use a single function, Apply(), with a parameter that is the type of the filter to apply*

---

## 6.2. Inheritance

In Section 5.6, when designing our simple game with three types of objects, I discussed options to deal with data types that have some members in common. Defining separate but somewhat similar types is one solution and defining an all-inclusive type that has all the possible members is another, both with their own problems with regards to program efficiency. Object-oriented programming offers a new way of dealing with such cases through **inheritance**. Imagine our simple game with the following data items and their members:

- Player. Animated with ability to keep track of life (decreased if hitting enemies) and score (increased when picking up score objects)
  - Data: coordinates, image, frames, current frame, life, score
  - Functions: Draw, Move, AdvanceAnimation
- Enemy. Non-animated moving in random direction
  - Data: coordinates, image, direction, speed
  - Functions: Draw, Move with bounce (change direction when hitting the side of the screen)
- Score. Stationary, disappearing after being picked up
  - Data: coordinates, image, visible (false when picked up)
  - Functions: Draw

Inheritance allows us to define a base (parent) class that includes all the basic and common members, and then define other classes as its derived (child) classes. We can still have some members in the base class that are not used by all child classes, but they are very common and we expect to be used by most future child classes. For example, `speed` and `Move()` are not used in Score but they are good members to have in the base class as we imagine that most classes will need them. Once the base class is defined, child classes can do three things:

1. Add new data members
2. Add new functions
3. Modify the implementation of existing functions

In our example, we can have the following classes:

- GameObject. Our base class for the main objects in the game[1]
  - Data: coordinates, image, speed (positive and negative to support direction), visibility
  - Functions: Draw, Move
- Player. Child of GameObject
  - New Data: frames, current frame, life, score
  - New Functions: AdvanceAnimation
  - Modified functions: Move and Draw
- Enemy. Child of GameObject
  - New Data:
  - New Functions: Move (new implementation)
- Score. Not necessary. We can use GameObject, or we could move visibility from base to a child Score class

Example 6.10c shows the C++ code for the new game. We still have a game class that manages the whole program.

---

[1] Note that here by "object" I mean entities that have a visible role in gameplay (game objects). There can be other possible classes in our program that are not based on GameObject and their instances are objects in programming sense but not in gameplay sense. For example, we can have a Game class or a Menu class, etc.

## Example 6.10.c

```
1.    #include "Main.h"
2.    #define SCREEN_W 640
3.    #define SCREEN_H 480
4.    #define PLAYER_W 48
5.    #define GROUND_LEVEL 410          //based on the map
6.    #define GRAVITY_ACCELERATION 1
7.    #define NUM_FRAMES 3
8.
9.    class GameObject
10.   {
11.   public:
12.       int x;
13.       int y;
14.       SDLX_Bitmap* shape;
15.       bool visible;
16.       int vy;
17.       int vx;
18.
19.       GameObject()
20.       {
21.           shape = NULL;
22.           x = y = vx = vy = 0;
23.           visible = false;
24.       }
25.       void Init(char* fname, int xx, int yy)
26.       {
27.           shape = SDLX_LoadBitmap(fname);
28.           x = xx;
29.           y = yy;
30.           visible = true;
31.       }
32.       void Move()
33.       {
34.           x += vx;
35.           y += vy;
36.       }
```

```
37.     void Draw(int offset = 0)
38.     {
39.         if (visible)
40.             SDLX_DrawBitmap(shape, x - offset, y);
41.     }
42.     bool Collision(GameObject* obj)
43.     {
44.         int distanceX = abs(x - obj->x);
45.         int distanceY = abs(y - obj->y);
46.         if (distanceX < 10 && distanceY < 10)
47.             return true;
48.         else
49.             return false;
50.     }
51. };
52.
53. class Player : public GameObject
54. {
55. public:
56.     //for jumping
57.     int y0;        //Y at the start of the jump
58.     int v0;        //VY at the start of the jump
59.     int t;         //jumping time
60.     bool jumping;
61.     //for animation
62.     SDLX_Bitmap* frames[NUM_FRAMES];
63.     int currentFrame;
64.
65.     Player()
66.     {
67.         y0 = v0 = t = 0;
68.         jumping = false;
69.         currentFrame = 0;
70.         for (int i = 0; i < NUM_FRAMES; i++)
71.             frames[i] = NULL;
72.     }
73.     void InitFrame(char* fname, int f, bool useMask=false, SDLX_Color*
    mask=NULL)
```

```
74.      {
75.          frames[f] = SDLX_LoadBitmap(fname, useMask, mask);
76.      }
77.      void AdvanceAnimation()
78.      {
79.          currentFrame++;
80.          if (currentFrame == NUM_FRAMES)
81.              currentFrame = 0;
82.      }
83.      void Move()
84.      {
85.          GameObject::Move();
86.
87.          if(vx != 0)
88.              AdvanceAnimation();
89.
90.          int gorund = GROUND_LEVEL-frames[0]->height;  //can change
    along the map
91.
92.          if (jumping)
93.          {
94.              int newy = y0 + v0 * t + GRAVITY_ACCELERATION * t * t / 2;
    //in jumping sy is the "initial" velocity
95.              if (newy > gorund)
96.              {
97.              y = gorund;
98.              jumping = false;
99.              }
100.             else
101.             {
102.             y = newy;
103.             }
104.             t++;
105.         }
106.     }
107.     void Draw(int offset = 0)
108.     {
109.         if (visible)
```

```
110.              SDLX_DrawBitmap(frames[currentFrame], x - offset, y);
111.      }
112.
113. };
114.
115. class Enemy : public GameObject
116. {
117. public:
118.     Enemy()
119.     {
120.
121.     }
122.     void Move()
123.     {
124.         vx = rand() % 5 - 2;
125.         vy = rand() % 5 - 2;
126.
127.         GameObject::Move();
128.     }
129. };
130. class Game
131. {
132. public:
133.     Player player;
134.     GameObject prize;
135.     Enemy enemy;
136.     int offset;
137.     SDLX_Bitmap* map;
138.     int ground_level;
139.     int end_wait;
140.     bool quit;
141.     bool win;
142.     bool left;
143.     bool right;
144.
145.     Game() //same as Init()
146.     {
147.         //init SDL library and graphics window
```

```
148.        SDLX_Init("Simple 2D Game", 640, 480, true);
149.
150.        //game map (background)
151.        map = SDLX_LoadBitmap("map.bmp", false, NULL); //do not use any
    mask for transparent color
152.
153.        //control variables
154.        quit = false;
155.        win = false;
156.        right = false;
157.        left = false;
158.        offset = 0;
159.        ground_level = 380;
160.        end_wait = 0;
161.
162.        //game objects
163.        player.Init((char*)"Player0.bmp", SCREEN_W / 2 - PLAYER_W / 2,
    GROUND_LEVEL-152);      //middle of screen
164.        SDLX_Color mask;
165.        mask.r = mask.b = 0;
166.        mask.g = 128;
167.        mask.a = 255;
168.        player.InitFrame((char*)"Player0.bmp", 0, true, &mask);
169.        player.InitFrame((char*)"Player1.bmp", 1, true, &mask);
170.        player.InitFrame((char*)"Player2.bmp", 2, true, &mask);
171.        enemy.Init((char*)"Enemy.bmp", 0.75 * SCREEN_W, GROUND_LEVEL-
    37);       //between player and end of screen
172.        prize.Init((char*)"Prize.bmp", 1.5 * SCREEN_W, GROUND_LEVEL-
    37);              //off screen
173.    }
174.    void Update()
175.    {
176.        //check if we hit the prize
177.        if (player.Collision(&prize))
178.        {
179.            quit = true;
180.            win = true;
181.            prize.visible = false;
```

```
182.            end_wait = 3000;
183.         }
184.         //check if we hit the enemy
185.         if (player.Collision(&enemy))
186.         {
187.             player.visible = false;
188.             quit = true;
189.             win = false;
190.             end_wait = 3000;
191.         }
192.
193.         //check if there is any movement
194.         //player controlled by keyboard
195.         SDLX_Event e;
196.         //keyboard events
197.         if (SDLX_PollEvent(&e))
198.         {
199.             if (e.type == SDL_KEYDOWN)
200.             {
201.             if (e.keycode == SDLK_ESCAPE)
202.                 quit = true;
203.             //update player
204.             if (e.keycode == SDLK_RIGHT)
205.                 right = true;
206.             if (e.keycode == SDLK_LEFT)
207.                 left = true;
208.             if (e.keycode == SDLK_SPACE && player.jumping == false)
209.             {
210.                 player.jumping = true;
211.                 player.y0 = player.y;
212.                 player.t = 0;
213.                 player.v0 = -25;
214.             }
215.             }
216.             if (e.type == SDL_KEYUP)
217.             {
218.             if (e.keycode == SDLK_RIGHT)
219.                 right = false;
```

```
220.            if (e.keycode == SDLK_LEFT)
221.                left = false;
222.            }
223.        }
224.
225.        //update player
226.        player.vx = 0;
227.        if (right)
228.            player.vx += 3;
229.        if (left)
230.            player.vx -= 3;
231.        player.Move();
232.
233.        //update enemy
234.        enemy.Move();
235.
236.        offset = player.x - (SCREEN_W / 2 - PLAYER_W / 2); //movement
    from original X
237.    }
238.    void Draw()
239.    {
240.        //note the order (last one will be drawn on top)
241.        SDLX_DrawBitmap(map, -1 * offset, 0);
242.        prize.Draw(offset);
243.        enemy.Draw(offset);
244.        player.Draw(offset);
245.
246.        //Update the screen
247.        SDLX_Render();
248.    }
249.    void Run()
250.    {
251.        while (!quit)
252.        {
253.            Update();
254.            Draw();
255.            SDLX_Delay(1);
256.        }
```

```
257.
258.          //done
259.          SDLX_Delay(end_wait);
260.      }
261. };
262.
263. int main(int argc, char* argv[])
264. {
265.     Game game;
266.     game.Run();
267.     return 0;
268. }
```

The syntax for defining inheritance in Python is similar:

```
class X(Y): # class X is a child of class Y
```

Note that when using inheritance, an instance of the child class has an instance of the parent class embedded in it. In other term, when we create a child object, we first create a parent object and then add the child to it[1]. Creating an object means calling its constructor, so the constructor of the parent is called before the constructor of the child class. This allows the parent members inherited by the child to be initialized and ready to use when the child constructor is called. The destructors of these classes, on the other hand, are called in reverse order. We clean up the child class first and then the parent.

The objects in Example 6.10c follow a similar aggregation as in Section 6.1.3, with an instance of class `Game` holding everything while instances of `GameObject` (so `Player` and `Enemy`) each include an Image. Aggregation is an example of relationships between objects and classes, that are essential in object-oriented programming. While aggregation shows how different objects are related based on membership[2], class hierarchies show how classes are related to each other based on inheritance.

---

[1] In C++, we can use a virtual base function that prevents unique parent objects among child classes. Instead of each child having its own instance of the parent, all child instances share the same instance of the parent, as long as they define the base class as virtual. `class X : public virtual Y`

[2] One being a member of the other

*Inheritance allows a child class to extend the parent class by (1) adding new data members, (2) adding new functions, and (3)modifying existing functions.*

## 6.2.1. Class Hierarchies

Inheritance allows us to define classes based on parent/child relationship. Like a family tree, such a relationship between classes results in a hierarchy with fewer, and more basic, classes growing into many classes that share the members of their parents (base classes) but add more members depending on their specific role. We call this tree structure of our classes a **class hierarchy**.

Let's consider a simple library management system as an example. While there are many activities going on in a library, for the sake of simplicity, let's assume we are dealing with two groups of data:

- Users
  - Staff
  - Patron
- Items
  - Printed
    - Book
    - Periodical
  - Digital
    - E-Book
    - Disc

As we saw in previous examples, our first step is to decide which objects and classes we have. Using inheritance, the list above gives us a good starting point for our main classes. We can add a library management system as the main application class that holds everything together. If we look at almost all our past examples, we will see that interactive programs follow a basic structure with an initialization part and a main loop which itself includes receiving input from user to change data (update) and showing results (draw). So, we can define a base class for our LMS as well, that is a template for all application or game classes. Exhibit 6.4 shows a simplified representation of our library program with users and books only.

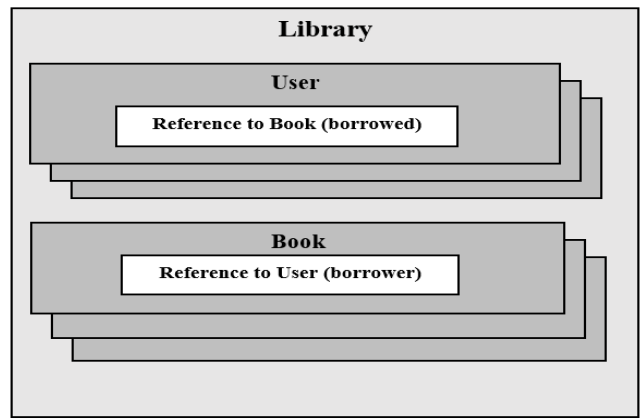

**Exhibit 6.4. Library Management System**

The key to mastering how to use inheritance, even the most complicated class hierarchies, in an algorithm/program is to understand that we can do three things when we create a child class:

1.  Add new data members
2.  Add new functions
3.  Override (change) existing functions

The above list not only shows what we can do in a child class, but also lets us decide if we need a child class. To decide if our program/algorithm should deal with a series of related entities as a single class or create a hierarchy and parent/child, we need to consider the above three possibilities and decide if it is necessary or advantageous enough to do them.

**New Class vs. a Type Data**

The library program has to deal with different types of users. There are actions that all users share, such as logging in and searching for an item. But they also have different abilities. For example, a staff can create new users and get a report on an item or a user. When a staff get s a report on an item, they can see who has borrowed it. When a patron, on the other hand, gets a report, they can only see if the item is available or not. For privacy reasons, they cannot have the borrower's name. Our algorithm for item report can be done in three different ways:

- We can have a single class called `User` and a variable called `type` in it that can be Staff and Patron. The `User` class has a `ItemReport()` function where if the user asking for report is Staff, they get more information. This solution results in:

  - A simpler class hierarchy
  - A more complex report function as it has to deal with user type

- We can have a base User class and a child called Staff. The User class represents the default user type (Patron) and the child class overrides its report function. This gives us more flexibility in defining different behaviours but adds a new class. Also, it assumes Staff have all the members of Patron, plus some new/overridden ones.

- We can have a base User class that is default and has minimum members. Then we derive two classes for Staff and Patron to give maximum flexibility and the most complex hierarchy, where we can add new types in future. Note that in this case, we don't expect the base class to be used for any instances as they all should belong to one of the user types[1].

There is no rule telling us which of these three designs to use. The guidelines, though, are the things we can do in the child class: if adding new members or changing existing ones significantly improves the re-use and maintenance of the program, then defining new child classes is preferred. In this case, here is the list of possible members for our classes:

- User

  - Data:
  - Function:

- Patron

- Staff

---

[1] In Section 6.3, I will show how to create abstract classes that cannot be used for any instances and are only for acting as a base.

From the members list above, note how some objects need to reference each other, particularly, patrons and items they borrow. Such an **object cross-referencing** is very common and there are different ways that the algorithm/program can achieve it. Let's take a look at the code and discuss them.

**Object Cross-referencing**

Example 6.11c shows the library program in C++. For the most part, the code and its algorithm are fairy straight-forward once we decide on the classes, objects, and their members. A good class hierarchy design makes it easy to build the program, as we have seen in other examples of the data-centred approach. Class data members are our data, and the class functions are where we initialize, use, and change them. One thing that is worth specific attention is how patrons and their borrowed items are linked.

In a library management system, we should be able to pick a patron and see a list of their borrowed items. Similarly, we should be able to pick an item and easily find out who has borrowed it. In other terms, we need these objects to reference or somehow access each other. There are three common ways to link an object to another:

1. One object can be a member of the other. For example, each member can have an array of items that they borrowed. This method is good in cases where the member object is only associated with, and used for, the owner object. In our simple game, we have a player character that can shoot bullets. These bullets can be members of the player class. They are not used by any other part of the game and so don't need to have an independent identity.

2. The objects can exist independently, but each has a specific identifier code (for example, an index to an array). These IDs are then used by the other object. In the library system, items do not belong to patrons, so it doesn't make sense to have them as members of patron class as in solution (1) above. Instead, a patron can have an array of integer IDs for the items they borrow. Similarly, each item can have an integer ID for its borrower. The borrower ID can be -1 or any other invalid value when the item is available[1].

3. The objects can exist independently, but each have a pointer/reference to each other. This pointer is implemented based on the programming language we are using. In C/C++, we have explicit pointers. In many other languages, a variable name is by default a pointer (or reference) when used for objects. So when we say something like "`item.borrower = patron`", `item.borrower` is not really an object but a reference to patron. This method is particularly good when there is no unique ID (e.g., no array) but it can also be used in any other case. It is the most versatile solution but has higher complexity.

---

[1] Using a known invalid value to show something is not used is a very common practice. A good example is initializing pointers to NULL in C/C++.

Example 6.11c uses solution #2. We have arrays of items and patrons, so it is easy to identify them using array index. Note that the ID doesn't need to be the array index itself, but something based on it. In our case, IDs start from 1000.

**Example 6.11c**

```
1.   #define MAX_BOOKS    3
2.   #define MAX_PATRONS 3
3.
4.   #define BOOK_TYPE       1
5.   #define PERIODICAL_TYPE 2
6.   #define EBOOK_TYPE      3
7.   #define DISC_TYPE       4
8.
9.   //=================================================================
10.
11.  //Base class for all objects
12.  class Entity
13.  {
14.  protected:
15.      int ID;
16.  public:
17.      Entity()
18.      {
19.          ID = -1;
20.      }
21.      Entity(int i)
22.      {
23.          ID = i;
```

```
24.        }
25.        int GetID()
26.        {
27.            return ID;
28.        }
29.        void Report()
30.        {
31.            cout << "\n";
32.            cout << "ID: " << ID << ", ";
33.        }
34.    };
35.    class Item : public Entity
36.    {
37.    protected:
38.        int type;
39.        int borrower;
40.
41.    public:
42.        Item()
43.        {
44.            //not initialized
45.            type = -1;
46.            borrower = -1;
47.        }
48.        Item(int i, int t) : Item()
49.        {
50.            //initialize
51.            ID = i;
52.            type= t;
53.        }
54.        void Report()
55.        {
56.            Entity::Report();
57.            cout << "Type: " << type << ", ";
58.            cout << "Borrower: " << borrower << ", ";
59.        }
60.        bool Available()
61.        {
```

```
62.          if(borrower == -1)
63.              return true;
64.          else
65.              return false;
66.      }
67.      int GetBorrower()
68.      {
69.          return borrower;
70.      }
71.      bool Borrow(int b)
72.      {
73.          if(borrower == -1 && b != -1)
74.          {
75.              borrower = b;
76.              return true;
77.          }
78.          else
79.              return false;
80.      }
81.      void Return()
82.      {
83.          borrower = -1;
84.      }
85. };
86. class PrintedItem : public Item
87. {
88. protected:
89.      int numPages;
90. public:
91.      PrintedItem()
92.      {
93.          numPages = -1;
94.      }
95.      PrintedItem(int i, int t, int n) : Item(i,t), numPages(n)
96.      {
97.      }
98.      void Report()
99.      {
```

```cpp
100.         Item::Report();
101.         cout << "Number of Pages: " << numPages << ", ";
102.     }
103. };
104. class Book : public PrintedItem
105. {
106. protected:
107.     char author[30];
108. public:
109.     Book()
110.     {
111.         author[0] = NULL;
112.     }
113.     Book(int i, int t, int n, char* name) : PrintedItem(i, t, n)
114.     {
115.         strcpy_s(author, 30, name);
116.     }
117.     void Report()
118.     {
119.         PrintedItem::Report();
120.         cout << "Author: " << author << ", ";
121.     }
122.     void OtherBooks()
123.     {
124.         cout << "other books by the sam author\n";
125.     }
126. };
127. class Periodical : public PrintedItem
128. {
129.
130. };
131. class DigitalItem : public Item
132. {
133.
134. };
135. class Ebook : public DigitalItem
136. {
137.
```

```cpp
138. };
139. class Disc : public DigitalItem
140. {
141.
142. };
143.
144. //====================================================================
145.
146. class User : public Entity
147. {
148. protected:
149.
150. public:
151.     User()
152.     {
153.     }
154.     User(int i) : User()
155.     {
156.         ID = i;
157.     }
158.     void Report()
159.     {
160.         Entity::Report(); //we don't really need this overridden
    function but maybe in future we add members to this class
161.     }
162. };
163. class Staff : public User
164. {
165.
166. };
167. class Patron : public User
168. {
169. protected:
170.     int item;
171. public:
172.     Patron()
173.     {
174.         item = -1;
```

```cpp
175.    }
176.    Patron(int i) : User(i), item(-1)
177.    {
178.    }
179.    bool Available()
180.    {
181.        if(item == -1)
182.            return true;
183.        else
184.            return false;
185.    }
186.    int GetItem()
187.    {
188.        return item;
189.    }
190.    bool Borrow(int i)
191.    {
192.        if(item == -1 && i != -1)
193.        {
194.            item = i;
195.            return true;
196.        }
197.        else
198.            return false;
199.    }
200.    void Return()
201.    {
202.        item = -1;
203.    }
204.    void Report()
205.    {
206.        User::Report();
207.        cout << "Borrowed Item: " << item << ", ";
208.    }
209. };
210.
211. //===================================================================
212.
```

```
213. class Library
214. {
215. protected:
216.     Book* books[MAX_BOOKS];
217.     Patron* patrons[MAX_PATRONS];
218.     int numBooks;
219.     int numPatrons;
220.
221. public:
222.     Library()
223.     {
224.         for (int i = 0; i < MAX_BOOKS; i++)
225.             books[i] = new Book(1000 + i, BOOK_TYPE, rand() % 900 +
    100, (char*)"name");
226.         numBooks = MAX_BOOKS;
227.         for (int i = 0; i < MAX_PATRONS; i++)
228.             patrons[i] = new Patron(1000 + i);
229.         numPatrons = MAX_PATRONS;
230.     }
231.     void Report()
232.     {
233.         cout << "\n\n=========================\n>> BOOKS:\n";
234.         for (int i = 0; i < MAX_BOOKS; i++)
235.         {
236.             if (books[i] != NULL)
237.                 books[i]->Report();
238.         }
239.         cout << "\n\n=========================\n>> PATRONS:\n";
240.         for (int i = 0; i < MAX_PATRONS; i++)
241.         {
242.             if (patrons[i] != NULL)
243.                 patrons[i]->Report();
244.         }
245.         cout << "\n\n=========================\n";
246.     }
247.     void RemoveBook(int id)
248.     {
249.         for (int i = 0; i < MAX_BOOKS; i++)
```

```
250.          {
251.              if (books[i] != NULL)
252.              {
253.                  if (books[i]->GetID() == id)
254.                  {
255.                      if (books[i]->Available() == false)
256.                      {
257.                          int id = books[i]->GetBorrower();
258.                          Patron* p = FindPatron(id);
259.                          p->Return();
260.                          books[i]->Return();
261.                      }
262.                      delete books[i];
263.                      books[i] = NULL;
264.                  }
265.
266.              }
267.          }
268.      }
269.      void AddBook()
270.      {
271.          for (int i = 0; i < MAX_BOOKS; i++)
272.          {
273.              if (books[i] == NULL) //empty location
274.                  books[i] = new Book(1000 + numBooks++, BOOK_TYPE,
         rand() % 900 + 100, (char*)"name");
275.          }
276.      }
277.      Book* FindBook(int id)
278.      {
279.          for (int i = 0; i < MAX_BOOKS; i++)
280.          {
281.              if (books[i] != NULL)
282.              {
283.                  if (books[i]->GetID() == id)
284.                  {
285.                      return books[i];
286.                  }
```

```
287.                }
288.            }
289.        return NULL;
290.    }
291.    void RemovePatron(int id)
292.    {
293.        for (int i = 0; i < MAX_PATRONS; i++)
294.        {
295.            if (patrons[i] != NULL)
296.            {
297.                if (patrons[i]->GetID() == id)
298.                {
299.                    if (patrons[i]->Available() == false)
300.                    {
301.                        int id = patrons[i]->GetItem();
302.                        Book* b = FindBook(id);
303.                        b->Return();
304.                        patrons[i]->Return();
305.                    }
306.                    delete patrons[i];
307.                    patrons[i] = NULL;
308.                }
309.
310.            }
311.        }
312.    }
313.    void AddPatron()
314.    {
315.        for (int i = 0; i < MAX_PATRONS; i++)
316.        {
317.            if (patrons[i] == NULL) //empty location
318.                patrons[i] = new Patron(1000 + numPatrons++);
319.        }
320.    }
321.    Patron* FindPatron(int id)
322.    {
323.        for (int i = 0; i < MAX_PATRONS; i++)
324.        {
```

```
325.            if (patrons[i] != NULL)
326.            {
327.                if (patrons[i]->GetID() == id)
328.                {
329.                    return patrons[i];
330.                }
331.            }
332.        }
333.        return NULL;
334.    }
335.    bool Borrow(int p, int b)
336.    {
337.        Patron* patron = FindPatron(p);
338.        Book* book = FindBook(b);
339.        if (patron != NULL && book != NULL)
340.        {
341.            if (patron->Available() && book->Available())
342.            {
343.                patron->Borrow(b);
344.                book->Borrow(p);
345.                return true;
346.            }
347.        }
348.        return false;
349.    }
350.    bool Return(int p, int b)
351.    {
352.        Patron* patron = FindPatron(p);
353.        Book* book = FindBook(b);
354.        if (patron != NULL && book != NULL)
355.        {
356.            if (patron->GetItem() == b && book->GetBorrower() == p)
357.            {
358.                patron->Return();
359.                book->Return();
360.                return true;
361.            }
362.        }
```

```
363.         return false;
364.     }
365.     void Run()
366.     {
367.         bool quit = false;
368.
369.         while (!quit)
370.         {
371.             int command;
372.             int p;
373.             int b;
374.             cout << "1-Report, 2-Borrow, 3-Return, 0-Exit \n";
375.             cin >> command;
376.             if (command == 0)
377.                 quit = true;
378.             else if (command == 1)
379.             {
380.                 Report();
381.             }
382.             else if (command == 2)
383.             {
384.                 cout << "Enter patron ID: ";
385.                 cin >> p;
386.                 cout << "Enter book ID: ";
387.                 cin >> b;
388.                 if (Borrow(p, b))
389.                     cout << "Dond\n";
390.                 else
391.                     cout << "Sorry\n";
392.                 Report();
393.             }
394.             else if (command == 3)
395.             {
396.                 cout << "Enter patron ID: ";
397.                 cin >> p;
398.                 cout << "Enter book ID: ";
399.                 cin >> b;
400.                 if (Return(p,b))
```

```
401.                     cout << "Dond\n";
402.                 else
403.                     cout << "Sorry\n";
404.                 Report();
405.             }
406.             else
407.                 cout << "Invalid command\n";
408.         }
409.     }
410. };
411. //==================================================================
412.
413. int main()
414. {
415.     Library lib;
416.     lib.Report();
417.
418.     lib.Borrow(1000, 1000);
419.     lib.Report();
420.
421.     lib.RemoveBook(1000);
422.     lib.Report();
423.
424.     lib.AddBook();
425.     lib.Report();
426.
427.     lib.Run();
428.
429.     return 0;
430. }
```

While examples such as our library program have relatively small class hierarchies, more complex applications and games will end up with much bigger hierarchies with many classes and multiple levels. Class hierarchies not only exist in object-oriented programs but also in **class libraries**, i.e., libraries of code provided specifically for object-oriented programs. For example, since C++ does not have graphics built-in abilities, there are many class libraries that add support for programmers. Some notable class libraries for C++ are:

- Microsoft Foundation Classes (MFC), a rather old one that is for Windows-based desktop programs with graphical user interfaces.
- OpenFrameworks, a multi-platform class library that offers various object-oriented programming tools such as graphics.
- Cocos2D, a class library for 2D games

The class hierarchy for these libraries can get quite big and complicated, as expected. But another feature of many class hierarchies is that instead of growing like a tree, they may also fold back branches together. This happens when a class is derived from multiple parents, as I discuss in the next section.

## 6.2.2. Multiple Inheritance

In our library system, users are staff and patrons. These two classes share a base but are distinct. We have no one who is both staff and patron. But what if we want to allow our staff to borrow items? Let's consider another example. In university courses, we have the instructor, students, and teaching assistants, who belong to the staff but are also students. A teaching assistant has some properties that are inherited from student class but also some that are inherited from staff class, parent to instructor, TA, and administrators. Such a class hierarchy where a class can be based on more than one parent demonstrate multiple inheritance shown in Exhibit 6.5.

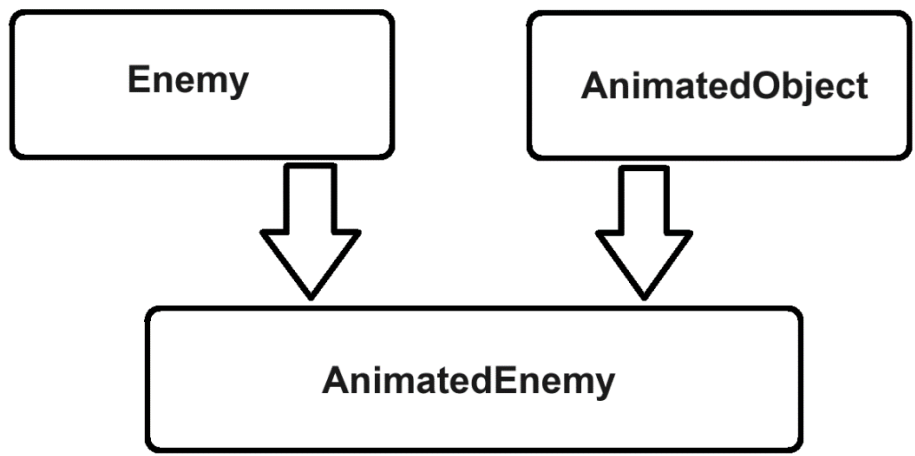

**Exhibit 6.5. Multiple Inheritance**

In Section 5.1.1, I discussed how an animated object can be defined as one with an array of images (frames). To demonstrate the use of multiple inheritance, let's consider our simple 2D game from Section 6.1.1. Imagine we are going to have an animated player and also some animated enemies. Note that I said "some," so we will have some other enemies who are not animated. The first step is to define an `AnimatedObject` class. In C++, the inheritance looks like the code below:

```
class AnimatedObject : public GameObject

{

public:

  //for animation

  SDLX_Bitmap* frames[NUM_FRAMES];

  int currentFrame;

  AnimatedObject()

  {

        currentFrame = 0;

        for (int i = 0; i < NUM_FRAMES; i++)

              frames[i] = NULL;

  }

  void InitFrame(char* fname, int f, bool useMask = false, SDLX_Color*
mask = NULL)

  {

        frames[f] = SDLX_LoadBitmap(fname, useMask, mask);

  }

  void AdvanceAnimation()

  {

        currentFrame++;

        if (currentFrame == NUM_FRAMES)

              currentFrame = 0;

  }

  void Move()

  {
```

```
        GameObject::Move();

        if (vx != 0)

            AdvanceAnimation();

    }

    void Draw(int offset = 0)

    {

        if (visible)

            SDLX_DrawBitmap(frames[currentFrame], x - offset, y);

    }

}
```

The `public` before `GameObject` tells the program that public and protected members of the base class (`GameObject`) will maintain their original access in the derived class[1]. This is called public inheritance and is, by far, the most common form. We can say `protected` or `private`, which mean public and protected members of the base class will become protected and private, respectively, in the derived class.

The `AnimatedObject` class can now be used to define new objects and also child classes of its own. For example, we can define `Player` class as a child of `AnimatedObject`:

```
class Player : public AnimatedObject

{

    //same as Example 6.10c but without animation code

    //that is now inherited from AnimatedObject

}
```

But we can't derive `Enemy` from `AnimatedObject` because not all enemies are animated. Instead, we can define two classes:

```
class Enemy : public GameObject
```

[1] Remember that private members of the base class are not accessible in the derived class, so they cannot become public or protected. They can only be accessed through original base class functions.

```
{
    //same as Example 6.10
}
class AnimatedEnemy : public AnimatedObject, public Enemy
{
public:
    AnimatedEnemy() { }
    void Move()        {
            //from Enemy
            vx = rand() % 5 - 2;
            vy = rand() % 5 - 2;

            //from base
            GameObject::Move();

            //from AnimatedObject
            if (vx != 0)
                    AdvanceAnimation();
    }
}
```

`AnimatedEnemy` uses multiple inheritance, so it inherits from both `Enemy` and `AnimatedObject`., receiving both enemy movement and animation features, respectively. Note that the `Move()` function now needs to do both operations in the base classes: random movement from Enemy and advancing animation from `AnimatedObject`. We could achive this by a simple base class call:

```
Void AnimatedEnemy::Move()
{
    Enemy::Move();           //random values for vx and vy
    AnimatedObject::Move(); //advance animation if vx not zero
```

```
}
```

Since the code for `Move()` in those two base classes call `GameObject::Move()`, calling both of those two functions could result in `GameObject:Move()` being called twice. So, we write the proper code a new overridden `AnimatedEnemy::Move().`

We can continue this process and define multiple types of objects, each with their own features, and then define our player, enemies, and scores as a combination of those features. One important thing to note here is what in object-oriented literature is usually referred to as **the dreaded diamond** problem. Remember that when using inheritance, a child class has a copy of its parent. So, if a Class D is based on classes B and C, which in turn are both based on A, then an instance of D will have two copies of A, with two copies of its members. In the example above, this means that `AnimatedEnemy` will have two sets of x, y, and other `GameObject` members[1].

---

*If Move() changes the coordinates, which pair of x/y will change? Which pair will be used by Draw()?*

*How else would two sets of x and y or other repeated members affect the class AnimatedEnemy?*

---

C++ resolves this issue by declaring the base classes as virtual, so only one copy of them will be used in inheritance.

```
class AnimatedObject : public virtual GameObject

class Enemy : public virtual GameObject
```

---

[1] We call this diamond, because a base class branches into two children and they merge back into a new child, forming a diamond shape.

**Classes with Selected Features**

Animation is only one example of features that we may want to have in our game objects. Here are some other common ones:

- Serializable: In programming, we use the term serialization as the ability of an object to load and save its own status. For example, for a game object, serialization can involve saving and loading location.
- Expiring: We may need some objects to disappear after a certain time.
- Highlighted/Glowing: Some objects may have special visual effects.
- Zoomable: We may need to enlarge some objects for better viewing.

These special features do not correspond to the actual object types we have in the program (player, enemy, score, etc.). They are characteristics that can be assigned to any of our main classes, in any combination. In other terms, these features are **cross-cutting** as they go across the regular class boundaries and can belong to any class[1]. When designing object-oriented programs and algorithms, our classes do not have to correspond to objects we create. Some classes may be used only as base class for others, through single or multiple inheritance. Base classes for cross-cutting features are examples of such base classes. They may not have everything that is needed to be an actual object in the program, but they add features to other classes.

Example 6.12c demonstrates the use of multiple features classes. In this example, the class `AnimatedEnemy` has animation and enemy movement as two features inherited from `AnimatedObject` and `Enemy`, while class `Player` only uses animation. Compared to Example 6.10c, this code has the following changes:

1. Add `AnimatedObject` class
2. Derive `Player` from `AnimatedObject` and remove animation code from it
3. Add `AnimatedEnemy` class
4. Add a new animated enemy to `Game` class
5. Move and draw the new enemy

---

[1] A more systematic way of dealing with cross-cutting concerns is to treat them as new category of modules instead of classes. These new modules are called Aspects and are the basis of Aspect-Oriented Programming.

*Example 6.12c*

```
1.   class Player : public AnimatedObject
2.   {
3.   public:
4.       //for jumping
5.       int y0;       //Y at the start of the jump
6.       int v0;       //VY at the start of the jump
7.       int t;        //jumping time
8.       bool jumping;
9.
10.      Player()
11.      {
12.          y0 = v0 = t = 0;
13.          jumping = false;
14.      }
15.      void Move()
16.      {
17.          AnimatedObject::Move();
18.
19.          int gorund = GROUND_LEVEL-frames[0]->height;  //can change
     along the map
20.
21.          if (jumping)
22.          {
23.              int newy = y0 + v0 * t + GRAVITY_ACCELERATION * t * t / 2;
     //in jumping sy is the "initial" velocity
24.              if (newy > gorund)
25.              {
26.              y = gorund;
27.              jumping = false;
28.              }
29.              else
30.              {
31.              y = newy;
32.              }
33.              t++;
34.          }
```

```
35.      }
36.  };
37.  class Game
38.  {
39.  public:
40.      Player player;
41.      GameObject prize;
42.      Enemy enemy;
43.      AnimatedEnemy animEnemy;
44.      int offset;
45.      SDLX_Bitmap* map;
46.      int ground_level;
47.      int end_wait;
48.      bool quit;
49.      bool win;
50.      bool left;
51.      bool right;
52.
53.      Game() //same as Init()
54.      {
55.          //init SDL library and graphics window
56.          SDLX_Init("Simple 2D Game", 640, 480, true);
57.
58.          //game map (background)
59.          map = SDLX_LoadBitmap("map.bmp", false, NULL); //do not use any
    mask for transparent color
60.
61.          //control variables
62.          quit = false;
63.          win = false;
64.          right = false;
65.          left = false;
66.          offset = 0;
67.          ground_level = 380;
68.          end_wait = 0;
69.
70.          //game objects
```

```
71.        player.Init((char*)"Player0.bmp", SCREEN_W / 2 - PLAYER_W / 2,
    GROUND_LEVEL-152);     //middle of screen
72.        SDLX_Color mask;
73.        mask.r = mask.b = 0;
74.        mask.g = 128;
75.        mask.a = 255;
76.        player.InitFrame((char*)"Player0.bmp", 0, true, &mask);
77.        player.InitFrame((char*)"Player1.bmp", 1, true, &mask);
78.        player.InitFrame((char*)"Player2.bmp", 2, true, &mask);
79.        enemy.Init((char*)"Enemy.bmp", 0.75 * SCREEN_W, GROUND_LEVEL-
    37);        //between player and end of screen
80.        prize.Init((char*)"Prize.bmp", 1.5 * SCREEN_W, GROUND_LEVEL-
    37);              //off screen
81.        //animated enemy
82.        animEnemy.Init((char*)"Player0.bmp", SCREEN_W / 3 - PLAYER_W /
    2, GROUND_LEVEL - 152);        //middle of screen
83.        animEnemy.InitFrame((char*)"Player0.bmp", 0, true, &mask);
84.        animEnemy.InitFrame((char*)"Player1.bmp", 1, true, &mask);
85.        animEnemy.InitFrame((char*)"Player2.bmp", 2, true, &mask);
86.    }
87.    void Update()
88.    {
89.        //check if we hit the prize
90.        if (player.Collision(&prize))
91.        {
92.            quit = true;
93.            win = true;
94.            prize.visible = false;
95.            end_wait = 3000;
96.        }
97.        //check if we hit the enemy
98.        if (player.Collision(&enemy))
99.        {
100.           player.visible = false;
101.           quit = true;
102.           win = false;
103.           end_wait = 3000;
104.        }
```

```
105.
106.          //check if there is any movement
107.          //player controlled by keyboard
108.          SDLX_Event e;
109.          //keyboard events
110.          if (SDLX_PollEvent(&e))
111.          {
112.              if (e.type == SDL_KEYDOWN)
113.              {
114.              if (e.keycode == SDLK_ESCAPE)
115.                      quit = true;
116.              //update player
117.              if (e.keycode == SDLK_RIGHT)
118.                      right = true;
119.              if (e.keycode == SDLK_LEFT)
120.                      left = true;
121.              if (e.keycode == SDLK_SPACE && player.jumping == false)
122.              {
123.                      player.jumping = true;
124.                      player.y0 = player.y;
125.                      player.t = 0;
126.                      player.v0 = -25;
127.              }
128.              }
129.              if (e.type == SDL_KEYUP)
130.              {
131.              if (e.keycode == SDLK_RIGHT)
132.                      right = false;
133.              if (e.keycode == SDLK_LEFT)
134.                      left = false;
135.              }
136.          }
137.
138.          //update player
139.          player.vx = 0;
140.          if (right)
141.              player.vx += 3;
142.          if (left)
```

```
143.            player.vx -= 3;
144.        player.Move();
145.
146.        //update enemy
147.        enemy.Move();
148.        animEnemy.Move();
149.
150.        offset = player.x - (SCREEN_W / 2 - PLAYER_W / 2); //movement
    from original X
151.    }
152.    void Draw()
153.    {
154.        //note the order (last one will be drawn on top)
155.        SDLX_DrawBitmap(map, -1 * offset, 0);
156.        prize.Draw(offset);
157.        enemy.Draw(offset);
158.        animEnemy.Draw(offset);
159.        player.Draw(offset);
160.
161.        //Update the screen
162.        SDLX_Render();
163.    }
164.    void Run()
165.    {
166.        while (!quit)
167.        {
168.            Update();
169.            Draw();
170.            SDLX_Delay(1);
171.        }
172.
173.        //done
174.        SDLX_Delay(end_wait);
175.    }
176. };
```

*Add scaled render as a feature (draw scaled).*

## Inheritance Ambiguity

When deriving a class from more than one base, it is possible to have functions that are implemented, possibly differently, in both base classes. If the child class overrides these functions, calls to that function will go to the child version of the function. But if there is no overridden version, it is not clear which parent's function needs to run.

```
class A {

public:

    A() {

            //some code

    }

    void Test() {

            //some code

    }

}

class B {

public:

    B() {

            //some code

    }

    void Test() {

            //some code

    }

}

class C : public A, public B {
```

```
    //members. No Test()

}

//somewhere in the code

C c;

c.Test();      //ambiguous call
```

in Python, this ambiguity is managed by the order of base classes. So, if the child is defined as `class C(A,B)`, then the members of A have priority over B. In C++, the order of base classes matters for calls to constructors and destructors, but not for other functions. As such, there are other solutions in languages like C++.

The first and most straight-forward solution is to override the common members[1]. In the above example, we add an overridden function `Test()` to class C. Any call to `c.Test()`[2] will result in the new function to be executed.

The second solution is to explicitly mention which version needs to run. Even without multiple inheritance, or with an overridden version, we can use this method to access base class functions. In C++, while accessing a member of an object is done though '.' , accessing a member of a class is done through ':'. Recall that if a class function is defined outside the class, we use `::` to identify it.

```
class A {

    void Test();        //No definition. Only declaration

    //some code

}

void A::Test()  {

}
```

Or to access static members:

---

[1] Note that the ambiguity can exist for data members too. If there is a member called X in both A and B, then c.X is ambiguous.

[2] Or any call to `Test()` from within C

```
class A {

   static int x;

   //some code

}

A::x = 0;
```

Similarly, if we want to identify which version of a member we are using, we can use ':: ' to show the class:

```
//outside class C

c.A::Test();

c.B::Test();

//inside class C

A::Test();

B::Test();
```

In Python, a non-typed language, we can simply say the function from which class needs to be called:

```
class Parent:

   def Test(self):

         #some code

class Child:

   def Test(self):

         #some code

      Parent.Test()      # or super().Test()
```

The child class can call the base version of a function by simply adding the name of the base class (or `super()`) before the function call.

The third solution to inheritance ambiguity is to use virtual base classes, as we saw in Example 6.12c. This solution only works when A and B are themselves based on another class (say X) and the common member is inherited from that original base. Since C has a copy of A and a copy of B, and each of those has a copy of X, C ends up with two copies of X and its members. Defining A and B with virtual base, makes them share a single copy of all X members, so there won't be any ambiguity.

```
class A : virtual public X

class B : virtual public X
```

The fourth possible solution to ambiguity is to provide no implementation for the shared functions in classes A and B. Providing no implementation for a function in C++ is done by a simple declaration of the member function with no code given:

```
class A {

    //other members not shown

    void Test();

};
```

In Python, we use the keyword `pass`:

```
class A:

    def Test():

            pass
```

This method works only when A and B never use the shared function. Having a function that is never used sounds pointless, but it can have a purpose. For example, it can be there to show that the child classes need to define and implement the function. These are commonly called virtual functions and I will discuss them in the next section.

## 6.3. Polymorphism

The examples discussed in Section 6.2 clearly show that when using inheritance, objects have more than one identity. They can act as an instance of the child class and also as the parent. All base functionality is there within the derived class and can be accessed, even if overridden. In object-oriented programming (OOP), the ability of an object to have different identities (act differently) based on class hierarchy is referred to as **Polymorphism**, which is considered the third principle of OOP, after encapsulation and inheritance.

The term polymorphism comes from Greek and means multiple shapes. It is used in many other fields such as biology and psychology with respect to species, cells, and even consciousness that can have different forms. In computer science, polymorphism goes beyond parent-child class relationship and applies to many other cases, even outside object-oriented programming. For example, it can include **function overloading**, i.e., having multiple functions with the same name but different implementation, which are distinguished based on their parameters:

```
void MyFunction(int i);

void MyFunction(int i, int j);

void MyFunction(float f);
```

Function overloading is a type of polymorphism that allows programmers to write different versions of a function based on what parameters are provided[1]. What in OOP is commonly referred to as polymorphism is, in fact, **subtyping**, which is another type of polymorphism.

In the previous section, I showed how, in C++ and Python, we can explicitly decide which version of a function needs to be called. C++, as a typed language, allows further control of polymorphism through the type of object for which the function is called. The correct version of the function is determined by the type of variable for which the function is called. While the type of actual object cannot change, different pointers of different types can point to the same object. For example, if class A and its child class B both have a function named `Test()`, pointers of type `A*` or `B*` can both point to the same instance of B (due to multiple identities) and so they can be used to call `Test()` resulting in `A::Test()` and `B::Test()`, respectively, as shown below:

```
class A {

public:

  A() {

        //some code
```

---

```
    }

    void Test() {

            //some code

    }

}

class B : public A {

public:

    B() {

            //some code

    }

    void Test() {

            //new code for overridden function

    }

}

//somewhere in the code

B b;

b.Test();              // calls B::Test()

B* pb = &b;            // a pointer to b object. Type B*

pb->Test();            // calls B::Test()

A* pa = (A*)&b;        // a pointer to b object. Type A*

pa->Test();            // calls A::Test()
```

While the type of the object `b` is `B` and cannot be changed, A* and B* pointers, `pa` and `pb`, can be defined that point to the same object but have different types[1]. This means that a call to `Test()` will result in different functions to be executed based on the pointer types.

---

[1] Note that these pointer types have to be the original object type and its parents. A pointer to child class cannot be used to point to an object of parent type. While child objects have a copy (identity) of their parents, parent objects cannot possibly have a copy of every child class.

*Polymorphism in OOP means that a child class can have its own identity or the parent's. So, it can still execute the parent's version of a function even after it overrides it.*

### 6.3.1. Virtual Functions and Abstract Classes

To give even more control to programmers in selecting the function, C++ supports **virtual functions**. If a member function is defined as virtual, any call will use the child version even if done using a pointer to parent. The parent version of the functions will only be called when the original object is an instance of the parent class.

```cpp
class A {

public:

  A() {

        //some code

  }

  virtual void Test() {

        //some code

  }

}

class B : public A {

public:

  B() {

        //some code for overridden function

  }

  virtual void Test() {

        //new code

  }
```

```
}
```

Since only the child version of a virtual function will be called, the parent class is allowed to not even have an implementation for a virtual function, assuming that its own instances don't use that function. The parent class can officially set the virtual function is not implemented:

```
virtual void Test() = 0;
```

Such a virtual function is called **pure virtual function** and a class that owns a pure virtual function is called an **abstract class**. Objects cannot be created as instance of an abstract class, which is used solely as base for other classes. In Java and C#, a special type of abstract class, called **interface**, is considered when no function has implementation[1]. In C++, interfaces exist simply as completely abstract classes. Interfaces and other abstract classes are very useful in defining class hierarchies and common elements for child classes. A base `Enemy` abstract class can define all the basic parts of an enemy in a game, when its child classes (`SimpleEnemy`, `IntelligentEnemy`, `AnimatedEnemy`, etc.) represent actual types of enemies.

Another use of abstract classes (and interfaces), as we saw earlier, is in multiple inheritance. Interfaces can define a variety of different features and abilities that an object can have. Then child classes can inherit/implement a combination of those interfaces.

Example 6.13c shows how abstract classes are used to define classes for our simple game of Example 6.12 using multiple inheritance and abstract classes. The code uses abstract base classes `AnimatedObject`, `RandomMover`, and `Enemy`, all based on `GameObject`. It then defines the following classes for the actual objects in the game:

- `Player` based on `AnimatedObject`
- `SimpleEnemy` based on `Enemy` and `RandomMover`
- `AnimatedEnemy` based on `Enemy`, `RandomMover`, and `AnimatedObject`

Note that the classes `AnimatedObject` and `RandomMover` can be the base (parent) for any class that require animation and random movement. They don't implement the `Move()` function which makes them abstract. `Enemy` is the base for all types of enemy objects.

***Example 6.13c***

---

[1] A child class based on an interface is said to "implement that interface."

```
1.  class AnimatedObject : public virtual GameObject  // add virtual to
    avoid "dreaded diamond" problem
2.  {
3.  public:
4.    //all members as before except for Move()
5.    // needs override
6.    virtual void Move() = 0;
7.  };
8.  class RandomMover : public virtual GameObject
9.  {
10. public:
11.     RandomMover() {}
12.     void RandomMove()
13.     {
14.         vx = rand() % 5 - 2;
15.         vy = rand() % 5 - 2;
16.     }
17.     // needs override
18.     virtual void Move() = 0;
19. };
20. class Enemy : public virtual GameObject
21. {
22. public:
23.     GameObject* target;
24.     Enemy()
25.     {
26.         target = NULL;
27.     }
28.     virtual void Move() = 0;
29. };
30. class SimpleEnemy : public Enemy, public RandomMover
31. {
32. public:
33.     SimpleEnemy() {}
34.     void Move()
35.     {
36.         RandomMove();
37.         GameObject::Move();
```

```
38.      }
39.   };
40.   class AnimatedEnemy : public Enemy, public AnimatedObject, public
      RandomMover
41.   {
42.   public:
43.      AnimatedEnemy()
44.      {
45.      }
46.      void Move()
47.      {
48.          RandomMove();
49.          GameObject::Move();
50.          if (vx != 0)
51.              AdvanceAnimation();
52.      }
53.   };
```

## 6.3.2. Rendering Multiple Types of GameObjects

When discussing our simple 2D game, I mentioned that games commonly used specialized classes such as `Physics` or `Render` to perform tasks that are common among all game objects. For example, regardless of what role an object has in the game, the same physics laws should be applied to it, or it will be drawn the same way (an image at a location). Also, those common tasks can have many complexities, options, and details that are (1) the same for all objects and (2) too much to have in the `GameObject` class. We may even need to change them independent of the game objects or use a library to perform them. Most game engines like Unity and Unreal, for example, have built-in classes for applying physics and drawing objects in different styles. All the programmer needs to do is to define the game objects.

Let's consider removing the `Draw()` function from `GameObject` and designing a `Render` class that draws all the objects in our simple game. This not only simplifies `GameObject`, but also allows further changes and complex options for drawing, all done in one place, independent of the `GameObject` and `Game` classes. Later, I will show an example of such changes by adding visual effects in drawing. For now, let's focus on getting the basic `Render` class done. Our initial attempt in defining a `Render` class may look like this:

```
class Render {

public:

    //Functions. Implementation not shown

    Render();

    void DrawPlayer(Player p);

    void DrawEnemy(Enemy e);

};
```

This design has two issues:

- It has specific functions for each class of object, which means `Render` class depends on the design of our `GameObject` child classes. In practice the whole point of having a Render class is to be independent and re-usable in any game with any classes as long as they are derived from `GameObject`.
- It passes the full objects instead of a reference, which takes time and memory. It also means changes to the object inside the function are not applied to the original object as the parameter is a copy of the object being drawn. In our case though, this is ok as draw is not changing the objects.

Both of these issues can be resolved by passing the objects as pointer to `GameObject` (base class). Due to polymorphism, all child objects can act as the parent, which has all the information needed for drawing (shape and location)[1].

```
class Render {

public:

    Render() {        //no initialization needed

    }

    void Draw(GameObject* obj)

    {
```

---

[1] Note that we introduced offset to support side scrolling in previous examples. It's been removed for simplicity in these examples.

```
        SDLX_DrawBitmap(obj->shape, obj->x, obj->y);

    }

};
```

Our `Game::Draw()` will call `Render::Draw()` and pass the address of objects to it for drawing. It will look something like this:

```
void Game::Draw()

{

        render.Draw(&prize);

        render.Draw(&enemy);

        render.Draw(&player);

}
```

This solution seems to have solved the problem except that it assumes `obj->shape` can be accessed (is public) and can always show the current shape of a game object. As I explained earlier, one of the key points of encapsulation is data hiding. Having a public variable like shape, assumes that the object uses a fixed shape and the outside code can access it. A game object class may need to have a dynamic shape (as in `AnimatedObject`) or require certain limitations on how to read or write the shape (private or protected). A better solution is to have a public `GetShape()` function in all game object classes that returns the current shape regardless of how the object manages its shape internally. Here are some examples of this function in game objects (note that they use a reference/pointer):

```
class GameObject {

    Image shape;

public:

    virtual void Image* GetShape {

        return shape;

    }

    //other members not shown

};

class AnimatedObject : public GameObject {
```

```
    Image frames[5];

    int current Frame;

public:

    virtual void Image* GetShape {

            return frames[currentFrame];

    }

    //other members not shown

};

//in Render::Draw()

            SDLX_DrawBitmap(obj->GetShape(), obj->x, obj->y);
```

Note that `GetShape()` is defined as `virtual`. If we don't do that, `Render::Draw()` will end up running the parent version of the `GetShape()`, which is not correct. The beauty of this solution is that as far as Render class is concerned, all objects are passed to it as GameObject. `Render` class doesn't need to know what child classes we have in our program based on `GameObject`. All objects are treated the same way by calling `GetShape()` that hides the details of how the object shape is implemented and decided. Being virtual guaranties that despite being passed as `GameObject`, the objects will end up using their own child implementation of `GetShape()`.

Example 6.14c shows the full code for the simple 2D game with Render class.

***Example 6.14c***

```
1.   //The base code is 6.13
2.   //GameObject and AnimatedObject changed as shown above
3.   class Render
4.   {
5.   public:
6.       Render()
7.       {       //no initialization needed
8.       }
9.       void Draw(GameObject* obj, int offset = 0)
10.      {
```

```
11.         SDLX_DrawBitmap(obj->GetShape(), obj->x - offset, obj->y);
12.     }
13. };
14. class Game
15. {
16. public:
17.     //for Example 6.14
18.     Render render;
19.
20.     void Draw()
21.     {
22.         //note the order (last one will be drawn on top)
23.         SDLX_DrawBitmap(map, -1 * offset, 0);
24.         render.Draw(&prize, offset);
25.         render.Draw(&enemy, offset);
26.         render.Draw(&animEnemy, offset);
27.         render.Draw(&player, offset);
28.         //prize.Draw(offset);
29.         //enemy.Draw(offset);
30.         //animEnemy.Draw(offset);
31.         //player.Draw(offset);
32.
33.         //Update the screen
34.         SDLX_Render();
35.     }
36. //everything else is the same as 6.13
37. }
```

## Visual Effects and Image Filters

Drawing can happen by simply using the 2D shape of the object as we did in previous examples, or it can have various **visual effects**. Let's add some image filters (Section 6.1.6, Example 6.9c) to the drawing process as an example of image processing and using virtual functions. To do so, we assume that `Render::Draw()` function has another parameter which specifies what effect to add. Our program has a base `Effect` class, and we can add as many child classes as we want. Note that a visual effect can be more than an image filter. When we apply a filter to an image, only the image pixels change. A visual effect, on the other hand, can change the surrounding area as well, as happens in glow or shine effect. To be able to extend to such effects, we call the base class `Effect`, instead of filter. But this example does not cover anything beyond image filters.

The overall flow of rendering includes the following steps:

- Create all `Effect` objects.
- Create a `Render` object.
- Create all game objects.
- Start the main loop in `Game` class.
- In the draw part of the loop:

  - Decide what effect to apply to each object.
  - Call `Render::Draw()` with object and effect.
  - In `Render::Draw()`, apply the effect and then draw the new image.

Since the `Effect` and `Render` classes have only one instance each, alternatively we can define their related functions as static and call them directly without creating an object. But I don't do that in this example, as creating objects allows future extensions, such as multiple renders and effects with different settings and stored information.

Example 6.15c shows the code for our game with visual effects. The implementation of effect classes is based on the `Filter` class of Example 6.9c. Note how `Render::Draw()` uses virtual functions to apply the correct effect even though the effect object is passed as base (an abstract class).

***Example 6.15c***

```
1.    class Effect  //abstract class as base
2.    {
3.    public:
4.        Effect()
5.        {
6.        }
7.        virtual void Apply(SDLX_Bitmap* img) = 0; //pure virtual
8.    };
9.    class GrayEffect : public Effect
```

```
10.  {
11.  public:
12.      GrayEffect()
13.      {
14.      }
15.      virtual void Apply(SDLX_Bitmap* img)
16.      {
17.          //image dimensions
18.          int w = img->width;
19.          int h = img->height;
20.
21.          //color
22.          SDLX_Color c;
23.
24.          //loop
25.          for (int i = 0; i < w; i++)
26.              for (int j = 0; j < h; j++)
27.              {
28.              SDLX_GetPixel(img, i, j, &c);
29.              if (SDLX_IsPixelMask(img, &c) == false)
30.              {
31.                  int gray = (c.b + c.g + c.r) / 3;
32.                  c.r = c.g = c.b = gray;
33.                  SDLX_PutPixel(img, i, j, &c);
34.              }
35.              }
36.      }
37.  };
38.  class BlurEffect : public Effect
39.  {
40.  public:
41.      BlurEffect()
42.      {
43.      }
44.      virtual void Apply(SDLX_Bitmap* img)
45.      {
46.          //similar to Gray and Example 6.9c
47.      }
```

```
48.  };
49.  class Render
50.  {
51.  public:
52.      Render()
53.      {       //no initialization needed
54.      }
55.      void Draw(GameObject* obj, int offset = 0)
56.      {
57.          SDLX_DrawBitmap(obj->GetShape(), obj->x - offset, obj->y);
58.      }
59.      void Draw(GameObject* obj, Effect* effect, int offset = 0)
60.      {
61.          //create and use a clone to not affect the original image
62.          SDLX_Bitmap* bmp = SDLX_CloneBitmap(obj->GetShape());
63.          //apply the effect to the clone
64.          effect->Apply(bmp);
65.          //draw clone
66.          SDLX_DrawBitmap(bmp, obj->x - offset, obj->y);
67.          //delete clone
68.          SDLX_DestroyBitmap(bmp);
69.      }
70.  };
71.  class Game
72.  {
73.  public:
74.      Render render;
75.      NoEffect noEffect; //empty effect for being consistent
76.      GrayEffect grayEffect;
77.      BlurEffect blurEffect;
78.
79.      void Draw()
80.      {
81.          //note the order (last one will be drawn on top)
82.          SDLX_DrawBitmap(map, -1 * offset, 0);
83.
84.          render.Draw(&prize, &noEffect, offset);
85.          render.Draw(&enemy, &grayEffect, offset);
```

```
86.        render.Draw(&animEnemy, &blurEffect, offset);
87.        render.Draw(&player, &noEffect, offset);
88.
89.        //Update the screen
90.        SDLX_Render();
91.    }
92. //everything else the same as before
93. };
```

Some other things to pay attention to in this code are:

- The `Apply()` function in `Effect` class only receives a shape, not the whole object.
- Applying the effect requires bypassing any mask (transparent) point. We call `SDLX_IsPixelMask` to check if the image is using a mask colour and if yes, if the current pixel has the mask value (line 29).
- `Render` class applies the filter (effect) to a clone of the object image (line 62). By using a clone, we save the original image in case we need it later.
- The `Game` class can use different effects for different objects (lines 84 to 87). For consistency of the code, we are suing a NoEffect class that applies no effect to the image, i.e., `Apply()` simply returns. Alternatively, we could call `Render::Draw()` with no filter.

## 6.4. Object-Oriented Algorithm Design

Looking at the examples in this chapter, we can see that the data-centred approach (Section 4.2.1) is followed but with some variations. Although some of my examples focused on specific aspects of this approach, designing algorithms that use objects and classes generally follows this modified data-centred process:

1. Visualize your program.
2. Identify the main objects in the program.
3. Identify the main classes, based on common structure of objects.

   - For example, if we have 10 enemies that all behave the same, we define one Enemy class.

4. For each class, identify data and functions.

- Functions can be necessary operations on their own (such as `Apply` in `Effect` class) or processing the class data (such as `AdvanceAnimation` in `AnimatedObject`).
- For each function of each class, identify any extra information that is needed in addition to class members and define them as local variables. Then ask three how questions (how to initialize, use, and change) to help define the algorithm in that function.

5. See if any global functions and variables are needed to bring everything together. The `main()` function is just one example.

Choosing objects and classes is the basis of this process and rather a tricky task. There is no rule on what the classes and their members should be, and different software designers make different design choices. The following options are good guidelines to help in deciding what can be a good candidate for a class:

- Entities that correspond to physical objects, for example player, enemy, house, book, and user.
- Collections of related data distinguished from other parts of the program, for example game, transaction, and database.
- Collections of related operations distinguished from other parts of the program, for example Filter, Render, Effect, and Physics.

While there is no unique design that works, with more practice, it will become easier to identify classes that result in an effective and efficient design. Ease of maintenance (debugging, improving, and extending), performance (speed and resources), and re-usability are among the factors that we consider when deciding which design is superior.

## Highlights

- Encapsulation can be used to define objects and classes, which are modules that combine related data and code.
- The most common way of initializing an object is in a special member function of the class called constructor that is called automatically when the object is created.

- Information hiding can involve making some members of the class private, so their definition and implementation is not accessible to outside code, which will become independent of how these members are created.

- To allow all instances of a class to share a data item, most object-oriented languages define two types of members: class members (vs. instance members) in Python and static members (vs. regular members) in C++ are examples of shared data between class instances.

- Inheritance allows a child class to extend the parent class by (1) adding new data members, (2) adding new functions, and (3) modifying existing functions.

- Polymorphism in OOP means that a child class can have its own identity or the parent's. So, it can still execute the parent's version of a function even after it overrides it.

- Object-oriented algorithm design is an extension of the data-centred approach that starts by identifying objects and classes.

# 7. Where do we go from here?

The goal of this book is to have a more methodical approach to algorithmic thinking, as the basis of software development. I started with this fundamental notion that programming and software design is more about thinking algorithmically than learning and using different programming languages. Once we train ourselves to think (and describe what the program must do) in terms of clear steps, writing the code is much easier. There is no single formula for thinking algorithmically and designing algorithms. But throughout the book, I introduced a toolbox that will facilitate the process. It included four groups of tools:

- The data-centred design approach, including its object-oriented extension. This is our basic method of designing algorithms.
- Selection, iteration, and modularization (both code and data). These are programming constructs that we use as the building block of any program/algorithm.
- Decomposition, pattern recognition, and abstraction. These are problem-solving skills that we use to come up with our algorithms.
- Many algorithmic patterns such as accumulation, search, sort, command processor, game loop, and object-oriented ones like information hiding and inheritance. These are the bases of many algorithms we commonly use.

I also used a variety of visualization and presentation tools for algorithms, such as pseudo-code, flow chart, class hierarchy, and object aggregation models. There are many more ways to describe the behaviour of a computer program. For example, Interaction Diagrams (Exhibit 7.1) are helpful to show how objects interact with each other. Frameworks such as Unified Modeling Language (UML) offer a wide range of tools for modeling software systems.

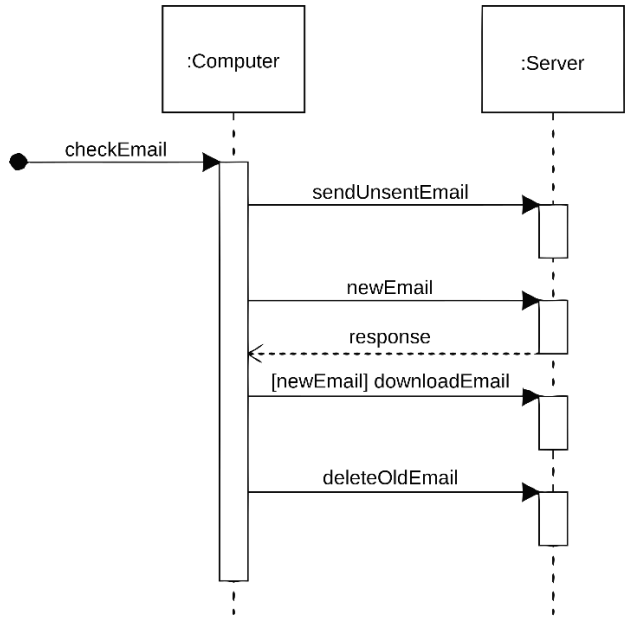

**Exhibit 7.1. Interaction Diagram showing the sequence of communication**

As you move forward with gaining more software design experience, you will improve your skills to use algorithmic thinking tools properly and effectively, and also explore more algorithmic patterns that can simplify your design. The human brain is structured to recognize and use patterns in order to make our processes more efficient. Once you are familiar with a particular pattern, you don't need to think about how to deal with it every time you encounter a similar one. A row of chairs in front of a desk with a board behind it is a common pattern for a classroom. Upon seeing this pattern, you quickly decide how to react to the familiar classroom environment. There is a chance that we are mistaken, but in the majority of cases, we are right, and the pattern-based approach saves us time. If the pattern is related to a predator, it may save us our lives. Algorithmic patterns (and in general, **Software Design Patterns**) play a similar role in making our software design process more efficient. In addition to personal experience, there are many good sources that can help you learn about these patterns such as:

- Software Design Patterns
- Game Design Patterns

Last but not least, larger software projects aim at systems rather than single programs. Such software systems consist of multiple programs and parts that are connected. These connections, similar to individual programs, follow patterns that we refer to as **Software Architecture**. Input-Process-Output (IPO) is a common architecture that every programmer has used even without noticing. In IPO, the program is divided into three parts: modules that deal with receiving input, those that process the information, and finally modules that display or otherwise present the output. IPO leads to layered architectural patterns such as web applications with front end, back end, and database. Other common architectural patterns include Client-Server and Model-View-Controller (MVC), as shown in Exhibits 7.2 and 7.3.

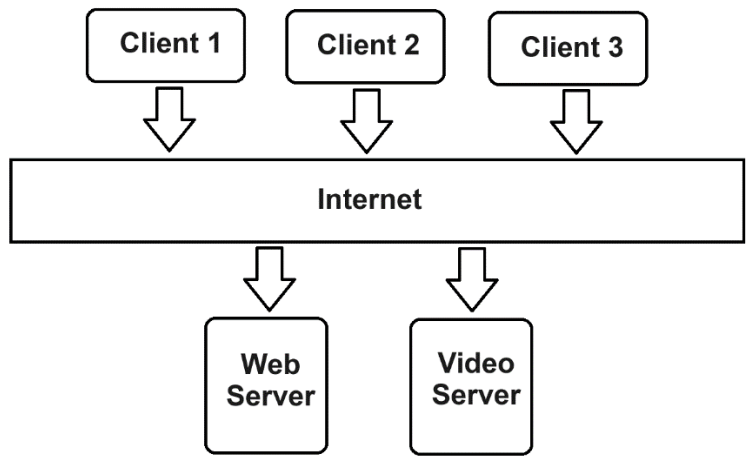

**Exhibit 7.2. Client-Server Architecture for online services**

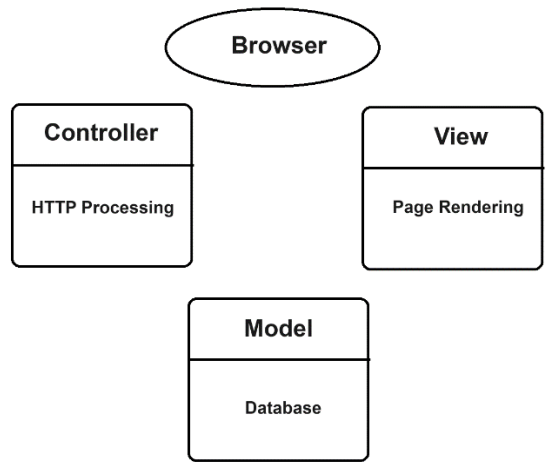

**Exhibit 7.3. MVC Architecture for a web-based application**

Understanding various presentation tools and software design and architectural patterns are further steps once you are more experienced with algorithmic thinking and have practiced common algorithmic patterns like those I discussed and many more. Remember that algorithmic thinking is not only about step-by-step action. It also involves arranging the data and code into modules for ands upon which those steps are taken.

So, remember to never think you know enough. Keep learning and enjoy thinking like a programmer!

# Appendix A. C/C++ Review

This appendix is a very quick review of C/C++ programming and related resources. For a more in-depth discussion of C/C++, see the followings:

- Anyone Can Code: Art and Science of Logical Creativity, Ali Arya
  - https://ali-arya.com/anyonecancode
- https://www.tutorialspoint.com/cprogramming
- http://www.cplusplus.com

Please find all the book examples with all their details, ready to compile and run, at https://ali-arya.com/anyonecancode. **Download ACC2-Examples.zip and look for README.txt in each of three folders: C for standard C/C++, SDL for graphics C/C++, and Python.**

## A.1. Installation

There are different tools that allow you to do C/C++ programming. All examples in this book are using Microsoft Visual Studio for Windows operating system, which is an Integrated Development Environment (IDE) and has a free version you can find and learn about here:

https://visualstudio.microsoft.com/vs/community

There are other tools for Windows, Mac, and Linux and also online C/C++ compilers such as:

https://www.onlinegdb.com/online_c_compiler

## A.2. Basic Programming

C is a typed programming language, meaning that data is defined as variables with a type, a name, and an optional value:

```
int a;
int b = 0;
```

```
float c;

char d;
```

Defined variables can be accessed in the code using their name. Code is defined as functions using a return type, a name, and am optional set of parameters. Each program needs to have at least a function called `main`. Data and code defined externally (outside current source file) have to be specified using `#include` directive.

Functions are accessed (called) using their names and passing parameters, if any is needed. If a function has a return value, it can be used as any other data. Comments are identified using `//` (the rest of the line) or `/*` and `*/` as a pair for a block of comments.

```
#include <stdio.h>   //for printf() function

void foo()

{

    int a = 0;

    int b = 2;

    printf("%d",a+b);

}

int bar(int a, int b)

{

    return a+b;

}

/* Required function

Program starts here */

int main()

{

    foo();

    int x = bar(7,8);

    return 0;

}
```

Keywords `if` and `switch/else` are used for selection, and `for` and `while` are used for iteration, as described in Chapter 4. Keyword struct is used to create a user-defined (complex) type while arrays are defined with [ ], as described in Chapter 5:

```
struct Student          //new type
{
   int grade1;
   int grade2;
};
Student s1;             //new variable
s1.grade1 = 0;          //access members
int grades[30];         //array
grades[0] = 0;          //access members
```

## A.3. Object-Oriented Programming

C++ is an object-oriented extension to C, meaning that it is almost backward compatible. The main difference is the introduction of classes, which are described in Chapter 6.

## A.4. Graphics

Standard C and C++ don't support graphics. Non-standard libraries have been introduced to add graphics capability to C/C++ programs. This book uses SDL (https://www.libsdl.org), which is not object-oriented and so works for both C and C++. Another popular one is OpenFrameworks (https://openframeworks.cc), which is object-oriented and so can only be used in C++ programs using classes. The first book in Anyone Can Code series (https://ali-arya.com/anyonecancode) describes both of them. The corresponding websites have the installation files and documents.

When using SDL, start by installing it from the website, then use my sample project or create a new empty Windows Desktop project and add the source files to it. Make sure the project properties are set correctly:

- Under Visual C++ Directories add SDL lib and include folders for External Include Directories and Library Directories

- Under Linker/Input, add sdl2.lib, dl2main.lib, and sdl2test.lib for Additional Dependencies.

Section 4.1.2 of this book introduces the basic structure of SDL programs and SDLX extension that I have developed to make programming with SDL easier. To use SDLX two files need to be added to your program: SDLX.cpp and SDLX.h.

# Appendix B. Python Review

This appendix is a very quick review of Python programming and related resources. For a more in-depth discussion, see the followings:

- Anyone Can Code: Art and Science of Logical Creativity, Ali Arya
    - https://ali-arya.com/anyonecancode
- https://www.tutorialspoint.com/python
- https://www.w3schools.com/python

Please find all the book examples with all their details, ready to compile and run, at https://ali-arya.com/anyonecancode. **Download ACC2-Examples.zip and look for README.txt in each of three folders: C for standard C/C++, SDL for graphics C/C++, and Python.**

## B.1. Installation

Using Python requires having its runtime environment installed on your computer, or accessing an online Python environment like **Replit** (https://replit.com/languages/python3). If you are using an IDE like Visual Studio, it may come with Python support. If not, you need to install Python (https://www.python.org/downloads). Once you download and install Python, you can set your IDE to use it:

https://learn.microsoft.com/en-us/visualstudio/python/installing-python-support-in-visual-studio?view=vs-2022

To test if you have Python installed, open a terminal (command prompt) and type Python (optionally, with a Python filename). You should see the Python command line where you can enter and run Python programs without any IDE.

## B.2. Basic Programming

Programming logic in Python is not much different from C/C++ but the syntax is quite different. Some of the main differences are listed here:

- Python is and interpreted language, not compiled like C/C++. This means any single line can be executed individually.
- There is no ; at the end of lines.
- Variables and functions have no explicit type. Don't mistake this with not having types. We just don't declare any variable or assign any type. The first time you use a variable name, Python runtime creates the variable of the right type based on the value you assign.
- Any line of code that is followed by a block of related code ends with a :. The related block is then defined by indention instead of { }. This can be very confusing for beginner Python programmers.
- There is no need to have functions, even a main. The program starts at the first line that is not part of a defined function. If you define functions, they won't run until they are called, just like in C/C++.
- Arrays (or lists) in Python are always dynamic and there are standard functions for processing them. In general, Python comes with a much larger set of built-in functions compared to C/C++.
- Python doesn't have pointer types. The language decides which variables are reference type and which ones are value type.

## B.3. Object-Oriented Programming

Python keyword `class` is used to create records and classes, same as C/C++ `struct` and `class`. As described in Sections 5.2 and Chapter 6, Python supports OOP, although not quite like C++.

## B.4. Graphics

Similar to C/C++, standard Python doesn't have graphics support. Libraries like PyGame add such capability. In this book, I use a very simple library called graphics.py. It is a single file that needs to be added to your projects and imported in your source files. It allows creating a graphic window and performing simple operations. It is written by John Zelle for use with the book "Python Programming: An Introduction to Computer Science" (Franklin, Beedle & Associates). The package is a wrapper around **Tkinter** library and should run on any platform where **Tkinter** is available. See Section 4.1.2 for an introduction.